\documentclass[fleqn,usenatbib]{mnras}

\usepackage{newtxtext,newtxmath}
\usepackage[T1]{fontenc}
\usepackage[utf8x]{inputenc}

\DeclareRobustCommand{\VAN}[3]{#2}
\let\VANthebibliography\thebibliography
\def\thebibliography{\DeclareRobustCommand{\VAN}[3]{##3}\VANthebibliography}

\usepackage{graphicx}	% Including figure files
\usepackage{amsmath}	% Advanced maths commands
\usepackage{amssymb}	% Extra maths symbols
\usepackage{hyperref}	% Extra maths symbols

\newcommand{\tbf}{\textbf}
\newcommand{\ti}{\textit}

\newcommand{\bea}{\begin{eqnarray}}
\newcommand{\be}{\begin{equation}}
\newcommand{\ben}{\begin{enumerate}}
\newcommand{\bi}{\begin{itemize}}
\newcommand{\eea}{\end{eqnarray}}
\newcommand{\ee}{\end{equation}}
\newcommand{\ei}{\end{itemize}}
\newcommand{\een}{\end{enumerate}}
\newcommand{\nn}{\nonumber}

\newcommand{\matC}{\mathbf C}

\def\bfk{\mbox{\bf k}}

\newcommand{\like}{L}
\newcommand{\prob}{P}
\newcommand{\probr}{P_r}

\newcommand{\pco}{\vek p_\mr{c}}
\newcommand{\pnu}{\vek p_\mr{n}}

\newcommand{\D}{\vek D}

\newcommand{\M}{\vek M}

\newcommand{\om}{\Omega_\mr m}
\newcommand{\omb}{\Omega_\mr b}
\newcommand{\sig}{\sigma_8}
\newcommand{\ns}{n_s}
\newcommand{\w}{w_0}
\newcommand{\wa}{w_a}

\newcommand{\mr}{\mathrm}

\newcommand{\vek}{\mathbf}

\title[Cosmology with WFIRST - Multi-Probe Strategies]{Cosmology with the Wide-Field Infrared Survey Telescope - Multi-Probe Strategies}

\author[Tim Eifler et al.]{Tim Eifler$^{1,2}$\thanks{E-mail:timeifler@arizona.edu}, 
Hironao Miyatake$^{2,3,4,5}$,
Elisabeth Krause$^{1,6}$,
Chen Heinrich$^{2}$,
\newauthor
Vivian Miranda$^{1}$,
Christopher Hirata$^{7}$, 
Jiachuan Xu$^{1}$,
Shoubaneh Hemmati$^{3}$, 
\newauthor
Melanie Simet$^{2,8}$,
Peter Capak$^{9}$,
Ami Choi$^{7}$,
Olivier Dor\'e$^{2,10}$,
Cyrille Doux$^{11}$,
\newauthor
Xiao Fang$^{1}$,
Rebekah Hounsell$^{11}$,
Eric Huff$^{2}$,
Hung-Jin Huang$^{1}$,
Mike Jarvis$^{11}$,
\newauthor
Dan Masters$^{2}$, 
Eduardo Rozo$^{6}$, 
Dan Scolnic$^{13}$,
David N. Spergel$^{14,15}$,
Michael 
\newauthor
Troxel$^{13}$,
Anja von der Linden$^{16}$,
Yun Wang$^{9}$,
David H. Weinberg$^{7}$,
Lukas Wenzl$^{17}$,
\newauthor
Hao-Yi Wu$^{7}$
\\
$^{1}$ Department of Astronomy/Steward Observatory, University of Arizona, 933 North Cherry Avenue, Tucson, AZ 85721-0065, USA \\
$^{2}$ Jet Propulsion Laboratory, California Institute of Technology, Pasadena, CA 91109, USA\\
$^{3}$ Institute for Advanced Research, Nagoya University, Nagoya 464-8601, Japan\\
$^{4}$ Division of Physics and Astrophysical Science, Graduate School of Science, Nagoya University, Nagoya 464-8602, Japan \\
$^{5}$ Kavli IPMU~(WPI), UTIAS, The University of Tokyo, Chiba 277- 8583, Japan\\
$^{6}$ Department of Physics, University of Arizona, 1118 E Fourth Str, AZ 85721, USA \\
$^{7}$ Center for Cosmology and AstroParticle Physics, The Ohio State University, 191 West Woodruff Avenue, Columbus, Ohio 43210, USA\\
$^{8}$ University of California Riverside, 900 University Ave, Riverside, CA 92521, USA \\
$^{9}$ IPAC, California Institute of Technology, Pasadena, CA 91125, USA\\
$^{10}$ California Institute of Technology, 1200 E. California Blvd., Pasadena, CA 91125, USA\\
$^{11}$ Department of Physics and Astronomy, University of Pennsylvania, Philadelphia, PA 19104, USA \\
$^{12}$ McWilliams Center for Cosmology, Department of Physics, Carnegie Mellon University, Pittsburgh, PA 15213, USA \\
$^{13}$Department  of  Physics,  Duke  University  Durham,  NC 27708, USA\\
$^{14}$Center for Computational Astrophysics, Flatiron Institute, NY NY 10010, USA\\
$^{15}$Department of Astrophysical Sciences, Princeton University, Princeton NJ 08544, USA\\
$^{16}$Department of Physics and Astronomy, Stony Brook University,Stony Brook, NY 11794, USA\\
$^{17}$Department of Astronomy, Cornell University, Ithaca, NY 14853, USA
}

\date{Accepted XXX. Received YYY; in original form ZZZ}

\pubyear{2015}

\begin{document}
\label{firstpage}
\pagerange{\pageref{firstpage}--\pageref{lastpage}}
\maketitle

\begin{abstract}
We simulate the scientific performance of the Wide-Field Infrared Survey Telescope (WFIRST) High Latitude Survey (HLS) on dark energy and modified gravity. The 1.6 year HLS Reference survey is currently envisioned to image 2000 deg$^2$ in multiple bands to a depth of $\sim$26.5 in Y, J, H and to cover the same area with slit-less spectroscopy beyond z=3. The combination of deep, multi-band photometry and deep spectroscopy will allow scientists to measure the growth and geometry of the Universe through a variety of cosmological probes (e.g., weak lensing, galaxy clusters, galaxy clustering, BAO, Type Ia supernova) and, equally, it will allow an exquisite control of observational and astrophysical systematic effects. In this paper we explore multi-probe strategies that can be implemented given WFIRST's instrument capabilities. We model cosmological probes individually and jointly and account for correlated systematics and statistical uncertainties due to the higher order moments of the density field. We explore different levels of observational systematics for the WFIRST survey (photo-z and shear calibration) and ultimately run a joint likelihood analysis in N-dim parameter space. We find that the WFIRST reference survey alone (no external data sets) can achieve a standard dark energy FoM of >300 when including all probes. This assumes no information from external data sets and realistic assumptions for systematics. Our study of the HLS reference survey should be seen as part of a future community driven effort to simulate and optimize the science return of WFIRST. 

\end{abstract}


% \renewcommand{\thefootnote}{\arabic{footnote}}
% \setcounter{footnote}{0}

%%%%%%%%%%%%%%%%%%%%%%%%%%%%%%%%%%%%%%%%%%%%%%%%%%%%%%%%%%%%%%%%%%%%%%%%%%%%%%
%%%%%%%%%%%%%%%%%%%%%%%%%%%%%%%%%%%%%%%%%%%%%%%%%%%%%%%%%%%%%%%%%%%%%%%%%%%%%%
\section{Introduction}
\label{sec:intro}
%%%%%%%%%%%%%%%%%%%%%%%%%%%%%%%%%%%%%%%%%%%%%%%%%%%%%%%%%%%%%%%%%%%%%%%%%%%%%%
%%%%%%%%%%%%%%%%%%%%%%%%%%%%%%%%%%%%%%%%%%%%%%%%%%%%%%%%%%%%%%%%%%%%%%%%%%%%%%

In the current $\Lambda$CDM paradigm cosmic acceleration is caused by the $\Lambda$-term in the Einstein field equations \citep{ein17}. In terms of physical interpretation, $\Lambda$ can be associated with the Universe's geometry or it can describe a new energy component of the universe, so-called dark energy.  In 1998 two teams \citep{rfc98,pag99} measured the energy density of $\Lambda$, $\Omega_\Lambda$, to be consistent with a value close to 0.7. To date, the science community lacks a convincing physics model for cosmic acceleration; constraining its properties and testing it against alternative theories is one of the main science drivers of ongoing and future surveys.

Major progress on this topic is made by the current (Stage 3) generation of photometric surveys, such as Kilo-Degree Survey (KiDS\footnote{http://www.astro-wise.org/projects/KIDS/}), the Hyper Suprime Cam (HSC\footnote{http://www.naoj.org/Projects/HSC/HSCProject.html}), the Dark Energy Survey (DES\footnote{www.darkenergysurvey.org/}) and spectroscopic surveys, such as the Baryon Oscillation Spectroscopic Survey (BOSS\footnote{http://www.sdss3.org/surveys/boss.php}). These low redshift constraints of the ($\Lambda$CDM) model can be contrasted with CMB measurements from the early Universe made e.g., by the Planck\footnote{https://sci.esa.int/web/planck} satellite, the Atacama Cosmology Telescope (ACT\footnote{https://act.princeton.edu/}), and the South Pole Telescope (SPT\footnote{https://pole.uchicago.edu/}). An emerging tension between these high and low redshift ($\Lambda$CDM) constraints may be indicative of new physics.

The potential tension between measurements, and with it the probability to discover new physics, increases with decreasing statistical uncertainty and better systematics control. With the advent of so-called Stage 4 surveys, e.g., the Dark Energy Spectroscopic Instrument \citep[DESI, ][]{DESI16}, the Prime Focus Spectrograph \citep[PFS, ][]{tec14}, the Large Synoptic Survey Telescope \citep[LSST\footnote{https://www.lsst.org/},][]{LSST19}, Euclid\footnote{https://sci.esa.int/web/euclid} \citep{laa11}, the Spectro-Photometer for the History of the Universe, Epoch of Reionization, and Ices Explorer \citep[SPHEREx\footnote{http://spherex.caltech.edu/},][]{dba14}, and the 4-metre Multi-Object Spectroscopic Telescope \citep[4MOST, ][]{4MOST19} the science community can expect an abundance of data to study the late-time Universe at increased precision. Similarly, the next generation of CMB surveys, such as the Simons Observatory \citep[SO, ][]{SO19} and CMB-S4 \citep{CMBS4} will enable us to contrast high and low redshift at increased precision and to combine information from both eras to increase the constraining power on cosmological models. 

The Wide-Field Infrared Survey Telescope \citep[WFIRST\footnote{https://wfirst.gsfc.nasa.gov/},][]{sgb15} is a successor mission to NASA's ground-breaking telescope endeavors such as the Hubble Space Telescope (HST\footnote{https://hubblesite.org/}), the Spitzer Space Telescope\footnote{http://www.spitzer.caltech.edu/}, and in the near future the James Webb Space Telescope (JWST\footnote{https://www.jwst.nasa.gov/}). WFIRST's science portfolio ranges from exoplanets to astrophysics to cosmology, building on a variety of standalone survey components: a microlensing survey, direct imaging of exoplanets, a supernovae survey, a guest observer program, and the High Latitude Survey (HLS). The latter is the main focus of this paper, in particular, we aim to quantify the HLS' constraining power on physics driving the late-time accelerated expansion of the Universe through a combination of multi-band imaging and spectroscopy. 
\begin{figure}
\includegraphics[width=8.3cm]{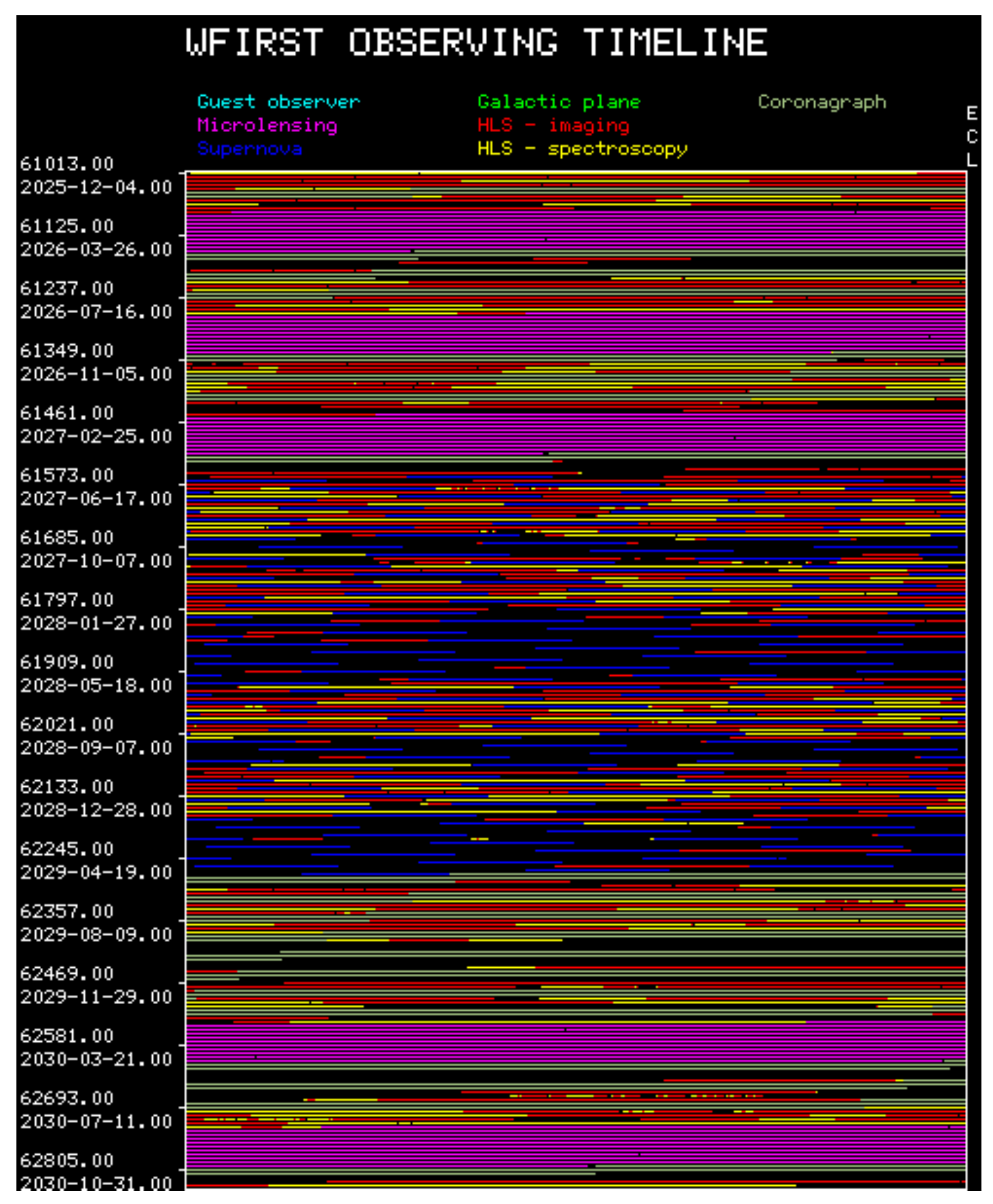}
\caption{An example WFIRST survey strategy as taken from the SDT 2015 report \citep{sgb15} and computed from the ETC v0.13 \citep{hgk12}. The individual survey components are colored into the timeline graphic: blue for the SN survey, Magenta for Microlensing, Red and Yellow for the HLIS and HLSS, respectively. The remaining time will be allocated as guest observer proposals to the science community.}
\label{fi:timeline}
\end{figure}

WFIRST is designed as a highly versatile missions that can flexibly react to findings of the aforementioned surveys. Its launch is planned for the mid 2020's into an L2 orbit with a nominal mission length of 5 years, however, this primary survey can be extended given that there are no consumables that prevent a 10+ year mission. The exact composition of the survey, i.e. the time allocation for the different science cases and the survey strategy within each science case is one of the most important topics that the community will discuss over the coming years prior to launch. 

Figure \ref{fi:timeline} shows an example WFIRST survey scenario composed of a 1.6 year High Latitude Survey (HLS), 6 months of SN observations distributed over 2 years, an  exoplanet and microlensing survey component, and a competed guest observer program that encompasses 25\% of the overall observing time. For the purpose of this paper we mainly focus on the HLS component, which can be divided further into the HLIS (High Latitude Imaging Survey) and the HLSS (High Latitude Spectroscopic Survey). 

The reference survey of the WFIRST HLS covers 2000 deg$^2$ with high-resolution, multi-band photometric imaging in four near infrared bands (HLIS) and deep grism spectroscopy (HLSS). This combination allows us to measure a variety of cosmological probes, e.g. weak lensing, galaxy clustering, galaxy clusters, redshift space distortions (RSD), Baryon Acoustic Oscillations (BAO). Together with the supernova survey, the reference HLS is designed to control systematics with minimal uncertainties; it will place tight constraints on the expansion history and structure growth in the Universe addressing questions about the nature of cosmic acceleration, neutrino physics, modified gravity, and dark matter.

%The WFIRST satellite is equipped with a primary mirror of 2.4m diameter, a wide-field instrument that can be used for imaging and spectroscopy, and a coronograph technology demonstration instrument to characterize the atmospheres of gaseous exoplanets of nearby stars. The WFC is composed of a 3x6 array, each containing 4k x 4k HgCdTe IR detectors, yielding an effective Field-of-View (FoV) of 0.281 deg$^2$. This FoV is about a factor of $\sim$200 larger than the instruments on HST (WFC3/IR) and JWST (NIRCam), enabling deep sky surveys of unprecedented size. Six imaging filters are used covering the $0.76-2.0$ $\mu$m range. The central wavelengths of filters are 0.87, 1.09, 1.30, 1.60, 1.88, 1.40 $\mu$m and spatial resolution per pixel is 0.11 arcsec. In addition, the WFC has a grism mode to perform slitless spectroscopy over a wavelength range of $1.35-1.89$ $\mu$m (at thermal background of 282K). The grism has a dispersion of $1.04-1.14$ nm per pixel and a spectral resolution $\lambda/\Delta \lambda \approx 622 - 871$ for 2 pixels~\citep{sgb15}. 

In this paper we develop a framework to simulate multi-probe strategies specifically for WFIRST. We outline the top-level concepts of combining cosmological probes including inference and covariance implementation in Sect. \ref{sec:multiprobe}, where we also show the main results of the paper, i.e. the WFIRST forecast that includes weak lensing, galaxy-galaxy lensing, galaxy clustering (photometric and spectroscopic), galaxy clusters number counts, cluster weak lensing, and SNIa. We consider subsets of this joint analysis and explore the impact of systematics in Sects. \ref{sec:3x2},  \ref{sec:clusters}, \ref{sec:HLSS}. We conclude in Sect. \ref{sec:conc}.

%%%%%%%%%%%%%%%%%%%%%%%%%%%%%%%%%%%
\section{Multi-Probe Likelihood Analyses}
\label{sec:multiprobe}
%%%%%%%%%%%%%%%%%%%%%%%%%%%%%%%%%%%
Contrasting and subsequently combining multiple probes is one of the most promising avenues to constrain cosmology: different probes are sensitive to different physics in the Universe, and they are affected differently by astrophysical uncertainties and observational systematics. Corresponding multi-probe strategies are relatively straightforward to implement if the observables are independent, e.g. when combining CMB temperature and polarization with BAO and SNIa, however, the story is much more complex when combining correlated probes. In the latter case one cannot simply combine the most sophisticated version of the single probe analyses a posteriori, but instead the analysis requires a joint covariances matrix that includes the statistical correlations and one must ensure the consistent modeling of systematics that affect the probes considered.

WFIRST's combination of spectroscopic and imaging instrumentation enables measuring a variety of LSS probes, such as weak lensing, galaxy clusters, galaxy clustering, and SNIa. The latter can be treated as independent information, though SN magnification in overdense regions could become non-negligible at some point in the future. The other probes however are tracers of the same underlying density field, where modes are significantly correlated due to nonlinear evolution of the late time density field. A corresponding likelihood analysis requires a multi-probe covariance matrix. 

\begin{figure}
 \includegraphics[width=8.5cm]{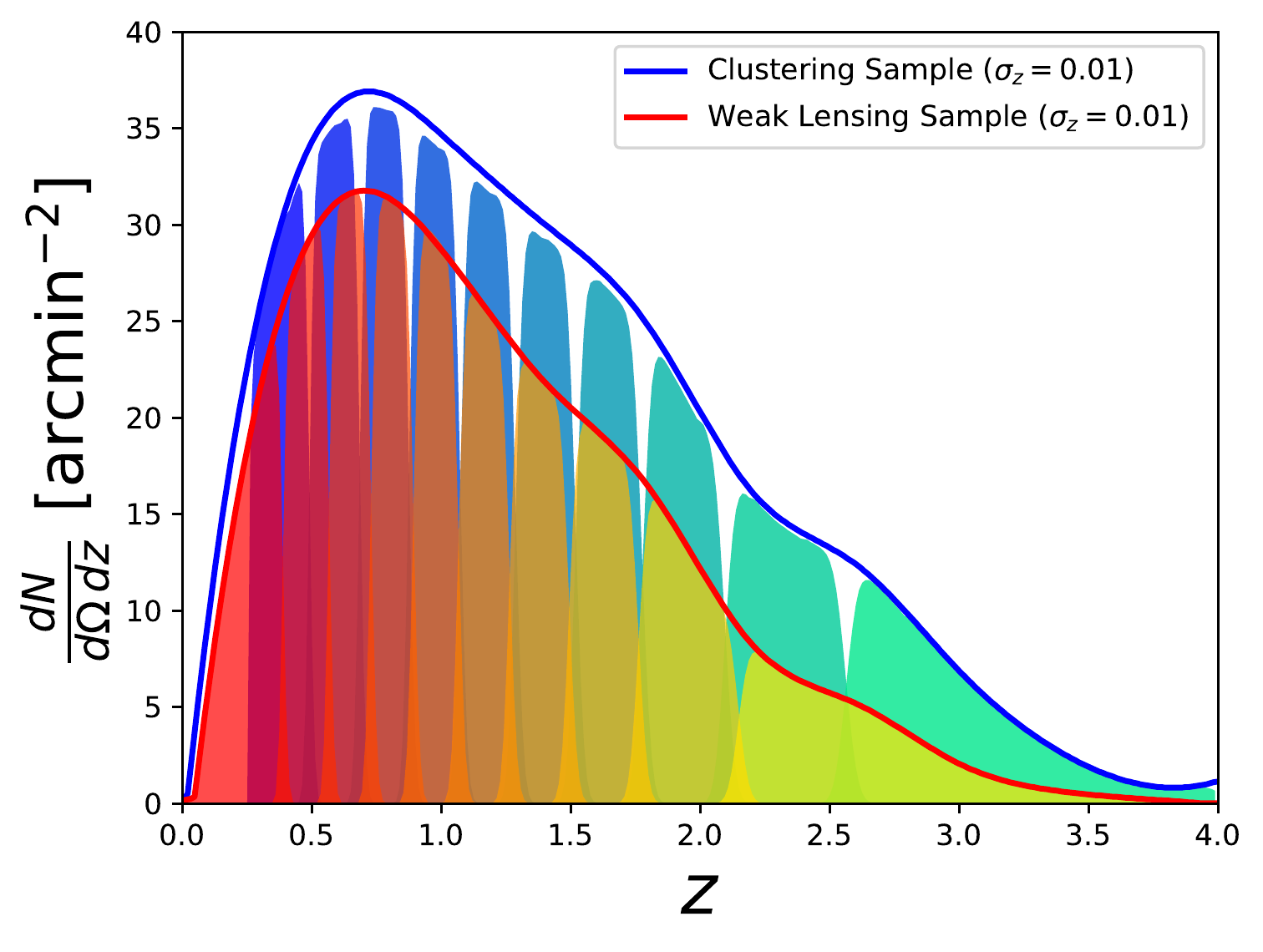}
 \includegraphics[width=8.5cm]{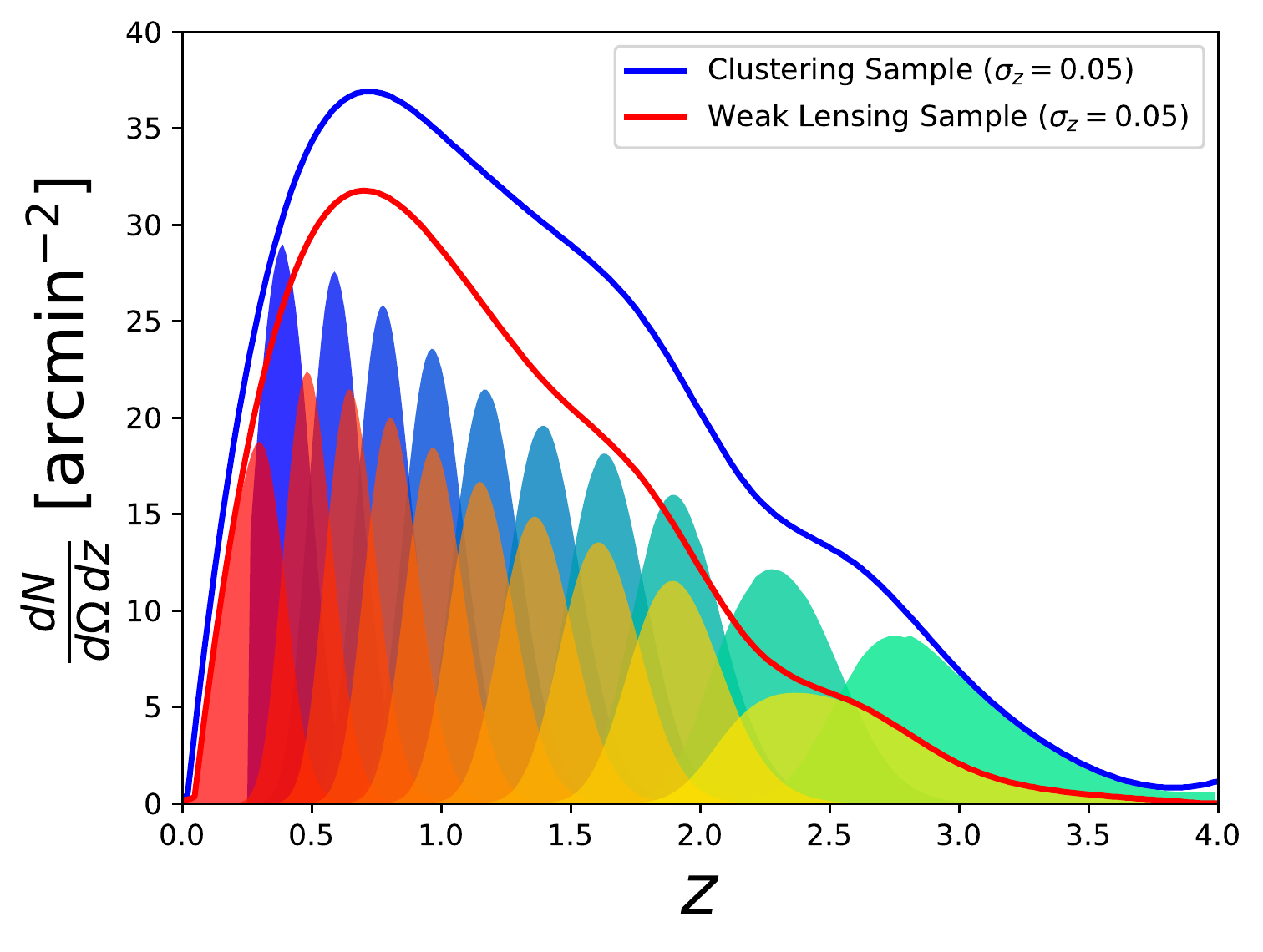}
 \caption{The redshift distributions for the lens and source galaxy sample for two different levels of photometric redshift precision, $\sigma_z$=0.01 and $\sigma_z$=0.05, respectively. These map onto our optimistic and pessimistic systematics scenarios considered for the HLIS.}
          \label{fi:zdistri}
 \end{figure}

%%%%%%%%%%%%%%%%%%%%%%%%%%%%%%%%%%%
\subsection{Analysis Choices}
\label{sec:choices}
%%%%%%%%%%%%%%%%%%%%%%%%%%%%%%%%%%%

 \begin{figure*}
 \includegraphics[width=13.5cm]{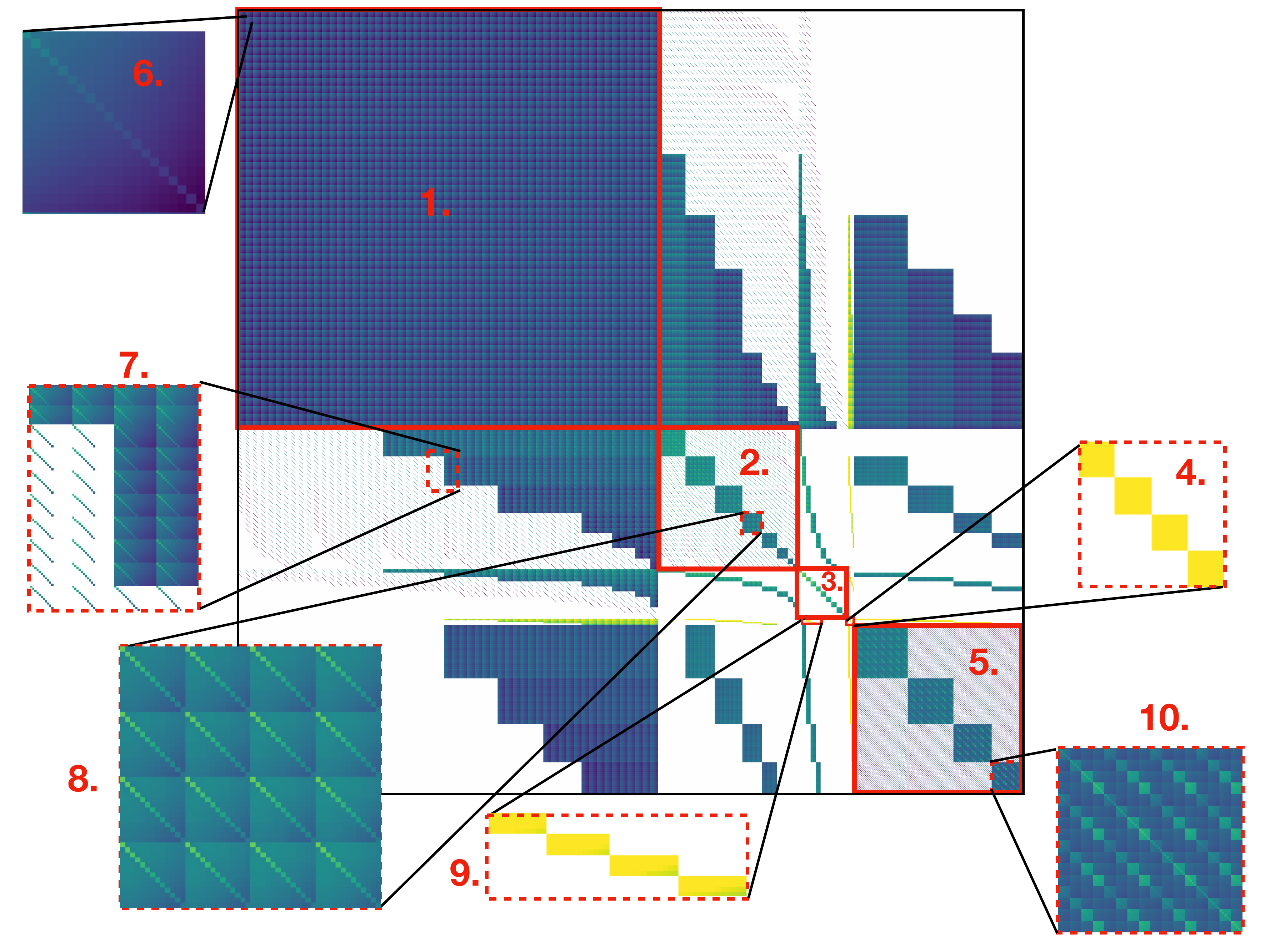}
 \caption{The multi-probe covariance matrix for the HLIS survey, calculated under the Limber approximation, where we have highlighted some parts of the matrix to illustrate the correlation structure: \tbf{(1)} depicts the cosmic shear covariance matrix, comprised of 55 tomographic combinations of source bins, each with 20 fourier $l$-bins. \tbf{(5)} shows one of the tomographic combinations, and the individual $l_1$, $l_2$ elements are clearly visible. \tbf{(2)} is the galaxy-galaxy lensing tomography covariance with \tbf{(8)} being the galaxy-galaxy combinations of the 4$^\mr{th}$ lens bin with all the non-overlapping source bins at higher redshifts. \tbf{(3)} is the clustering auto-probe matrix with 10 tomographic bins. \tbf{(4)} corresponds to the cluster number counts auto-probe matrix, which is comprised of 4 cluster redshift bins each with 4 richness bins (hardly distinguishable within in the 4 yellow squares). \tbf{(5)} is the auto-probe covariance of the cluster weak lensing part of the data vector, which uses the 4 cluster redshift bins as lens bins and the source sample as source bins. \tbf{(10)} zooms into the covariance of the 4$^\mr{th}$ cluster redshift bin, which again is split into 4 richness bins, all of which are then correlated with the highest 4 source galaxy redshift bins. One can see that the diagonal structure consists of 16 blocks that are each composed of 5x5 elements. The latter correspond to the covariance of the 5 cluster weak lensing $l$-bins, which range from $l \in [4000-15000]$. Zoom-in box \tbf{(6)} is a zoom into the first tomographic bin combination cosmic shear covariance matrix, \tbf{(7)} shows the cross-probe covariance of cosmic shear and galaxy-galaxy lensing. The impact of the $k_\mr{max}$ scale cuts causes the blocks to be non-quadratic. The Limber approximation leads to non-Gaussian terms only for specific combinations of lens and source tomographic bins (all 3 source bins need to be behind the lens bin). \tbf{(9)} is the cross-probe covariance of galaxy clustering and cluster number counts, which only has non-zero elements when both probes overlap in redshift, i.e. in the range $z \in [0.2-1.2]$. The shape of the yellow rectangles is determined by the number of $l$-bins used in the clustering data vector, i.e. 20, and the number of richness bins in cluster number counts, i.e. four.}
          \label{fi:covstruct}
 \end{figure*}

\begin{figure*}
\includegraphics[width=8.5cm]{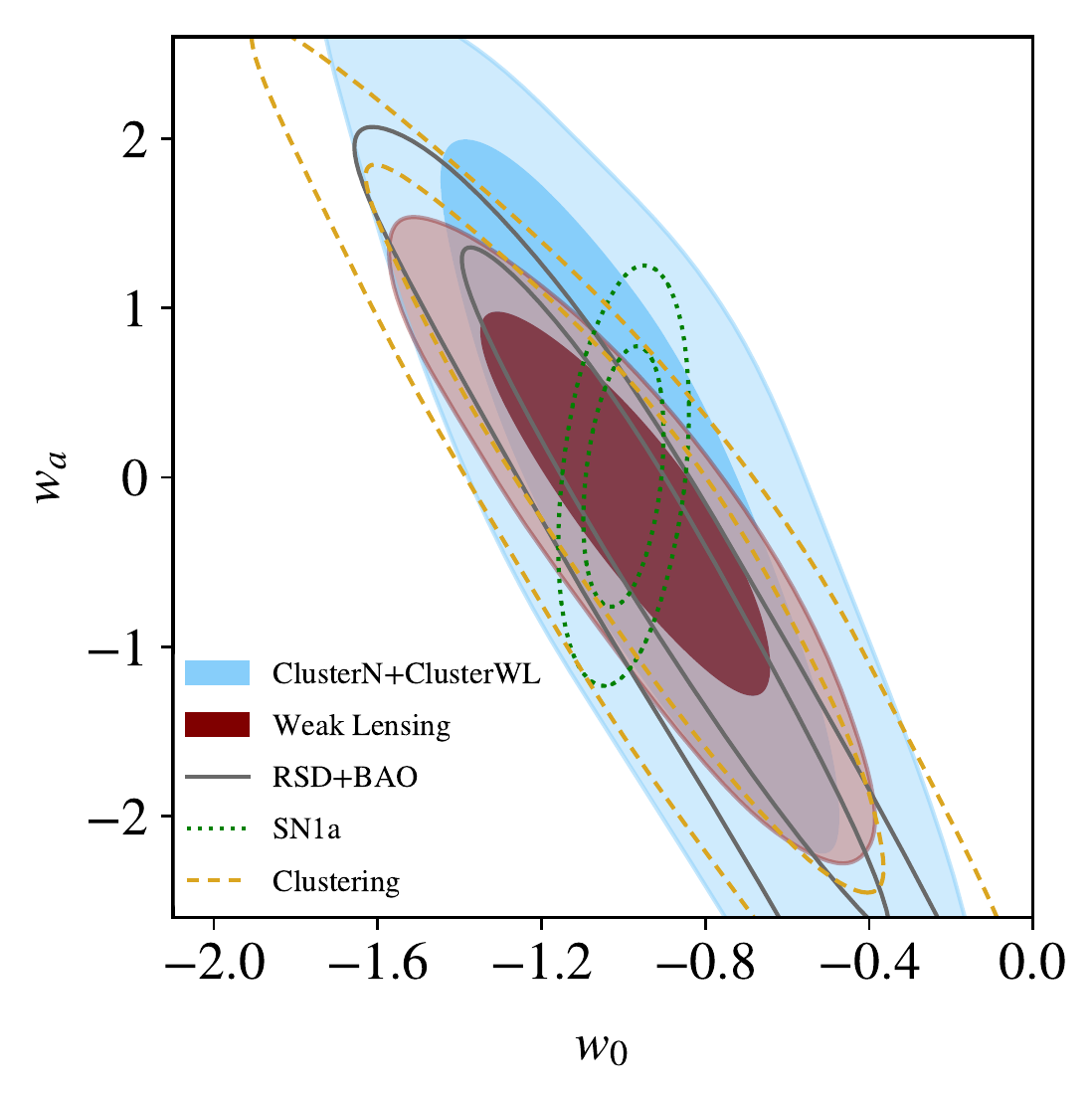}
\includegraphics[width=8.5cm]{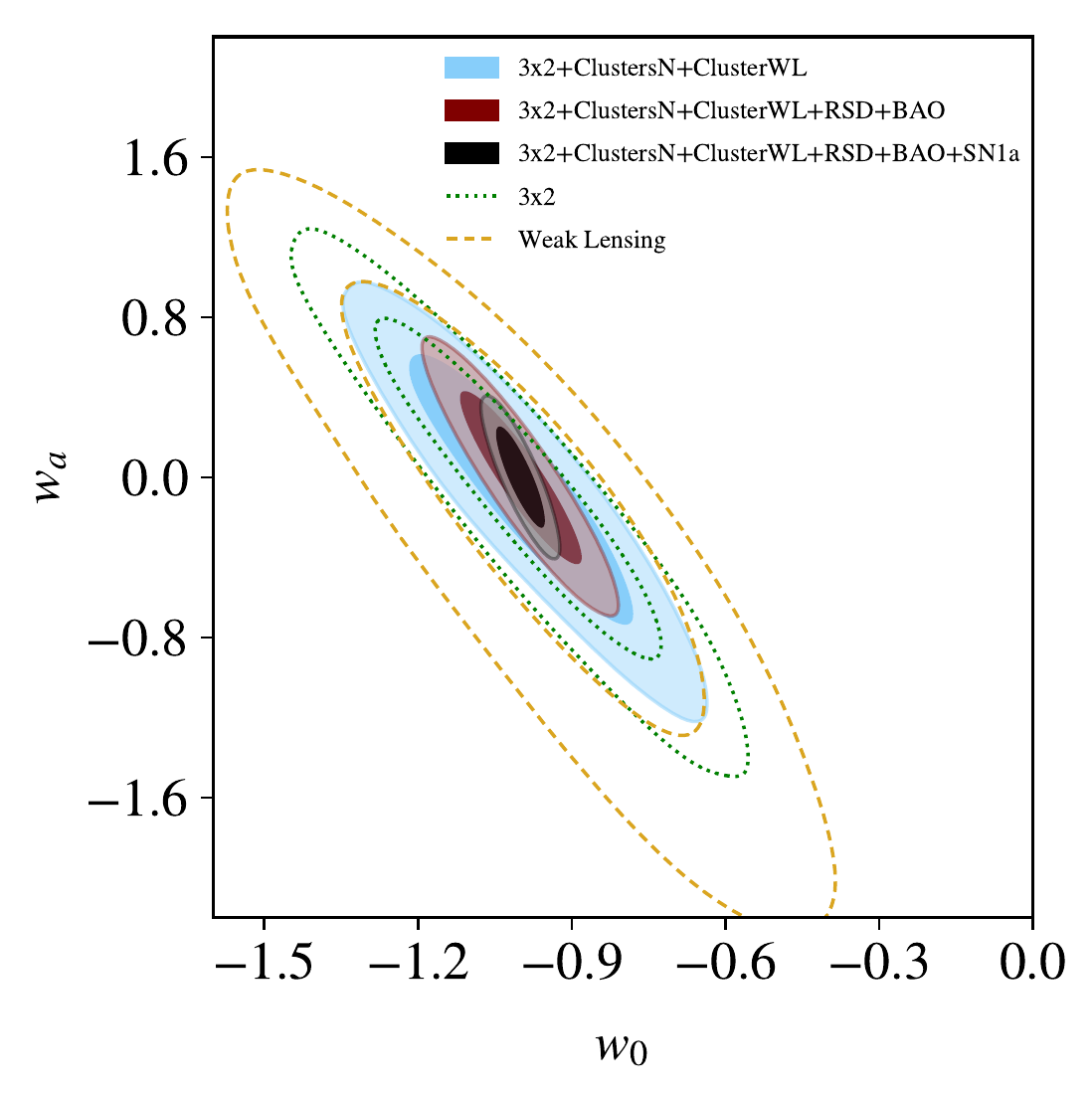}
\caption{\ti{Left:} Individual probes considered in this analysis, i.e. weak lensing, galaxy clustering, galaxy cluster number counts calibrated through cluster weak lensing, redshift space distortions power spectra including the BAO scale, and SNIa. \ti{Right:} Multi-probe analyses starting from weak lensing only, then adding clustering and galaxy-galaxy lensing (3x2), then adding cluster number counts and cluster weak lensing, then adding RSD and BAO information, and lastly adding in SNIa based on the findings of \citep{hdf18}. The FoMs for the individual and multi-probe chains can be found in Table \ref{tab:indi_multi_fom}.}
\label{fi:indi_vs_multi}
\end{figure*}

Designing a multi-probe analysis for the galaxies observed with the WFIRST reference survey can be broadly split into the following steps:
\begin{enumerate}
\item \tbf{Choose broad categories of cosmological probes} that are to be combined: For our WFIRST reference survey these are weak lensing, galaxy clusters, galaxy clustering (photometric and spectroscopic).
\item \tbf{Define specific probe combinations and summary statistics} that make up the data points of the data vector, which in our case are one-point functions and two-point functions that represent the corresponding probes. We do not consider higher-order correlation functions.
\item \tbf{Define the galaxy samples} that are associated with the aforementioned probes. We use the WFIRST exposure time calculator (ETC) \citep{hgk12} to compute realistic survey scenarios for WFIRST's coverage of area and depth in a given band. We fix the time per exposure and vary the number of exposures to build up depth over the survey area of a given scenario. For the HLS Reference Survey this area is 2,000 deg$^2$. The total survey time for a given number of exposures includes a simple prescription for overheads and is correct to approximately 10$\%$. 

In order to obtain accurate redshift distributions we closely follow \cite{hcm19} in applying the ETC results to the CANDELS data set \citep{Candels,candels2}, which is the only data set available that is sufficiently deep in the near-infrared to model WFIRST observations. The ETC has a built-in option to obtain a weak lensing catalog based on an input catalog of detected sources. The criteria for galaxies to be considered suitable for weak lensing are $S/N$>18 (J+H band combined, matched filter), ellipticity dispersion $\sigma_\epsilon<0.2$, and resolution factor R>0.4, where we used the \cite{bej02} convention (i.e. $\epsilon=(a^2-b^2)/(a^2+b^2)$ instead of $(a-b)/(a+b)$).

We apply these selections to the CANDELS catalog and obtain our source sample for the WFIRST HLS 4 NIR band survey. For the lens sample we select CANDELS galaxies with $S/N>10$ in each of the 4 WFIRST bands. Our WFIRST analysis assumes LSST photometry from the ground, hence we further down-select both samples by imposing a $S/N>5$ cut in each LSST band except for u-band (we note that 50\% of our galaxy sample has $S/N>5$ in the u-band as well). 

The resulting number densities for the HLS are 
\bea
\label{eq:nbar}
\bar{n}_{\mathrm{source}} = N_{\mathrm{source}}/\Omega_{\mathrm{s}} &=& 51\;\mathrm{galaxies/arcmin^2} \\ \,
\bar{n}_{\mathrm{lens}} = N_{\mathrm{lens}}/\Omega_{\mathrm{s}} &=& 66\;\mathrm{galaxies/arcmin^2}\,.
\eea
where $\Omega_{\mathrm{s}}$ is the WFIRST survey area. We impose a $z_\mr{min} = 0.25$ for the lens sample and define 10 tomographic bins for each sample such that $\bar{n}^i_{x}=\bar{n}_{x}/10$. These tomographic bins are then convolved with a Gaussian distribution, which is further described in Sect. \ref{sec:systematics}. 

We consider two different Gaussian photo-z scenarios: an optimistic variation with mean zero and narrow width of $\sigma_z=0.01$ and a more pessimistic scenario with  broader kernel of $\sigma_z=0.05$. The resulting redshift distributions are depicted in Fig. \ref{fi:zdistri}.

\renewcommand{\arraystretch}{1.3}
\begin{table}
\caption{FoMs for individual and multi-probe chains depicted in Fig. \ref{fi:indi_vs_multi}. Note that 3x2 includes cosmic shear.}
\begin{center}
\begin{tabular*}{0.45\textwidth}{@{\extracolsep{\fill}}| l c c |}
\hline
\multicolumn{3}{|c|}{\tbf{Multi-probe FoM summary}} \\
\hline
Probe & Individual & Cumulative \\  
Cosmic shear & 9.8  & 9.8   \\
3x2 & 23.46 & 23.46   \\
Clusters & 3.86  & 31.56   \\
RSD+BAO & 8.19  & 89.54   \\
SNIa & 24.62  & 300.11   \\
\hline
\end{tabular*}
\end{center}
\label{tab:indi_multi_fom}
\end{table}
\renewcommand{\arraystretch}{1.0}

\begin{itemize}
\item \ti{Source galaxy sample,} for which we require position, photometric redshift, and galaxy shape measurements.
\item \ti{Lens galaxy sample,} for which we require position and photometric redshift  measurements.
\item \ti{Galaxy clusters,} for which we require position, photometric redshift and optical richness estimates for galaxy clusters that are identified in the overall galaxy catalog.
\item \ti{Spectroscopic galaxy sample,} which requires measurements of positions and spectroscopic redshifts.
\end{itemize}
\item \tbf{Define exact analysis choices:} Given that we are looking at 2-point functions as summary statistics, we need to decide on the exact auto and cross-galaxy samples that constitute a cosmological probe. Further, we need to define the exact binning within each probe, in particular which angular scales and tomographic redshift binning are considered. The decision tree for these choices is complex and takes into account our ability to accurately model physics and systematics at specific angular scales and redshifts, and in particular our ability to model the correlations across all data points in the covariance matrix. For the WFIRST data vector that we use to simulate the HLS Reference Survey, we choose: 
\begin{itemize}
\item \ti{Source galaxies -- cosmic shear:}
In terms of angular binning we universally choose 25 logarithmically spaced Fourier mode bins ranging from $l_{\rm{min}} = 30$ to $l_{\rm{max}} = 15000$ for all two-point functions in our data vector, however we impose different scale cuts for the different probes. The idea of universal binning across probes is driven by the desire to avoid computing cross-covariances of probes with different $l$-binning. For the cosmic shear part of the data vector we impose a scale cut of $l_{\rm{max}} \rm{(cosmic \, shear)} = 4000$, which leaves 20 bins that carry information. The ten tomographic bins translate into 55 auto-and cross power spectra.
\item \ti{Lens galaxies -- photometric clustering:}
The redshift distribution for the lens sample is further detailed  in Sect. \ref{sec:systematics} and divided into 10 tomographic bins. We exclude $l-$bins, if scales below $R_\mr{\min}=2\pi/k_\mr{max}=21 \mr{\;Mpc/h}$ contribute to the Limber integral (see Eq. (\ref{eq:projected})), which imposes a redshift dependent scale cut in the $l$-binning.  

\item \ti{Lens $\times$ source galaxies -- photometric galaxy-galaxy lensing:}
The galaxy-galaxy lensing part of the data vector assumes the lens galaxy sample as foreground and the source galaxy sample as background galaxies; we only consider source-lens combinations where the source bin is fully behind the lens bin in redshift. We again impose a cut-off at $R_\mr{\min}=21 \mr{ Mpc/h}$.

\item \ti{Galaxy cluster number counts:} This is the one one-point function we include in our data vector. We split our cluster sample into four cluster redshift bins (0.4-0.6,0.6-0.8,0.8-1.0,1.0-1.2) and 4 cluster richness bins between $\lambda_{\mathrm{min}} = 40$ and $\lambda_{\mathrm{max}} = 220$ in each redshift bin.
\item \ti{Galaxy clusters $\times$ source galaxies -- cluster weak lensing:} In order to calibrate the cluster mass--richness relation (Eq.~\ref{eq:cl:mor}), we consider the stacked weak lensing signal from all combinations of cluster redshift and richness bins with source galaxies, with the restriction that source galaxies are located at higher redshift than the galaxy clusters. Specifically, we use the cluster lensing power spectrum in the angular range $4000 < l <15000$, which corresponds mostly to the 1-halo cluster lensing signal. 

\item \ti{Spectroscopic $\times$ spectroscopic -- spectroscopic galaxy clustering:} While our analysis considers all cross-covariance terms for the 5 cosmological probes above, WFIRST's spectroscopic clustering is treated as an independent probe whose cosmological information is determined separately and added a posteriori. This is an approximation, however the derivation of a 2D+3D joint covariance is beyond the scope of this paper and deferred to future work. 
Our spectroscopic clustering data vector is comprised of 3D power spectrum fourier modes $P(k, \mu)$ and we select 100 logarithmic bins ranging from $k_\mr{min}=0.001$ to $k_\mr{max}=0.3$ h/Mpc, 10 linearly spaced $\mu$ bins from $0-1.0$, and 7 density weighted redshift bins that start at $0.83$ and range out to 3.7. This data vector captures both the BAO and RSD information.
\end{itemize}
\end{enumerate}

%%%%%%%%%%%%%%%%%%%%%%%%%%%%%%%%%%%
\subsection{Inference, Likelihoods, Covariances}
\label{sec:covs}
%%%%%%%%%%%%%%%%%%%%%%%%%%%%%%%%%%%

Given the data vector $\D$, we sample the joint parameter space of cosmological $\pco$ and nuisance parameters $\pnu$ using the \texttt{emcee\footnote{\url{https://emcee.readthedocs.io/en/stable/}}} \citep{fhg13}, which is based on the affine-invariant sampler of \cite{gow10}. At each step we compute the posterior using Bayes' theorem 
\be
\label{eq:bayes}
\prob(\pco, \pnu|\D) \propto \underbrace{\probr (\pco, \pnu)}_{\mr{SNIa}} \; \underbrace{\like (\D| \pco, \pnu)}_{\mr{HLS}}. \\
\ee
$\probr (\pco, \pnu)$ denotes the prior probability which in our case is based on the WFIRST SNIa survey forecast from \citep{hdf18}. Specifically, we reran the ``Imaging: Allz (optimistic)'' scenario \citep[c.f. Sect. 5.4 and Table 13 in][]{hdf18} centered it on the fiducial cosmology of our analysis (see Table \ref{tab:3x2params}). We did not include any information from CMB or BAO experiments, which explains the different contours compared to \citep{hdf18}.

The cosmological information from the HLS enters our simulations through the second term in Eq. (\ref{eq:bayes}), i.e. the likelihood,  $\like (\D| \pco, \pnu) = N \, \times \, \exp( -\frac{1}{2} \chi^2(\pco, \pnu))$. We assume that the errors of this data vector are distributed as a multi-variate Gaussian 
\be
\like (\D| \pco, \pnu) = N \, \times \, \exp \Big( -\frac{1}{2} \big[ \chi_\mr{HLIS}^2(\pco, \pnu) + \chi_\mr{HLSS}^2(\pco, \pnu)\big] \Big)\,,
\ee
which is composed of two $\chi^2= (\D -\M)^t \, \matC^{-1} \, (\D-\M) $ terms reflecting our approximation that the cosmological information from HLSS and HLIS is independent. We note that future work should explore correlations between HLIS and HLSS and develop a joint covariance matrix for these measurements. $N$ is a normalization constant. 

Based on the analysis choices (probes, redshifts, scales) described in Sect. \ref{sec:choices} we compute the data vectors and covariance matrices for HLIS and HLSS at the fiducial cosmology and systematics parameters (see Tables \ref{tab:3x2params}, \ref{tab:clusterparams}, \ref{tab:specparams}, for the different probes). In case of the HLSS survey the covariance matrix is diagonal and further described in Sect. \ref{sec:HLSS}, in case of the HLIS the matrix has significant off-diagonal terms. 

Figure \ref{fi:covstruct} illustrates the structure of the matrix with the auto-probe matrices denoted as numbers 1-5 corresponding to cosmic shear (1), galaxy-galaxy lensing (2), galaxy clustering (3), cluster number counts(4), and cluster weak lensing (5). Calculation of the individual terms of the covariance can be found in the Appendix (Eqs: A2-A14) of \citet{kre17}. 

Since this covariance matrix is calculated analytically and not estimated from either simulations or data, it can be considered noise-free and is easily invertible. It does not inherently limit the number of data points that can enter our analysis, which would be the case if the covariance were computed from a limited set of realizations \citep[see e.g.,][for details on these constraints]{tjk13,dos13}.

%%%%%%%%%%%%%%%%%%%%%%%%%%%%%%%%%%%%%%%%%%%%%%%%%%
%%%%%%%%%%%%%%%%%%%%%%%%%%%%%%%%%%%%%%%%%%%%%%%%%%
\section{Cosmic Shear and Galaxy Clustering}
\label{sec:3x2}
%%%%%%%%%%%%%%%%%%%%%%%%%%%%%%%%%%%%%%%%%%%%%%%%%%
%%%%%%%%%%%%%%%%%%%%%%%%%%%%%%%%%%%%%%%%%%%%%%%%%%

We start exploring WFIRST multi-probe analyses by looking at the HLIS weak lensing and photometric galaxy clustering probes, which when combined with galaxy-galaxy lensing form a so-called 3x2pt analysis. Here, we summarize the computation of angular (cross) power spectra for the different probes and the computation of galaxy cluster number counts. We use capital Roman subscripts to denote observables, $A,B\in \left\{ \kappa,\delta_{\mathrm{g}},\delta_{\lambda_\alpha}\right\}$, where $\kappa$ references lensing, $\delta_{\mathrm{g}}$ the density contrast of (lens) galaxies. The density contrast of galaxy clusters in richness bin $\alpha$, $\delta_{\lambda_\alpha}$, will be considered in Sect. \ref{sec:clusters}.

%%%%%%%%%%%%%%%%%%%%%%%%%%%%%%%%%%%%%%%%%%%%%%%%%%
\subsection{Modeling of observables}
\label{observables1}
%%%%%%%%%%%%%%%%%%%%%%%%%%%%%%%%%%%%%%%%%%%%%%%%%%
 \renewcommand{\arraystretch}{1.3}
\begin{table}
\caption{Fiducial parameters, flat priors (min, max) for cosmology and galaxy bias, and Gaussian priors ($\mu$, $\sigma$) for observational systematics. We consider optimistic and pessimistic scenarios in this paper, which is indicated in the corresponding sections of the table.}
\begin{center}
\begin{tabular*}{0.45\textwidth}{@{\extracolsep{\fill}}| c c c |}
\hline
\hline
Parameter & Fiducial & Prior \\  
\hline 
\multicolumn{3}{|c|}{\tbf{Survey}} \\
$\Omega_{\mathrm{s}}$ & 2,000 deg$^2$ & fixed\\
$n_{\mathrm{source}}$ & 51 gal/arcmin$^2$ & fixed\\
$n_{\mathrm{lens}}$ & 66 gal/arcmin$^2$& fixed \\
$\sigma_\epsilon$ &0.26& fixed\\
\hline 
\multicolumn{3}{|c|}{\tbf{Cosmology}} \\
$\om$ & 0.3156 &  flat (0.1, 0.6)  \\ 
$\sig$ & 0.831 &  flat (0.6, 0.95)  \\ 
$\ns$ & 0.9645 & flat (0.85, 1.06)  \\
$\w$ &  -1.0 &   flat (-2.0, 0.0)   \\
$\wa$ &  0.0 &  flat (-2.5, 2.5)   \\
$\omb$ &  0.0492 &  flat (0.04, 0.055)  \\
$h_0$  & 0.6727 &  flat (0.6, 0.76)   \\
\hline
\multicolumn{3}{|c|}{\tbf{Galaxy bias (tomographic bins)}} \\
$b_\mr{g}^i$ & 1.3 + i $\times$ 0.1  & flat (0.8, 3.0) \\
\hline
\multicolumn{3}{|c|}{\tbf{Lens photo-z (optimistic)}} \\
$\Delta_\mr{z,lens}^i $ & 0.0 & Gauss (0.0, 0.002) \\
$\sigma_\mr{z,lens} $ & 0.01 & Gauss (0.01, 0.002) \\
\multicolumn{3}{|c|}{\tbf{Lens photo-z (pessimistic)}} \\
$\Delta_\mr{z,lens}^i $ & 0.0 & Gauss (0.0, 0.02) \\
$\sigma_\mr{z,lens} $ & 0.05 & Gauss (0.05, 0.02) \\

\hline
\multicolumn{3}{|c|}{\tbf{Source photo-z (optimistic)}} \\
$\Delta_\mr{z,source}^i $ & 0.0 & Gauss (0.0, 0.002) \\
$\sigma_\mr{z,source}$ &0.01 & Gauss (0.01, 0.002) \\
\multicolumn{3}{|c|}{\tbf{Source photo-z (pessimistic)}} \\
$\Delta_\mr{z,source}^i $ & 0.0 & Gauss (0.0, 0.02) \\
$\sigma_\mr{z,source}$ &0.05 & Gauss (0.05, 0.02) \\
\hline
\multicolumn{3}{|c|}{\tbf{Shear calibration (optimistic)}} \\
$m_i $ & 0.0 & Gauss (0.0, 0.002)\\
\multicolumn{3}{|c|}{\tbf{Shear calibration (pessimistic)}} \\
$m_i $ & 0.0 & Gauss (0.0, 0.01)\\
\hline
\end{tabular*}
\end{center}
\label{tab:3x2params}
\end{table}
\renewcommand{\arraystretch}{1.0}

\begin{figure*}
\includegraphics[width=5.8cm]{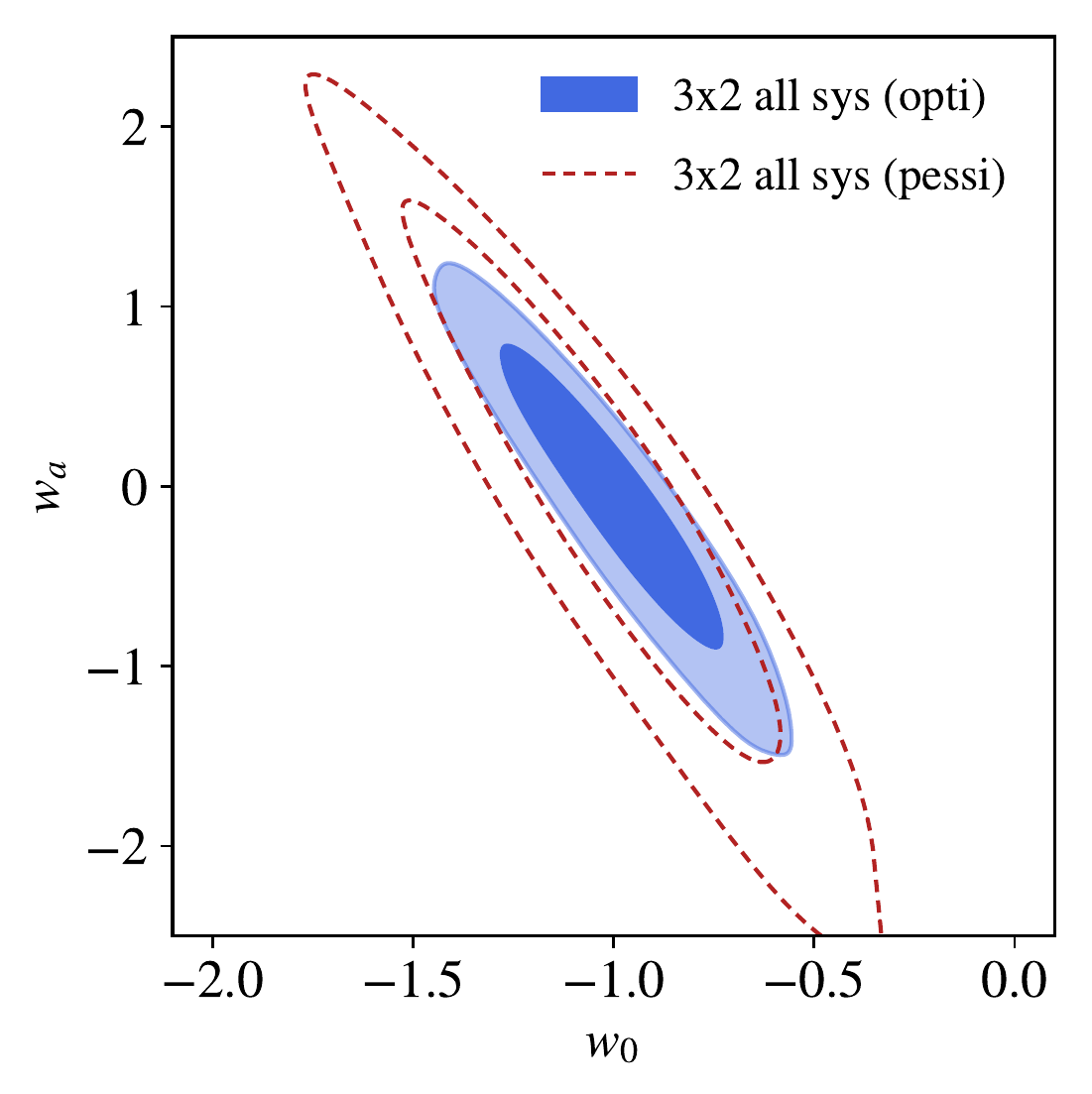}
\includegraphics[width=5.8cm]{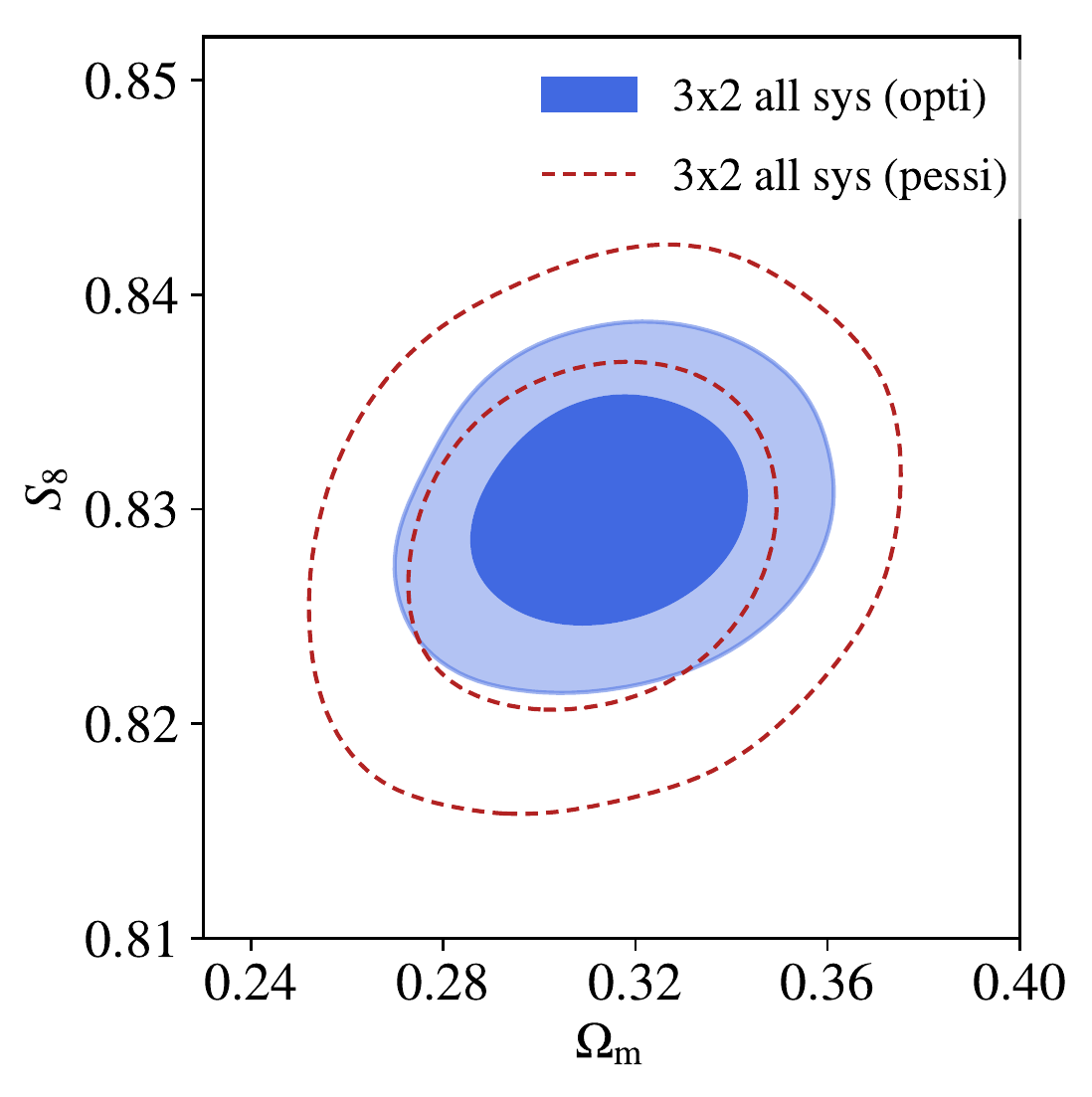}
\includegraphics[width=5.8cm]{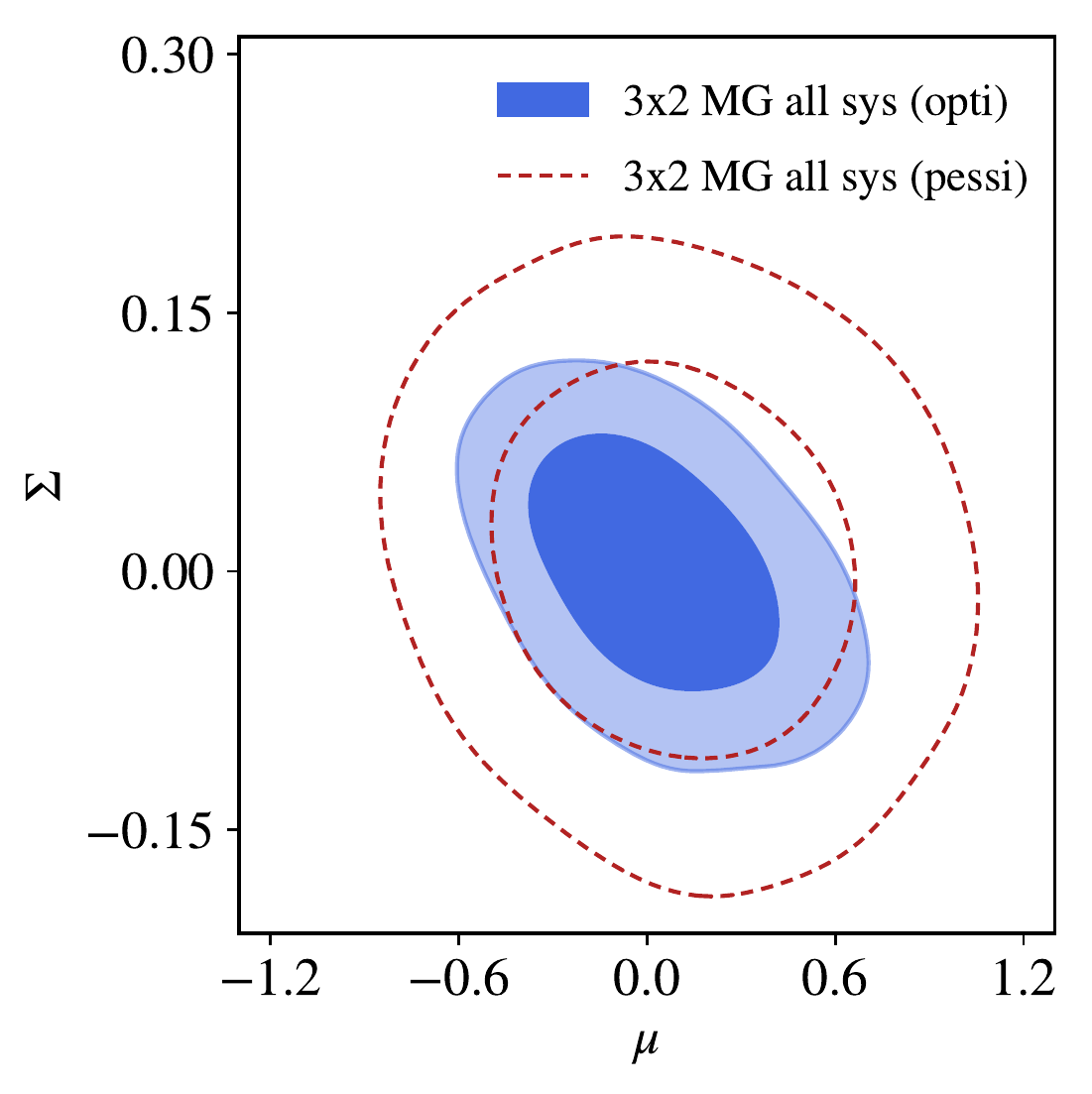}
\caption{ Constraining power on dark energy equation of state parameters $w_0$ and $w_a$ (\textit{left}), $\om$ and $S_8$ (\textit{middle}), and on modified gravity parameters $\Sigma_0$ and $\mu_0$ for optimistic and pessimistic systematics scenarios for a 3x2 analysis. Note that the likelihood analysis in the left two panels assume GR to be the correct theory, only in the right panel we vary $\Sigma_0$ and $\mu_0$. The relative loss in information depicted here is quantified as FoMs in Table \ref{tab:3x2_science_fom}.}
\label{fi:3x2_DE}
\end{figure*}

\renewcommand{\arraystretch}{1.3}
\begin{table}
\caption{FoMs for optimistic and pessimistic systematics scenarios for the science cases depicted in Fig. \ref{fi:3x2_DE}}
\begin{center}
\begin{tabular*}{0.45\textwidth}{@{\extracolsep{\fill}}| l c c |}
\hline
\multicolumn{3}{|c|}{\tbf{3x2 different science cases FoM summary}} \\
\hline
Science case & optimistic & pessimistic \\  
Dark Energy & 23.46  &  7.88 \\
Modified Gravity & 22.20  &  9.49  \\
\hline
\end{tabular*}
\end{center}
\label{tab:3x2_science_fom}
\end{table}
\renewcommand{\arraystretch}{1.0}

We calculate the angular power spectrum between redshift bin $i$ of observable $A$ and redshift bin $j$ of observables $B$ at projected Fourier mode $l$, $C_{AB}^{ij}(l)$, using the Limber and flat sky approximations \citep[we refer to e.g.][for the potential impact when analyzing data]{2019arXiv191111947F}:
\be
\label{eq:projected}
C_{AB}^{ij}(l) = \int d\chi \frac{q_A^i(\chi)q_B^j(\chi)}{\chi^2}P_{AB}(l/\chi,z(\chi)),
\ee
where $\chi$ is the comoving distance, $q_A^i(\chi)$ are weight functions of the different observables 
given in Eqs.~(\ref{eq:qg}-\ref{eq:qkappa}), and $P_{AB}(k,z)$ the three dimensional, probe-specific power spectra detailed below. 
The weight function for the projected galaxy density in redshift bin $i$,$q_{\delta_{\mathrm{g}}}^i(\chi)$, is given the normalized comoving distance probability of galaxies in this redshift bin
\begin{equation}
\label{eq:qg}
q_{\delta_{\mathrm{g}}}^i(\chi) =\frac{n_{\mathrm{lens}}^i(z(\chi)) }{\bar{n}_{\mathrm{lens}}^i}\frac{dz}{d\chi}\,,
\end{equation}
with $n_{\mathrm{lens}}^i(z)$ the redshift distribution of galaxies in (photometric) galaxy redshift bin $i$ (c.f. Eq.~{\ref{eq:photoz}}), and $\bar{n}_{\mathrm{lens}}^i$ the angular number densities of galaxies in this redshift bin (c.f. Eq.~{\ref{eq:nbar}}). 
For the convergence field, the weight function $q_\kappa^{i}(\chi)$ is the lens efficiency, 
\begin{equation}
\label{eq:qkappa}
q_\kappa^{i}(\chi) = \frac{3 H_0^2 \Omega_m }{2 \mathrm{c}^2}\frac{\chi}{a(\chi)}\int_\chi^{\chi_{\mr h}} \mr d \chi' \frac{n_{\mathrm{source}}^{i} (z(\chi')) dz/d\chi'}{\bar{n}_{\mathrm{source}}^{i}} \frac{\chi'-\chi}{\chi'} \,,
\end{equation}
with $n_{\mathrm{source}}^{i} (z)$ the the redshift distribution of source galaxies in (photometric) source redshift bin $i$ (Eq.~{\ref{eq:photoz}}), $\bar{n}_{\mathrm{source}}^i$ the angular number densities of source galaxies in this redshift bin (Eq.~{\ref{eq:nbar}}), and $a(\chi)$ the scale factor.

The three-dimensional power spectra $P_{AB}(k,z)$ can be expressed through the matter density power spectrum $P_{mm}(k,z)$. For the purpose of this section $P_{mm}(k,z)$ corresponds to the density power spectrum $P_{\delta \delta}(k,z)$, where we use the \citet{tsn12} fitting formula to model nonlinear evolution. Noting that $P_{AB} = P_{BA}$, we describe the different cases in Eqs. (\ref{eq:Pkappa},\ref{eq:Pg},\ref{eq:Pcluster}).
For $A = \kappa$, this is trivial,
\be
\label{eq:Pkappa}
P_{\kappa B}(k,z) = P_{mB}(k,z) \,.
\ee
For quantities related to the galaxy density, we note that we only consider the large-scale galaxy distribution, where it is valid to assume that the galaxy density contrast on these scales can be approximated as the non-linear matter density contrast times an effective galaxy bias parameter $b_g(z)$
\be
\label{eq:Pg} P_{\delta_{\mathrm{g}} B}(k,z)  = b_g(z) P_{mB}(k,z) \, .
\ee
 
%%%%%%%%%%%%%%%%%%%%%%%%%%%%%%%%%%%%%%%%%%%%%%%
\subsection{Modified Gravity modeling}
\label{sec:modfield_gravity_parametrization}
%%%%%%%%%%%%%%%%%%%%%%%%%%%%%%%%%%%%%%%%%%%%%%%
Since there is no compelling model of modified gravity, we adopt phenomenological modified gravity parameters $(\mu_0, \Sigma_0)$ which we define similar as e.g., \cite{shp13}. 

In this parameterization the expressions for the Newtonian potential $\Psi$ and the curvature potential $\Phi$ that govern the perturbed  Friedmann-Robertson-Walker metric 
\be
ds^2=(1+2\Psi)dt^2-a^2(t)(1-2\Phi)dx^2,
\ee
are altered. Within general relativity $\Psi=\Phi$ holds. The $(\mu, \Sigma)$ parameters give additional freedom to the Newtonian gravitational potential $\Psi$ experienced by non-relativistic particles and the lensing potential $(\Phi+\Psi)$ experienced by relativistic particles, specifically
\bea
\Psi(k,a) &=& [1+\mu(a)] \Psi_{\rm GR}(k,a),\\
\Psi(k,a)+\Phi(k,a) &=& [1+\Sigma(a)] (\Psi_{\rm GR}(k,a)+\Phi_{\rm GR}(k,a))\,.
\eea

We assume that $\mu(a)$ and $\Sigma(a)$ are both  scale independent. Furthermore, since their  motivation was to explain the dark energy phenomenon, we assume that the modified gravity parameters scale with the dark energy density, i.e.,
\bea
\mu(a) &=& \mu_0\frac{\Omega_\Lambda(a)}{\Omega_{\Lambda}},\\
\Sigma(a) &=& \Sigma_0\frac{\Omega_\Lambda(a)}{\Omega_{\Lambda}},
\eea
where $\Omega_\Lambda$ is the present day dark energy density. Note that in the case of general relativity, $\mu_0 = \Sigma_0 = 0$.

The $\mu_0$ parameter modifies the growth of linear density perturbation such that
\be
  \label{eq:growth_deriv_mu_a}
  \delta^{\prime\prime}+\left(\frac{2}{a}+\frac{\ddot{a}}{\dot{a}^2} \right)\delta^\prime-\frac{3\Omega_m}{2a^2} \left[1+\mu\left(a\right) \right] \delta = 0,
\ee
which changes the growth function, and consequently the density-density power spectrum $P_{\delta \delta}$ and all projected power spectra described in Eq.~(\ref{eq:projected}).

The $\Sigma_0$ parameter only affects lensing related quantities, which in a 3x2pt analysis means the galaxy-shear and shear-shear power spectrum. Specifically,  Eq. (\ref{eq:projected}) is modified as 
\be
\label{eq:projected2}
C_{AB}^{ij}(l) = \int d\chi \frac{q_A^i(\chi)q_B^j(\chi)}{\chi^2} \left[1+\Sigma\left(\chi\right)\right]^k P_{AB}(l/\chi,z(\chi)),
\ee
where the exponent $k=2$ if $A=B=\kappa$, $k=0$ if $A=B=\delta_\mr g$, and $k=1$ if either $A=\kappa$ or $B=\kappa$.  

%%%%%%%%%%%%%%%%%%%%%%%%%%%%%%%%%%%%%%%%%%%%%%%
\subsection{Systematics}
\label{sec:systematics}
%%%%%%%%%%%%%%%%%%%%%%%%%%%%%%%%%%%%%%%%%%%%%%%

\begin{figure*}
\includegraphics[width=5.8cm]{plots/plot_3x2_sys_opti_vs_pessi_v2.pdf}
\includegraphics[width=5.8cm]{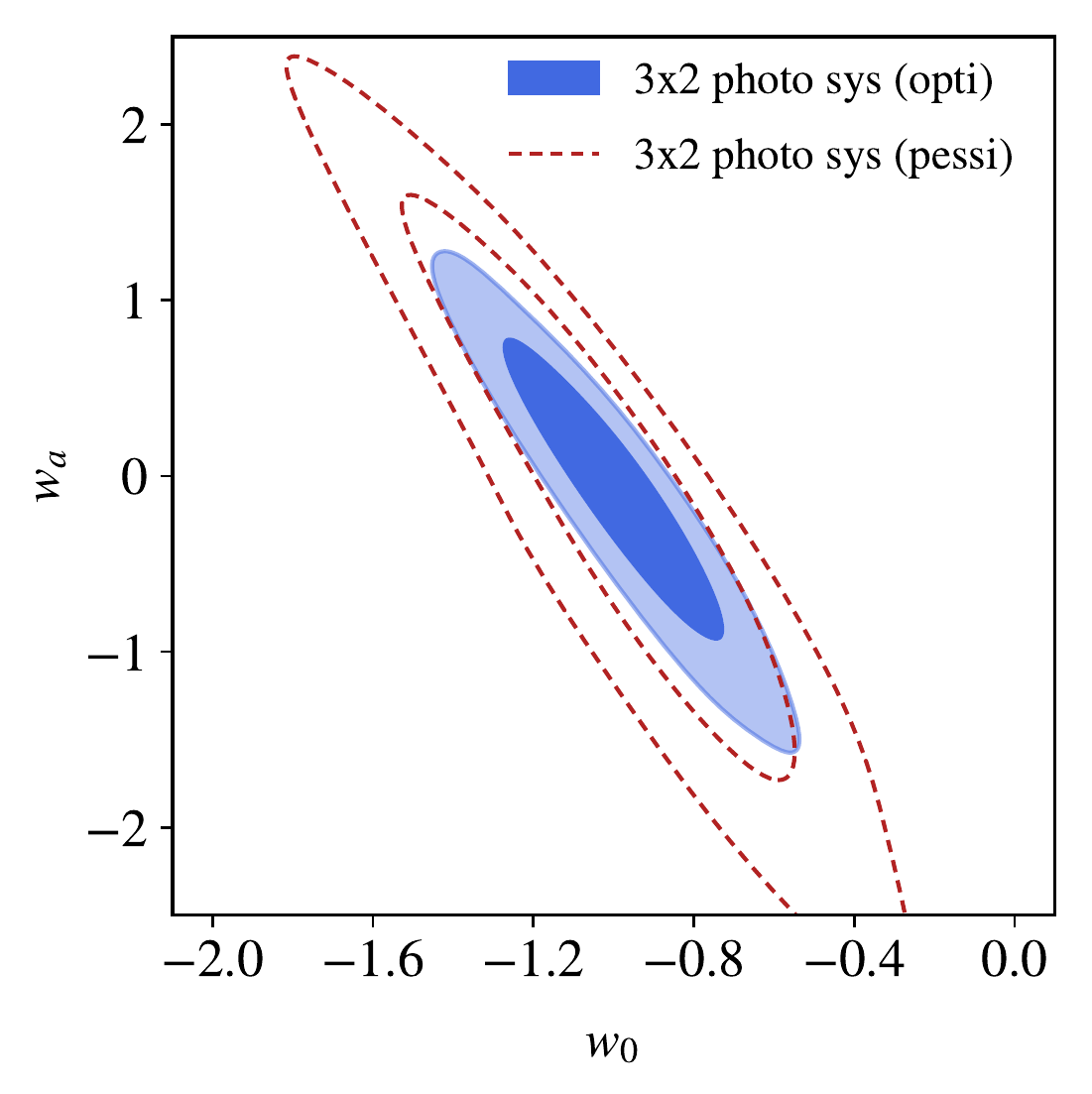}
\includegraphics[width=5.8cm]{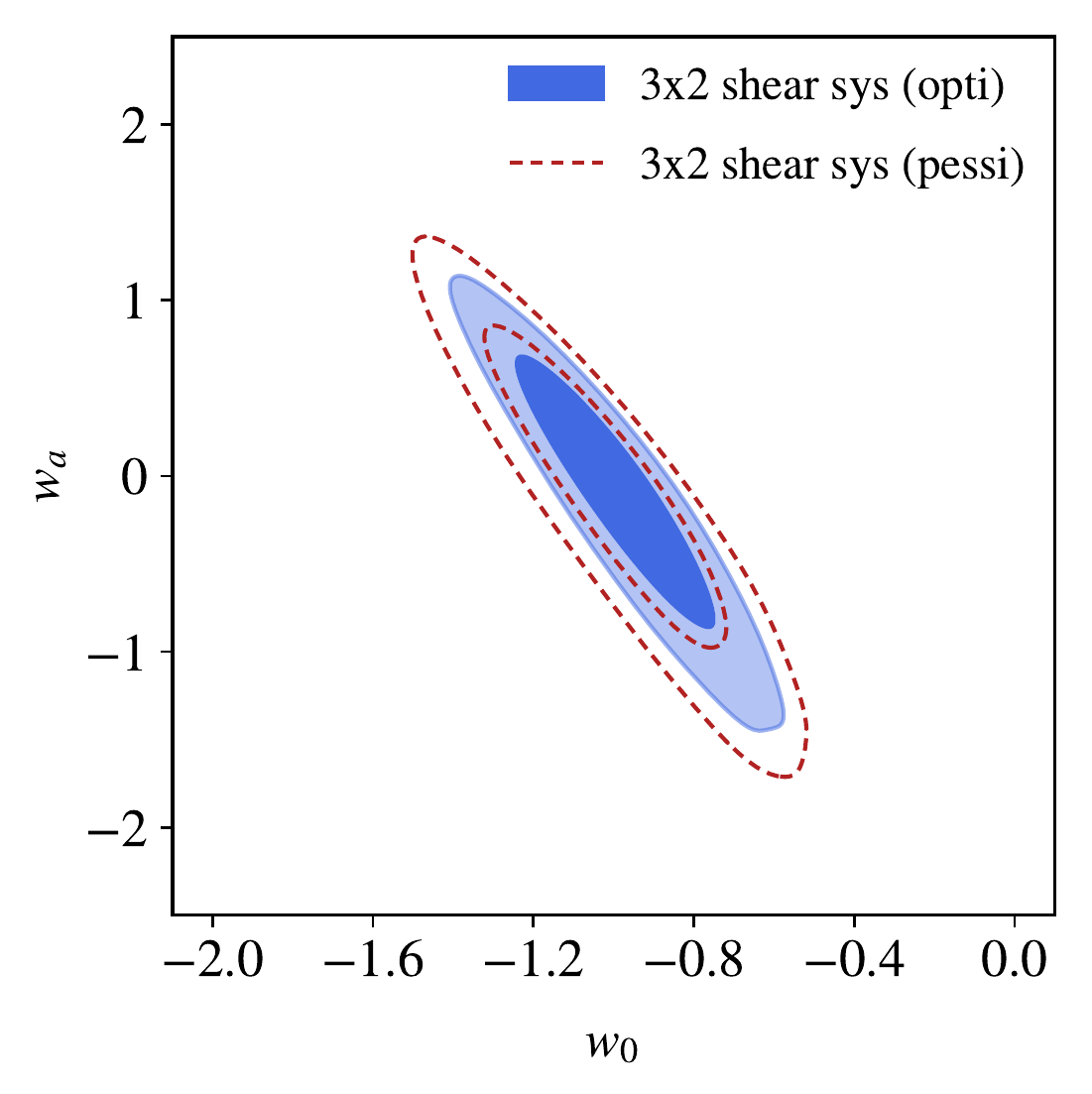}
\caption{Study of systematic effects for a 3x2 dark energy analysis. On the very left we again show Fig. \ref{fi:3x2_study} as a baseline. The center panel shows the difference when only considering photo-z uncertainties and the right panel shows results when only considering shear calibration uncertainties. There are two main findings: 1) In the optimistic scenario, shear calibration and photo-z uncertainties are equally (un)important; 2) In the pessimistic case, we find that photo-z uncertainties are a significantly larger contribution to the systematics budget compared to shear calibration.}
\label{fi:3x2_study}
\end{figure*}

\renewcommand{\arraystretch}{1.3}
\begin{table}
\caption{FoMs for optimistic and pessimistic systematics for shear and photo-z calibration depicted in Fig. \ref{fi:3x2_study}}
\begin{center}
\begin{tabular*}{0.45\textwidth}{@{\extracolsep{\fill}}| l c c |}
\hline
\multicolumn{3}{|c|}{\tbf{Systematics impact FoM summary}} \\
\hline
Systematic & optimistic & pessimistic \\  
shear+photo-z & 23.46  & 7.88   \\
photo-z & 23.56  & 7.00   \\
shear calibration  & 26.95  &  16.88  \\
\hline
\end{tabular*}
\end{center}
\label{tab:3x2_sys_fom}
\end{table}
\renewcommand{\arraystretch}{1.0}

We parameterize uncertainties arising from systematics through nuisance parameters, which are summarized with their fiducial values and priors in Table \ref{tab:3x2params}. Our default likelihood analysis includes the following systematics:

\paragraph*{Photometric redshift uncertainties} 
The true redshift distribution as measured from the CANDELS data (c.f. Fig.~\ref{fi:zdistri}) is convolved with a Gaussian photometric redshift uncertainty model to obtain the distribution within tomographic bin $i$
\be
\label{eq:photoz}
n^i_x(z_{\mathrm{ph}}) = \int_{z^i_{\mathrm{min},x}}^{z^i_{\mathrm{max},x}} dz \, n_{\mathrm{x}}(z) \, p^i\left(z_{\mathrm{ph}}|z,x\right)\,,
\ee
where $p\left(z_{\mathrm{ph}}|z,x\right)$ is the probability distribution of $z_{\mathrm{ph}}$ at given true redshift $z$ for galaxies from population $x$ 
\be
\label{eq:redbin}
p^i\left(z_{\mathrm{ph}}|z,x\right) = \frac{1}{\sqrt{2\pi}\sigma_{z,x}(1+z)}
\exp\left[-\frac{\left(z-z_{\mathrm{ph}} - \Delta^i_{z,x}\right)^2}{2\left(\sigma_{z,x}(1+z)\right)^2}\right]\,.
\ee

The resulting Gaussian tomographic bin is parameterized through scatter $\sigma_z(z)$ and bias between $z-z_{\mathrm{ph}}$, i.e. $\Delta^i_z(z)$. The bias $\Delta^i_z(z)$ has fiducial value of zero; the fiducial value for $\sigma_z$ is assumed to be the same for the lens and source sample and we choose $\sigma_z=0.01$ for the optimistic and $\sigma_z=0.05$ for the pessimistic scenario. The resulting distributions are shown in Fig.~\ref{fi:zdistri}.

In this analysis we only consider Gaussian photometric redshift uncertainties, which are characterized by scatter $\sigma_z(z)$ and bias $\Delta_z(z)$. While these may in general be arbitrary functions, we further assume that the scatter can be described by the simple redshift scaling $\sigma_{z,x}(1+z)$ and allow one (constant) bias parameter $\Delta^i_{z,x}$ per redshift bin.
For our 10 lens and source galaxy redshift bins, this model results in 22 parameters describing photo-z uncertainty, 10 photo-z bias, and one photo-z scatter parameter for each lens and source sample.   

\paragraph*{Linear galaxy bias} is described by one nuisance parameter per tomographic lens galaxy redshift bin, which is marginalized over using conservative flat priors in a likelihood analysis. The fiducial values of galaxy bias in lens bin $i$ follow the simple description $1.3+i \times 0.1$. We note that the actual fiducial value is not important for the constraining power; important is the range over which we marginalize (flat priors from 0.8-3.0) and the fact that we use one free parameter per redshift bin instead of a parameterized redshift evolution.

Future efforts should investigate several aspects of galaxy bias: 1) perturbative or simulation based parameterizations that allow the analyst to push to smaller scales; 2) improved parameterizations, in particular such that parameterize the redshift evolution with fewer parameters; 3) informative priors.

\paragraph*{Multiplicative shear calibration} is modeled using one parameter $m^i$ per redshift bin, which affects cosmic shear and galaxy-galaxy lensing power spectra via
\bea
\nonumber C_{\kappa \kappa}^{ij}(l) \quad &\longrightarrow& \quad (1+m^i) \, (1+m^j) \, C_{\kappa \kappa}^{ij}(l), \\
C_{\delta_{\mathrm{g}}\kappa}^{ij}(l) \quad &\longrightarrow& \quad (1+m^j) \, C_{\delta_{\mathrm{g}}\kappa}^{ij}(l),
\eea
where the cluster lensing power spectra are affected analogously to the galaxy-galaxy lensing spectra. We marginalize over each $m^i$ independently with Gaussian priors (10 parameters). Similar to the photo-z scenarios we are looking at optimistic and pessimistic prior information shear calibration \cite[which can come from either simulations or external data such as in][]{ske17}.

 \paragraph*{Other systematics} In this paper we only consider observational uncertainties (and galaxy bias), but neglect astrophysical systematics most notably baryonic physics uncertainties \citep[e.g.,][]{dsb11,shs11,zsd13,ekd15,crd18,hem19,cmj19} and uncertainties in modeling intrinsic alignment of galaxies \citep[e.g.,][]{his04, mhi06, job10, smm14, tri14,tsm15, bvs15, ccl15,keb16, vcs19, bmt19, sbt19}. In the context of 3x2pt analyses for WFIRST and LSST, we explore the impact of baryonic physics and intrinsic alignment in a companion paper \citep{esk20}.

%%%%%%%%%%%%%%%%%%%%%%%%%%%%%%%%%%%%%%%%%%%%%%%%%%
\section{Galaxy Clusters}
\label{sec:clusters}
%%%%%%%%%%%%%%%%%%%%%%%%%%%%%%%%%%%%%%%%%%%%%%%%%%
This section summarizes the halo model for galaxy cluster observables employed in this analysis. We consider galaxy clusters stacked in bins of optical richness, $\lambda_\alpha$, and relate their properties to dark matter halos using the probability distribution function $p(\ln \lambda|M,z)$, which describes the probability that a dark matter halo of mass $M$ at redshift $z$ hosts a cluster with richness $\lambda$. We will specify and explain our specific choice of cluster mass observable relation (MOR) further in Sect. \ref{sec:mor}.
Throughout this paper we define halo properties using the overdensity  $\Delta = 200$, which is defined with respect to the mean matter density, and employ the \citet{trk10} fitting function for the halo mass function.

%%%%%%%%%%%%%%%%%%%%%%%%%%%%%%%%%%%%%%%%%%%%%%%%%%
\subsection{Modeling of observables}
%%%%%%%%%%%%%%%%%%%%%%%%%%%%%%%%%%%%%%%%%%%%%%%%%%
%%%%%%%%%%%%%%%%%%%%%%%%%%%%%%%%%%%%%%%%%%%%%%%%%%
\paragraph*{Cluster Number Counts} The expected cluster count in richness bin $\alpha$, with $\lambda_{\alpha,\mathrm{min}}< \lambda < \lambda_{\alpha,\mathrm{max}}$, and redshift bin $i$ with $z^i_{\lambda,\mathrm{min}}<z<z^i_{\lambda,\mathrm{max}}$ is given by
\begin{equation}
\mathcal{N}^i(\lambda_\alpha) = \Omega_{\mathrm s}\int_{z^i_{\lambda,\mathrm{min}}}^{z^i_{\lambda,\mathrm{max}}}dz \frac{d^2V}{dz d\Omega} \int dM \frac{dn}{dM}\int_{\lambda_{\alpha,\mathrm{min}}}^{\lambda_{\alpha,\mathrm{max}}} d\ln\lambda\, p(\ln\lambda | M,z)\,,
\label{eq:N}
\end{equation}
where $d^2V/dzd\Omega$ is the comoving volume element, $dn/dM$ the halo mass function in comoving units for which we omitted the redshift dependence.

\paragraph*{Galaxy cluster weak lensing}
%%%%%%%%%%%%%%%%%%%%%%%%%%%%%%%%%%%%%%%%%%%%%%%%%%
\begin{figure}
\includegraphics[width=8.5cm]{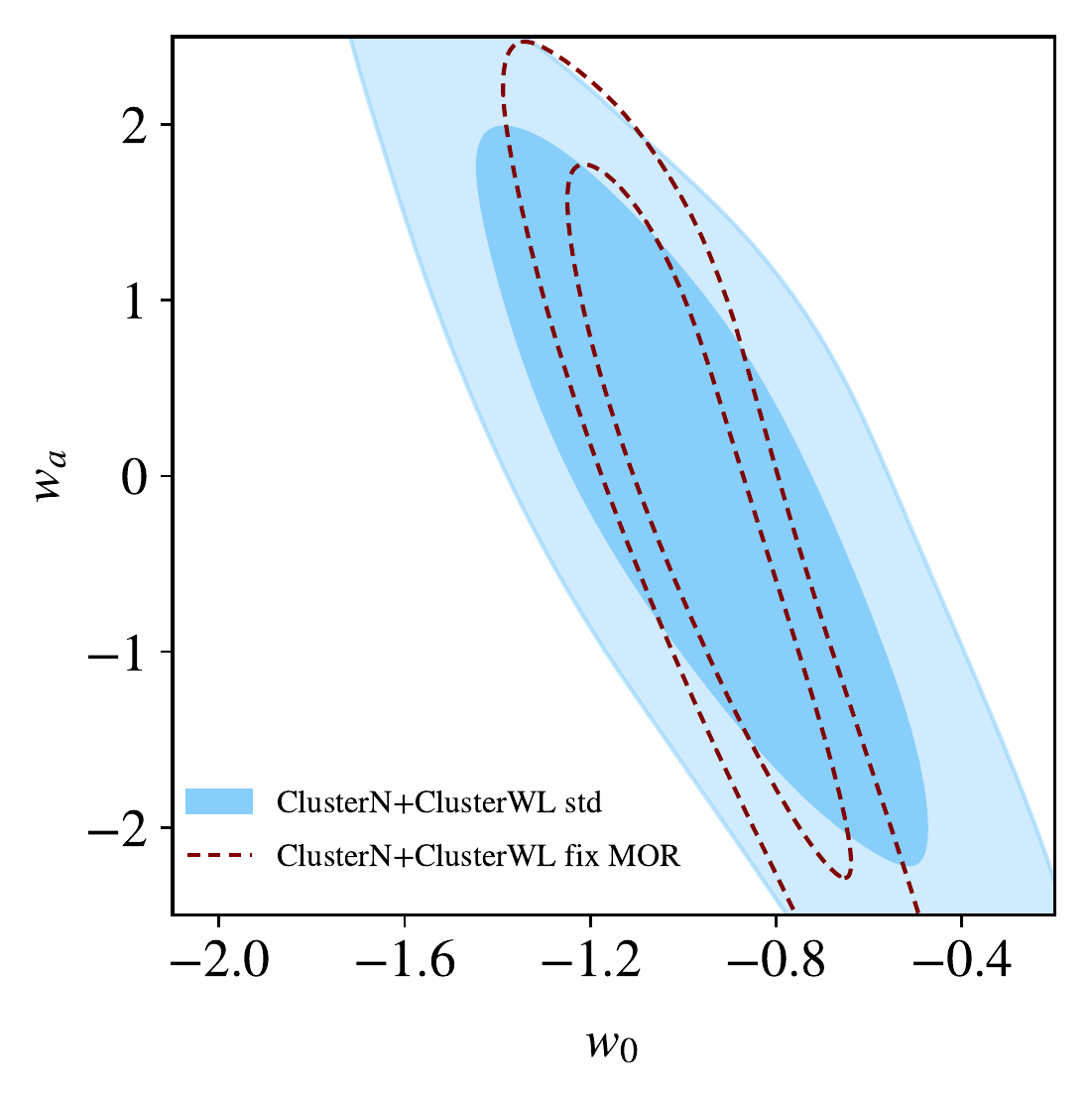}
\includegraphics[width=8.5cm]{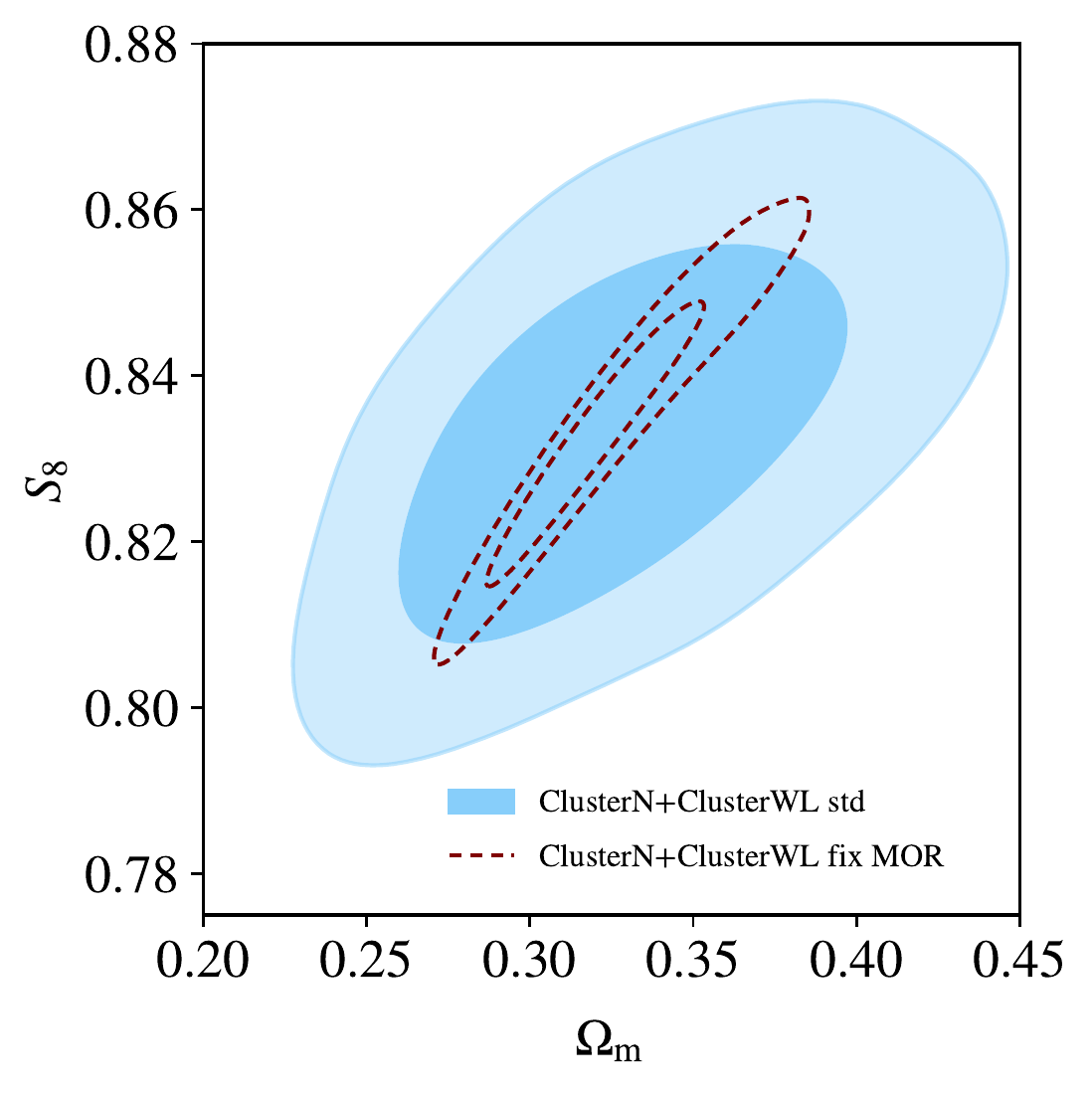}
\caption{Impact on the cosmological constraints from a joint cluster number counts and cluster weak lensing analysis when knowing the mass-observable relation perfectly. We show the equation of state parameters $w_0$, $w_a$ (\textit{upper panel}) and the combination $\om$ and $S_8=\sigma_8 \times (\om /0.315)^{0.35}$ (\textit{lower panel}).}
\label{fi:MOR_study}
\end{figure}

Starting again from the Limber and flat-sky expression for projected power spectra, i.e. Eq. (\ref{eq:projected})
\be
\label{eq:projected3}
C_{AB}^{ij}(l) = \int d\chi \frac{q_A^i(\chi)q_B^j(\chi)}{\chi^2}P_{AB}(l/\chi,z(\chi)),
\ee
we can express the weight function for the projected cluster density similar to Eqs. (\ref{eq:qg}, \ref{eq:qkappa})
\begin{equation}
\label{eq:qlambda}
q_{\delta_{\lambda_\alpha}}^i(\chi) =  \Theta\left(z(\chi)-z^i_{\lambda,\mathrm{min}}\right)\Theta\left(z^i_{\lambda,\mathrm{max}}-z(\chi)\right)\frac{dV}{d\chi d\Omega}\,,
\end{equation}
with $\Theta(x)$ the Heaviside step function. Note, that we neglect variations of the cluster selection function within redshift bins, as well as uncertainties in the cluster redshift estimate.

Within the halo model, the cross power spectrum between cluster centers and matter density contrast can be written as the usual sum of two- and one-halo term,
\begin{eqnarray}
\label{eq:Pcluster}
\hspace{-0.7cm} P_{\delta_{\lambda_\alpha} m}(k,z) &\approx& b_{\lambda_\alpha}(z) P_{\mathrm{lin}}(k,z) \nn \\
&&\hspace{-1.5cm} + \frac{ \int dM  \frac{dn}{dM}\frac{M}{\bar{\rho}} \tilde{u}_{\mathrm{m}}(k,M) \int_{\ln \lambda_{\alpha,\mathrm{min}}}^{\ln \lambda_{\alpha,\mathrm{max}}} d\ln \lambda \,p(\ln \lambda | M,z)}{\int dM \frac{dn}{dM}  \int_{\ln \lambda_{\alpha,\mathrm{min}}}^{\ln \lambda_{\alpha,\mathrm{max}}} d\ln \lambda\, p(\ln \lambda | M,z)},
\end{eqnarray}
with $P_{\mathrm{lin}}(k,z)$ the linear matter power spectrum. The mean linear bias of clusters in richness bin $\alpha$ reads 
\begin{equation}
\label{eq:blambda}
b_{\lambda_\alpha}(z) =  \frac{ \int dM  \frac{dn}{dM} b_{\mathrm{h}}(M)\int_{\ln \lambda_{\alpha,\mathrm{min}}}^{\ln \lambda_{\alpha,\mathrm{max}}} d\ln \lambda \,p(\ln \lambda | M,z)}{\int dM   \frac{dn}{dM}\int_{\ln \lambda_{\alpha,\mathrm{min}}}^{\ln \lambda_{\alpha,\mathrm{max}}} d\ln \lambda\, p(\ln \lambda | M,z)}\,,
\end{equation}
where $b_{\mathrm{h}}(M)$ the halo bias relation, for which we use the fitting function of \citet{trk10}. The Fourier transform of the radial matter density profile within a halo of mass $M$, $\tilde{u}_{\mathrm{m}}(k,M)$, is modeled assuming the Navarro-Frenk-White (NFW) profile \citep{nfw97} with the \citet{bhh11} mass-concentration relation $c(M,z)$.

%%%%%%%%%%%%%%%%%%%%%%%%%%%%%%%%%%%%%%%%%%%%%%%%%%
\subsection{Systematics}
\label{sec:mor}
%%%%%%%%%%%%%%%%%%%%%%%%%%%%%%%%%%%%%%%%%%%%%%%%%%

\paragraph*{Cluster mass-observable relation}
We chose to implement the MOR scatter defined in \cite{mon19} and further extend their parameterization to account for possible redshift dependence in the scatter of the mass-richness relation. 

Specifically, we assume a log-normal distribution with mass- and redshift-dependent mean and scatter $\sigma_{\ln \lambda|M}$
\be
p(\ln \lambda| M,z) = \frac{1}{\sqrt{2\pi}\sigma_{\ln\lambda|M,z}} \exp \left[-\frac{(\ln\lambda-\left\langle\ln\lambda\right\rangle(M))^2}{2\sigma_{\ln\lambda|M,z}^2}\right]\,.
\ee
The mean relation is defined as
\begin{equation}\label{eq:cl:mor}
\left\langle \ln \lambda \right\rangle (M,z|A, B, C) = A + B\ln\left(\frac{M}{M_{\rm pivot}}\right)+C\ln\left(1+z\right),
\end{equation}
with normalization $A$, slope $B$, redshift dependence $C$, and the pivot mass $M_{\rm pivot}=3\times10^{14}\ M_{\odot}/h$. The mass- and redshift dependent MOR scatter is defined as
\begin{equation}\label{eq:cl:morscat}
\sigma_{\ln \lambda|M}(M, z| \sigma_0, q_M, q_z)=\sigma_0+q_M\ln\left(\frac{M}{M_{\rm pivot}}\right)+q_z\ln\left(1+z\right).
\end{equation}
We assume fiducial values for $(A, B, \sigma_0, q_M)=(3.207, 0.993, 0.456, 0.0)$, which correspond to the findings in \cite{mon19}. For the redshift-dependent MOR parameters which are newly introduced in this paper ($C$ and $q_z$) we assume fiducial values of 0.

\renewcommand{\arraystretch}{1.3}
\begin{table}
\caption{Fiducial parameters, flat priors (min, max), and Gaussian priors centered on the fiducial value with the $\sigma$ given in brackets.}
\begin{center}
\begin{tabular*}{0.45\textwidth}{@{\extracolsep{\fill}}| c c c |}
\hline
\multicolumn{3}{|c|}{\tbf{Cluster Mass Observable Relation scenarios}} \\
\hline
Parameter & Fiducial &  Prior \\  
\hline
 $A$ & 3.207 & Gauss (3.207,0.045)  \\
 $B$ & 0.993 & Gauss (0.993,0.045) \\
 $C$ & 0.0 & Gauss (0.0,0.3)\\
 $\sigma_0$ &0.456 & Gauss (0.456,0.045) \\
 $q_M$ & 0.0  & Gauss (0.0,0.03) \\
 $q_z$ &0.0  & Gauss (0.0,0.1) \\
\hline
\end{tabular*}
\end{center}
\label{tab:clusterparams}
\end{table}
\renewcommand{\arraystretch}{1.0}

Our fiducial priors for $\sigma_0$ and $q_M$ are from the posterior distributions derived in \cite{mon19}, i.e., a Gaussian prior centered at the fiducial values described above and with the width of $0.045$ and $0.03$, respectively, and a prior for $q_z$ is centered at 0 with the broader width of $0.1$. 

We note that this is conservative, since prior information on the MOR is expected to grow substantially in the coming years, near-term with the full HSC survey, which will be one of the deepest imaging surveys yielding the most stringent constraints on galaxy cluster physics before the LSST and WFIRST era. 

For example, the full HSC survey will have 20,000 optically-selected clusters with a mean galaxy density of background sources of 20~arcmin$^{-2}$. Scaling the product of the number of clusters and the source number density in \cite{mon19}, 8,000~clusters and 1~arcmin$^{-2}$, respectively, to the product of these numbers for the full HSC survey, translates into a factor of 7 improvement on the priors of the MOR under the assumption that we can translate optical richness as measured in HSC into the realm of NIR WFIRST measurements. 

In Fig. \ref{fi:MOR_study} we investigate the gain in constraining power for a perfectly known MOR, i.e. when fixing all the parameters in Table \ref{tab:clusterparams} to their fiducial values. The gain in information from blue contours to red serves as an upper limit for this particular choice of MOR parameterization. We note that we expected a larger improvement when assuming perfect knowledge of the MOR but we note that the redshift scaling in Eq. (\ref{eq:cl:mor}) is likely the reason to diminish the science return on dark energy. 

Studying the most promising cluster MOR parameterization to optimize the cluster cosmology component of the WFIRST survey further will be important future work as the mission preparation progresses. 

\paragraph*{Other systematics} We do not consider galaxy cluster mis-centering, assembly bias and stochasticity in this paper but instead postpone studies of these effects to future work. 

%%%%%%%%%%%%%%%%%%%%%%%%%%%%%%%%%%%%%%%%%%%%%%%%%%
%%%%%%%%%%%%%%%%%%%%%%%%%%%%%%%%%%%%%%%%%%%%%%%%%%
\section{The High Latitude Spectroscopic Survey}
\label{sec:HLSS}
%%%%%%%%%%%%%%%%%%%%%%%%%%%%%%%%%%%%%%%%%%%%%%%%%%
%%%%%%%%%%%%%%%%%%%%%%%%%%%%%%%%%%%%%%%%%%%%%%%%%%
In this section, we study the trade space of area versus depth for the High Latitude Spectroscopic Survey, starting from a baseline survey of 2000 deg$^2$ and a wavelength range of 1.05-1.85 microns. The section is split into two parts, where the first part focuses on dark energy parameter constraints using MCMC and the second part is a Fisher analysis of how well WFIRST will be able to measure the BAO scale $s$ and the parameter combination $f \sigma_8$ for RSD. 
The assumptions and systematics modeling differ slightly but are clearly explained in each subsection.

%%%%%%%%%%%%%%%%%%%%%%%%%%%%%%%%%%
\subsection{Dark energy forecasts}
\label{sec:grsmethod}
%%%%%%%%%%%%%%%%%%%%%%%%%%%%%%%%%%
\renewcommand{\arraystretch}{1.3}
\begin{table}
\caption{HLSS survey parameters.}
\begin{center}
\begin{tabular*}{0.45\textwidth}{@{\extracolsep{\fill}}| c c c |}
\hline
\multicolumn{3}{|c|}{\tbf{HLSS survey params ($\Omega_\mr s$ = 2000 deg$^2$)}} \\
\hline
Redshift & Comoving volume  &  Galaxy density \\
(density weighted) & ($10^9$ Mpc/h)$^3$   &  (h/Mpc)$^3$\\
\hline
 0.84 & 0.041 & 0.003803 \\
 1.28 & 0.062 & 0.002845 \\
 1.75 & 0.085 & 0.001182 \\
 2.28 & 0.111 & 0.000503 \\
 2.75 & 0.133 & 0.000195 \\
 3.26 & 0.158 & 0.000069 \\
 3.71 & 0.180 & 0.000025 \\
 \hline
\end{tabular*}
\end{center}
\label{tab:specsurveyparams}
\end{table}
\renewcommand{\arraystretch}{1.0}

\begin{figure*}
\includegraphics[width=5.5cm]{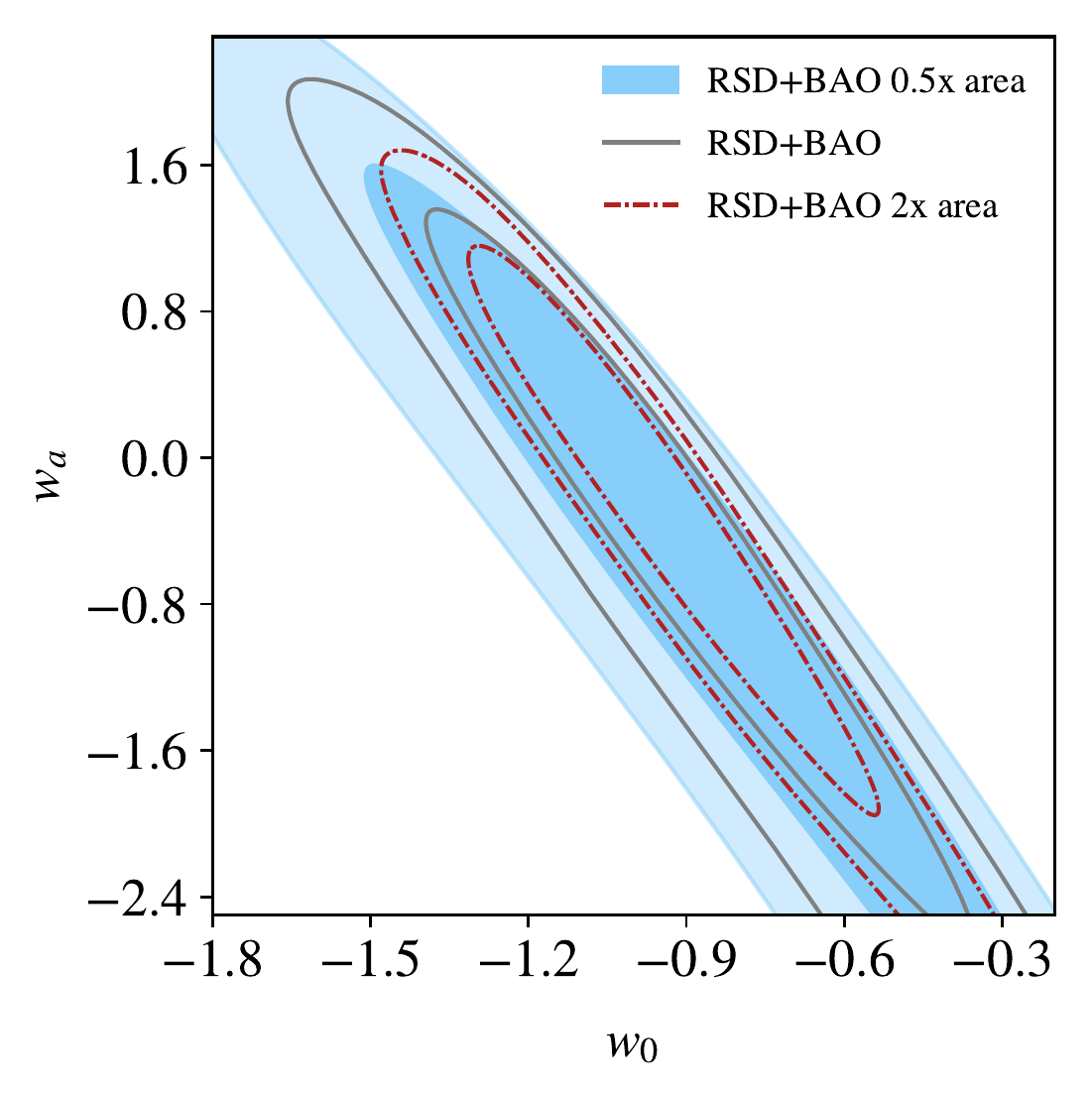}
\includegraphics[width=5.5cm]{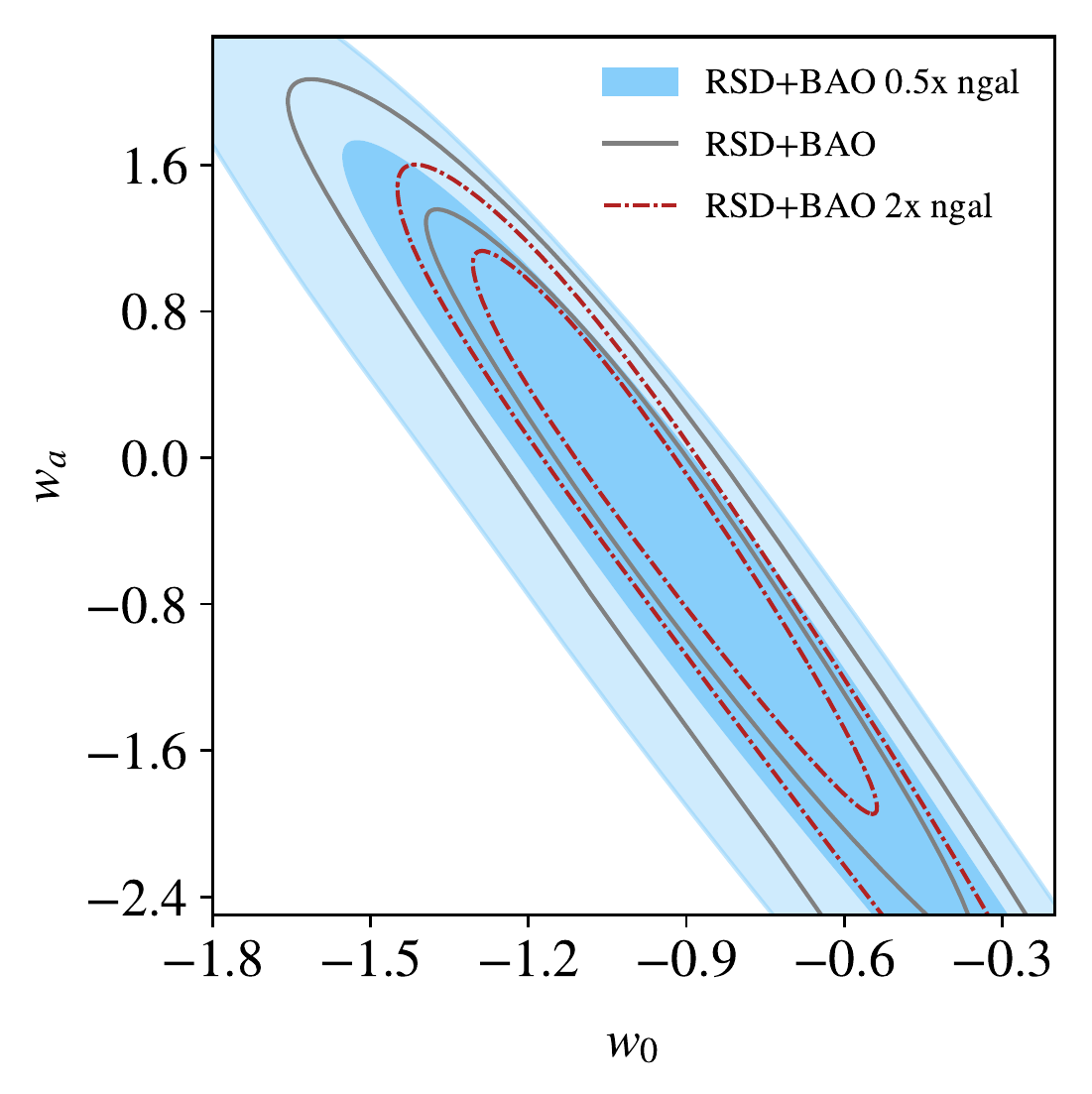}
\includegraphics[width=5.5cm]{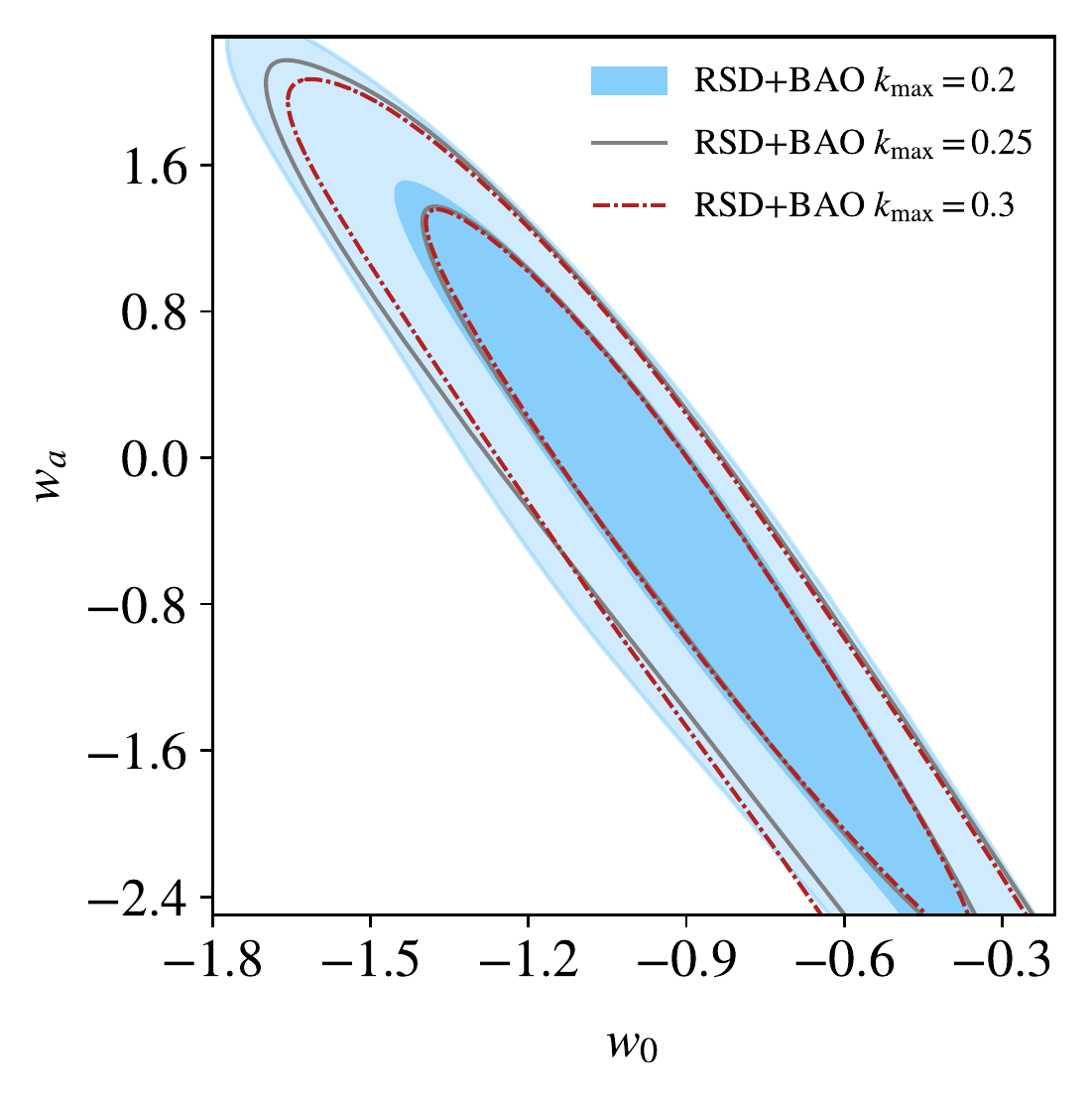}
\caption{The impact of variations in area, depth, and scales to which we assume to be able to model $P_\delta(k)$ for the HLSS part of the reference survey (0.6 months). We summarize the FoMs in Table \ref{tab:GRS_fom}.}
\label{fi:GRS_study}
\end{figure*}

\renewcommand{\arraystretch}{1.3}
\begin{table}
\caption{Spectroscopic Survey: Fiducial parameters, flat priors (min, max), and Gaussian priors centered on the fiducial value with the $\sigma$ given in brackets.}
\begin{center}
\begin{tabular*}{0.45\textwidth}{@{\extracolsep{\fill}}| c c c |}
\hline
\multicolumn{3}{|c|}{\tbf{HLSS systematics parameters}} \\
\hline
Parameter & Fiducial &  Prior \\  
\hline
 $b_1$ & 1.55 & [0.6-4.2] \\
 $b_2$ & 1.87 & [0.6-4.2] \\
 $b_3$ & 2.22 & [0.6-4.2] \\
 $b_4$ & 2.62 & [0.6-4.2] \\
 $b_5$ & 2.97 & [0.6-4.2] \\
 $b_6$ & 3.38 & [0.6-4.2] \\
 $b_7$ & 3.72 & [0.6-4.2] \\
 $\sigma_p (i)$ & 290 km/s & Gaussian (290, 50) \\
 $k_*$ & 0.24 h/Mpc & Gaussian (0.24, 0.024) \\
 $\sigma_{r,z}$ & 0.001 & Gaussian (0.001,0.0001) \\
 $P_\mr{shot}$ & 0.0 & [-0.001,0.001] \\

\hline
\end{tabular*}
\end{center}
\label{tab:specparams}
\end{table}
\renewcommand{\arraystretch}{1.0}

We use the WFIRST exposure time calculator (ETC) version 16 of \citet{hgk12} to compute galaxy densities and redshift distributions for our baseline scenario (c.f. Table \ref{tab:specsurveyparams}) and then consider doubling (halving) the survey area, doubling (halving) the galaxy number density, and decreasing the minimum scale which we include in our analysis (see Fig. \ref{fi:GRS_study}). 

Following \citep{see03,wch13} we model the cosmological information from redshift space distortions (RSD) and Baryon Acoustic Oscillations (BAO) through features in the observed power spectrum  
\bea
\label{eq:Pg_seo}
P_{g}(k^{\rm ref}_{\perp},k^{\rm ref}_{\parallel}) &=&
\frac{\left[D_A^{\rm ref}(z)\right]^2  H(z)}{\left[D_A(z)\right]^2 H^{\rm ref}(z)}
\, b^2 \left( 1+\beta\, \mu^2 \right)^2 \\
&\times& \left[ \frac{G(z)}{G(z=0)}\right]^2 \, P_m(k, z=0) \, e^{-k^2\mu^2 \sigma^2_{r,z}} + P_{\rm shot}\,,
\nonumber
\eea
where we assume that the 3D Fourier mode $\bf k$ can be decomposed into a line-of-sight $k_\parallel$ and a transverse $k_\perp$ component with $\mu=k_\parallel/|\bf k|$ as the cosine of the angle between the 3D vector and the line-of sight. The  arguments for the observed power spectrum $k^{\rm ref}_{\perp}$ and $k^{\rm ref}_{\parallel}$ are computed at a reference cosmology, indicated through the superscript $^\mr{ref}$. In order to relate the observed power spectrum to the true underlying power spectrum a correction factor  
$\left([D_A^{\rm ref}(z)]^2  H(z)\right)/\left([D_A(z)]^2 H^{\rm ref}(z)\right)$
which accounts for the volume difference between the two cosmologies is introduced. The $P_\mr{shot}$ term describes residual uncertainties that remain after subtracting the shot noise term computed from the inverse number density of galaxies. These residuals occur, e.g., because of galaxy clustering bias \citep{sel00}.  
Equation (\ref{eq:Pg_seo}) accounts for residual redshift uncertainty in our measurement, e.g. from fitting emission lines, through the damping factor $e^{-k^2\mu^2 \sigma^2_{r,z}}$. Following \cite{wch13} we consider the dewiggeled power spectrum 
\be
P_m(k, z=0) = P_0 \, k^{n_s} \, T^2_{\rm dw}(k) \,,
\ee
where $P_0$ defines the normalization of the linear power spectrum at redshift zero, $n_s$ is the spectral index, and the (dewiggeled) transfer function $T^2_{\rm dw}(k,z)$ is given by
\bea
T^2_{\rm dw}(k,z)&=& T^2_{\rm nw}(k) + \left[T^2(k)-T^2_{\rm nw}(k)\right] e^{-g_\mu k^2 /(2k_*^2)}\nonumber \\
&\equiv & T^2_{\rm nw}(k) + T^2_{\rm BAO}(k) e^{-g_\mu k^2 /(2k_*^2)} \,,
\label{eq:dwtransfer}
\eea
where $g_{\mu} (k,z) = 1 - \mu^2+ \mu^2[(1+f_g(z))^2 - 1]$ \citep[c.f.]{esw07} and $f_g(z)$ being the linear growth factor.

The BAO transfer function is defined as the difference between the linear matter transfer functions with and without baryons, and the exponential damping due to nonlinear effects is only applied to the transfer function associated with BAO. The uncertainty in nonlinear effects that are still present in the power spectrum even after reconstruction \citep{see07,pxe12} is paramterized through 
\be 
k_*^{-1}=8.355\, \mr{Mpc}/h \, \frac{\sigma_8}{0.8} \, p_\mr{NL} \,.
\ee
In case no reconstruction algorithm is applied nonlinear effects in structure growth, galaxy bias, and redshift space distortions are fully present and $p_\mr{NL}=1.0$. We assume an optimistic reconstruction algorithm in line with \cite{wch13} of $p_\mr{NL}=0.5$, which corresponds to $k_*=0.24  h/\mathrm{Mpc}$. We allow for uncertainty in the reconstruction algorithm through varying $k_*$ and marginalize over a Gaussian prior with 10\% uncertainty in the fiducial value.

The dewiggled model characterized through Eq. \ref{eq:dwtransfer} will break down on small scales where RSD couples with the damping factor but has been shown to work well on quasi-linear scales \citep{abf08}.

We bin the observable power spectrum linearly in $k$ (100 bins between $k_\mr{min}=0.001$ and $k_\mr{max}=0.3$) and $\mu$ (10 bins between 0 and 1) and assume 7 bins in redshift (c.f. Table \ref{tab:specsurveyparams}). We model the fractional error of said power spectrum as detailed in \cite{see03}
\be
\label{eq:GRS_error}
\sigma(k,\mu)=2 \pi \sqrt{\frac{2}{V_\mr{survey} k^2 \Delta k \Delta \mu}} \left(\frac{1+nP(k, \mu)}{n} \right) \,,
\ee
where $n$ refers to the galaxy number density within a given redshift bin, which again are specified in Table \ref{tab:specsurveyparams}.

Figure \ref{fi:GRS_study} shows the variation of the WFIRST and BAO and RSD measurements on $w_0$ and $w_a$. We again use the \texttt{emcee} sampler to cover the parameter space; each chain is $>3$M steps and, in addition to the cosmological parameters mentioned in Table \ref{tab:3x2params}, we sample the 11 systematics parameters specified in Table \ref{tab:specparams}. Specifically, we account for uncertainties in the level of shot noise $P_\mr{shot}$ (1 parameter), uncertainties in galaxy bias modeling parameterized through one free parameter $b_i$ in each redshift bin (7 parameters), uncertainties in redshift measurements $\sigma^2_{r,z}$ (1 parameter), uncertainties in modeling peculiar velocities $\sigma_p$ in each redshift bin (7 parameters), uncertainty in residual nonlinear effects $k_*$ (1 parameter). 

Figure \ref{fi:GRS_study} shows the change in constraining power when increasing/decreasing the survey area (\ti{left}), increasing/decreasing the number density of galaxies (\ti{middle panel}) and when changing our fiducial $k_\mr{max}$ from 0.3 to 0.25 and 0.2. Note that the observing time is not held fixed in the left and middle panel (as opposed to the calculations in  Sect.~\ref{sec:fisher}), which means that when considering twice the area in the left panel this implies doubling the observing time compared to reference HLSS survey. We summarize the FoMs in Table \ref{tab:GRS_fom} and find that the difference for different $k_\mr{max}$ is negligible, and that there is a slight preference for going deeper compared to going wider.
\begin{figure}
\includegraphics[width=8cm]{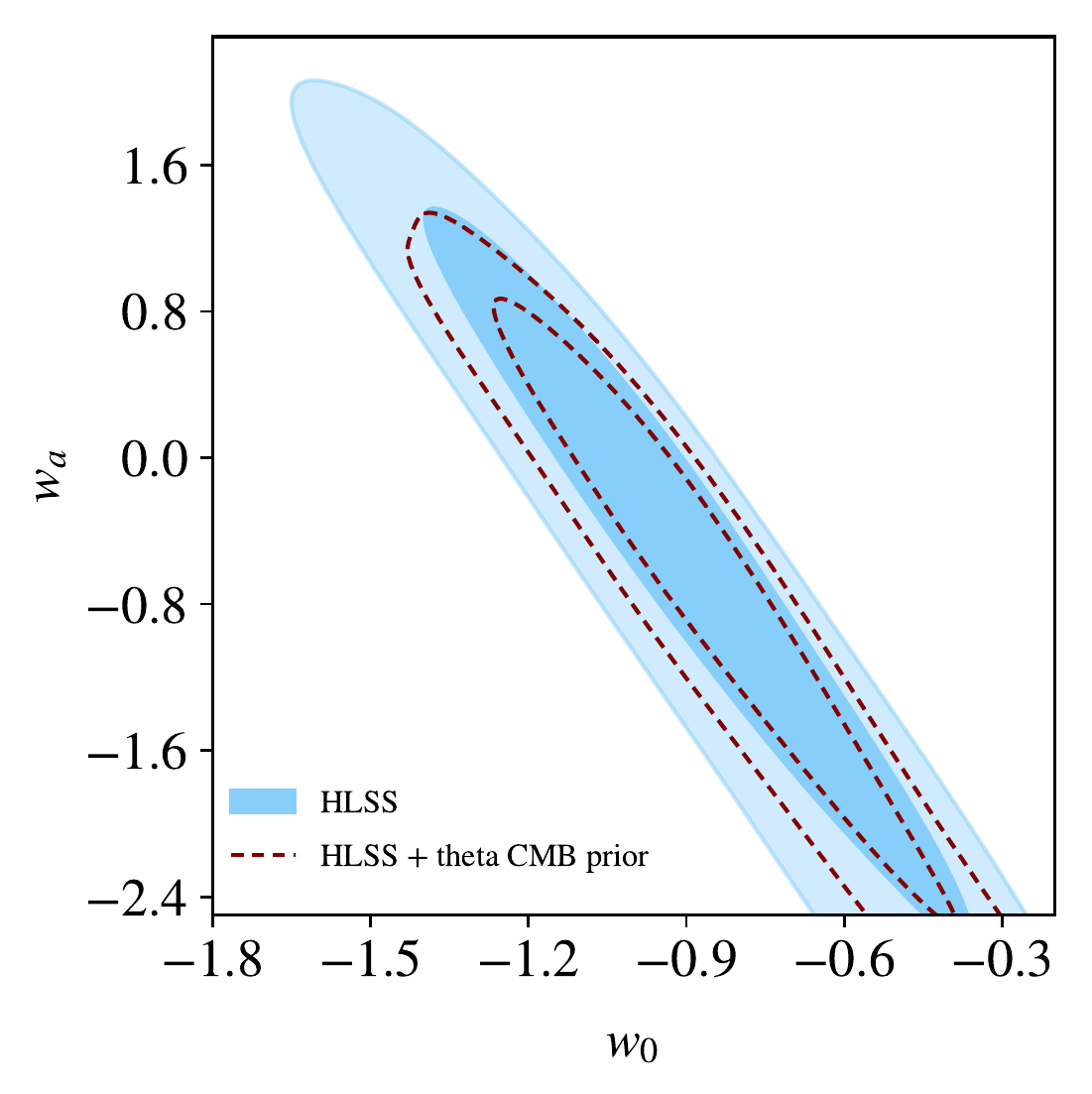}
\caption{ We see the gain in constraining power when assuming that the scale of the BAO feature in the CMB is known at Planck precision \citep[see Eq. 9 in ][]{Planckcosmo18}.}
\label{fi:GRS_prior_study}
\end{figure}

\renewcommand{\arraystretch}{1.3}
\begin{table}
\caption{FoM for chains depictd in Fig. \ref{fi:GRS_study}.}
\begin{center}
\begin{tabular*}{0.45\textwidth}{@{\extracolsep{\fill}}| c c c c |}
\hline
\multicolumn{4}{|c|}{\tbf{HLSS FoM summary}} \\
\hline
Area & 2000 deg$^2$ & 4000 deg$^2$ &  1000 deg$^2$ \\  
FoM & 8.19  & 14.34  & 5.33 \\
\hline
Galaxy density & reference & 2 x ref & 0.5 x ref \\  
FoM & 8.19  & 14.60 & 4.74 \\
\hline
$k_\mr{max}$ & 0.3 & 0.25 & 0.2 \\  
FoM & 8.19  & 7.79 & 6.68 \\
\hline
\end{tabular*}
\end{center}
\label{tab:GRS_fom}
\end{table}
\renewcommand{\arraystretch}{1.0}

We note that including an absolute measurement of the BAO scale imprinted in the CMB would notably increase the information compared to the HLSS survey alone. In Fig. \ref{fi:GRS_prior_study} we include information from \citep{Planckcosmo18} on the acoustic angular scale $\theta_{*}=r_{*}/(1+z) D_a$, where $r_{∗}$ is the comoving sound horizon at recombination and $D_a$ is the comoving angular diameter distance to the CMB. The combined likelihood of Planck TT, TE, EE, low-E measurements gives $\theta_{*}=0.0104109 \pm 0.0000030$, which we re-center to our fiducial cosmolgy and use as a prior in Fig. $\ref{fi:GRS_prior_study}$.

% If in addition, galaxy bias uncertainties can be reduced to a Gaussian with $\sigma=0.2$ instead of the wide flat priors we assume in our standard analysis, we can shrink the uncertainties even more c.f. Fig. \ref{fi:BAO} lower panel). 

%%%%%%%%%%%%%%%%%%%%%%%%%
\subsection{BAO scale and RSD measurement Fisher forecasts}
\label{sec:fisher}
%%%%%%%%%%%%%%%%%%%%%%%%%
\begin{figure}
	\includegraphics[width=0.45\textwidth]{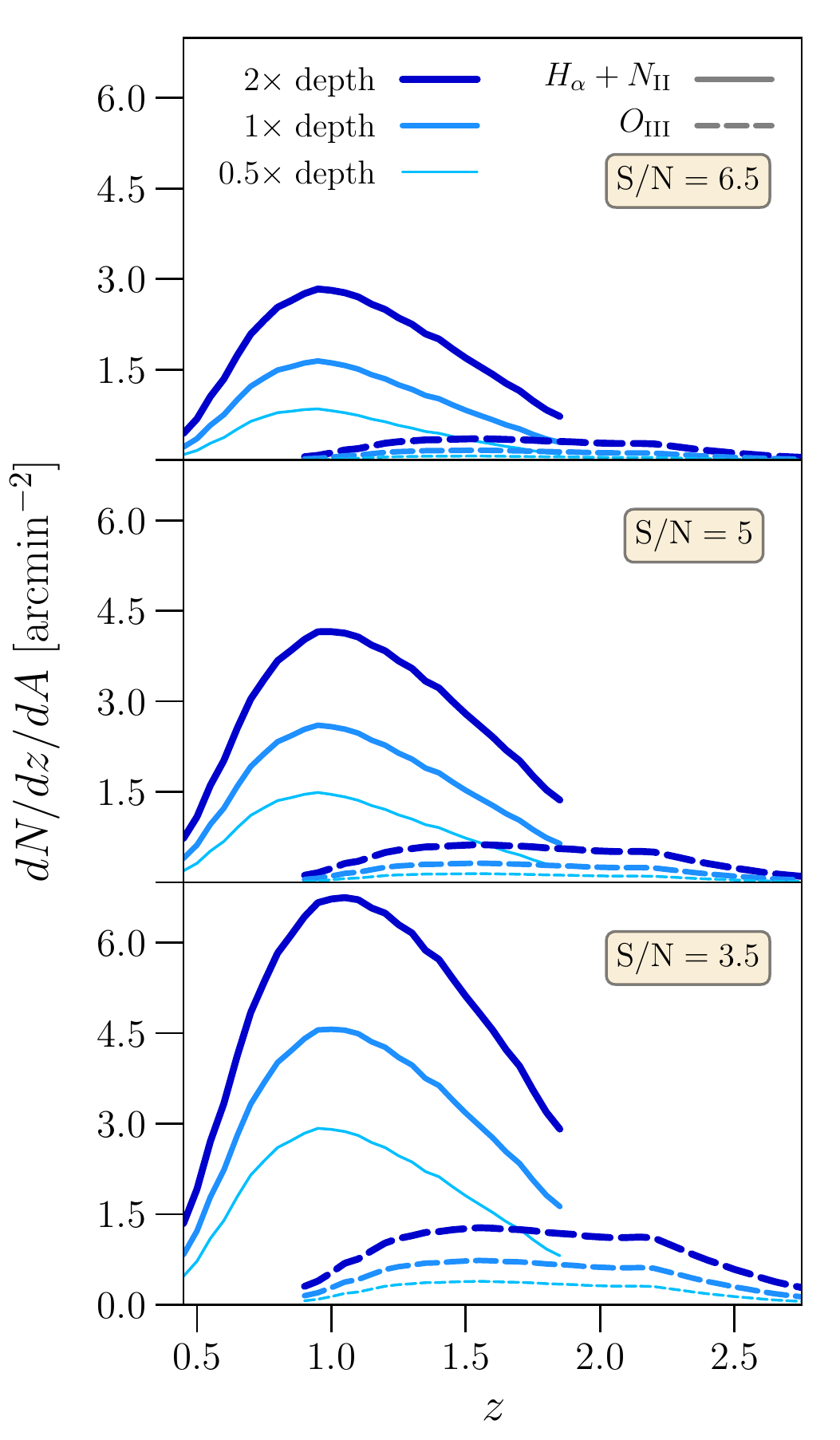}
\caption{$dN/dz/dA$ for the two galaxy populations used in the BAO and RSD forecast: H$\alpha\,$+[\,N$_{\rm II}$] (solid lines) and [O$_{\rm III}$] (dashed lines). The various curves used in the trade-off study are shown: (1) from top to bottom panel, varying the $S/N$ cutoff from 6.5, 5 to 3.5; (2) inside each panel, varying survey depth from 2x, 1x to 0.5x the fiducial depth with decreasing thickness. These curves are used as input for the forecast results shown in Fig.~\ref{fi:grs}.}
\label{fig:dndz}
\end{figure}

\begin{figure*}
\includegraphics[width=.45\textwidth]{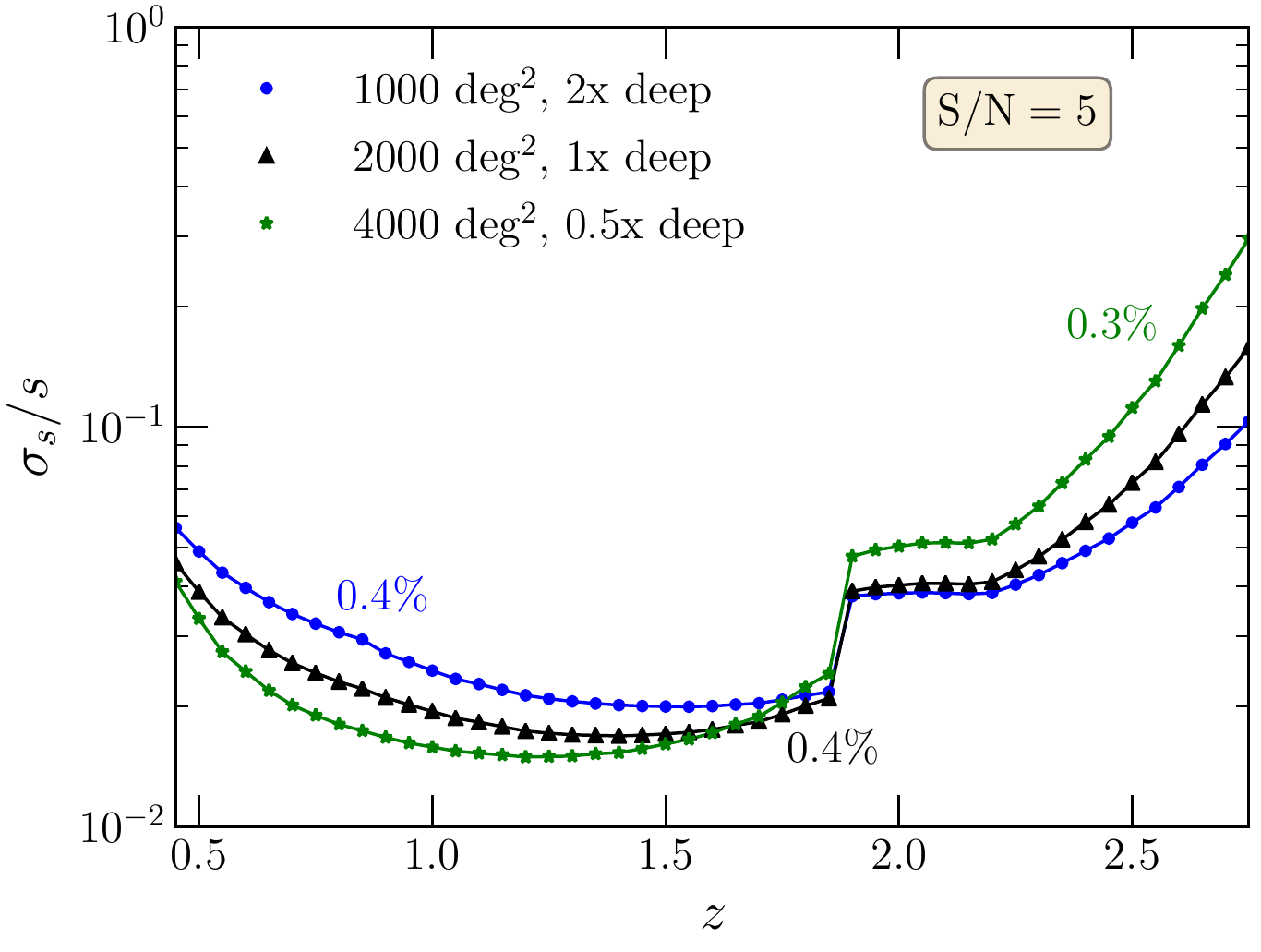}
\includegraphics[width=.45\textwidth]{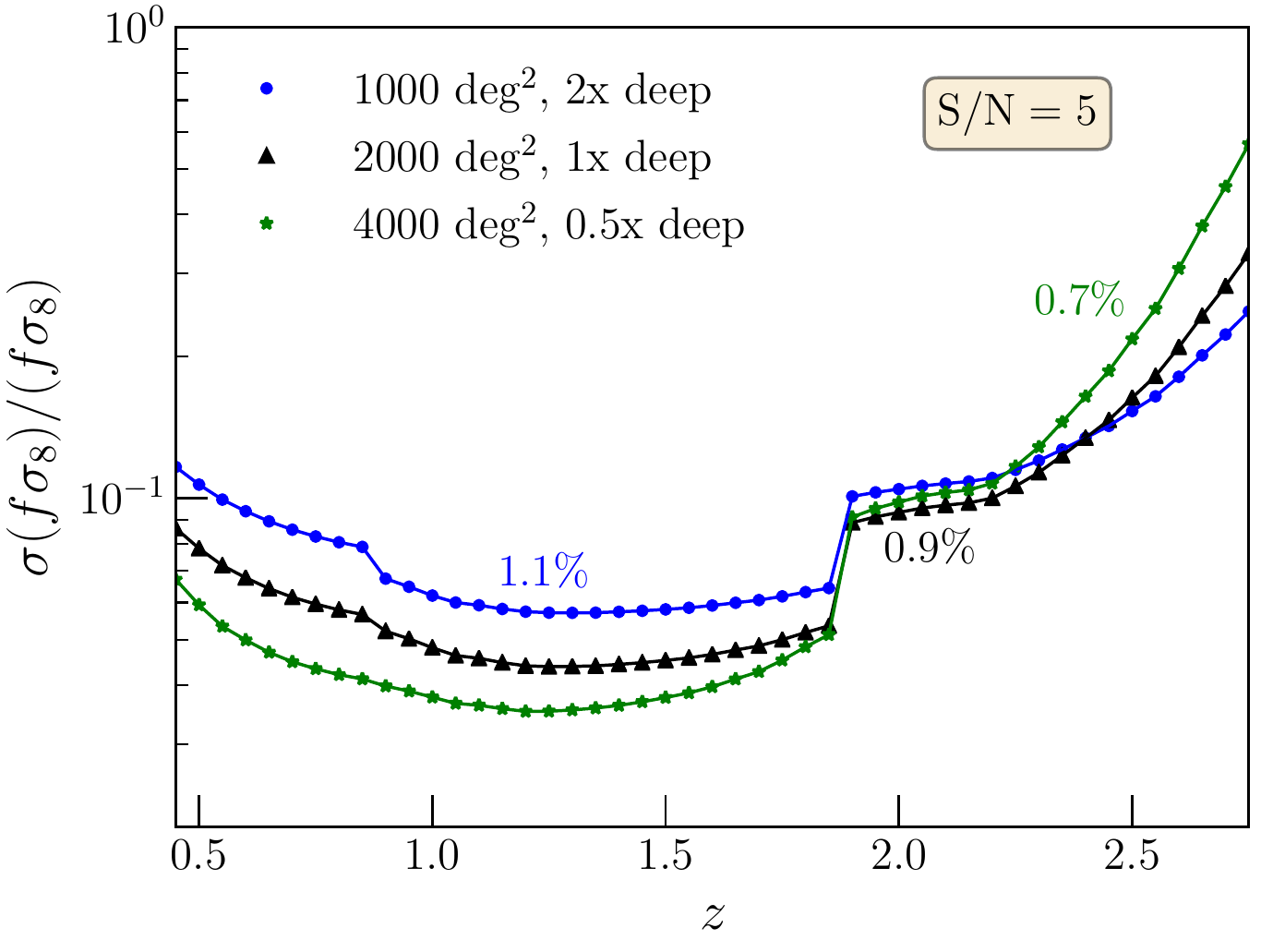}
\includegraphics[width=.45\textwidth]{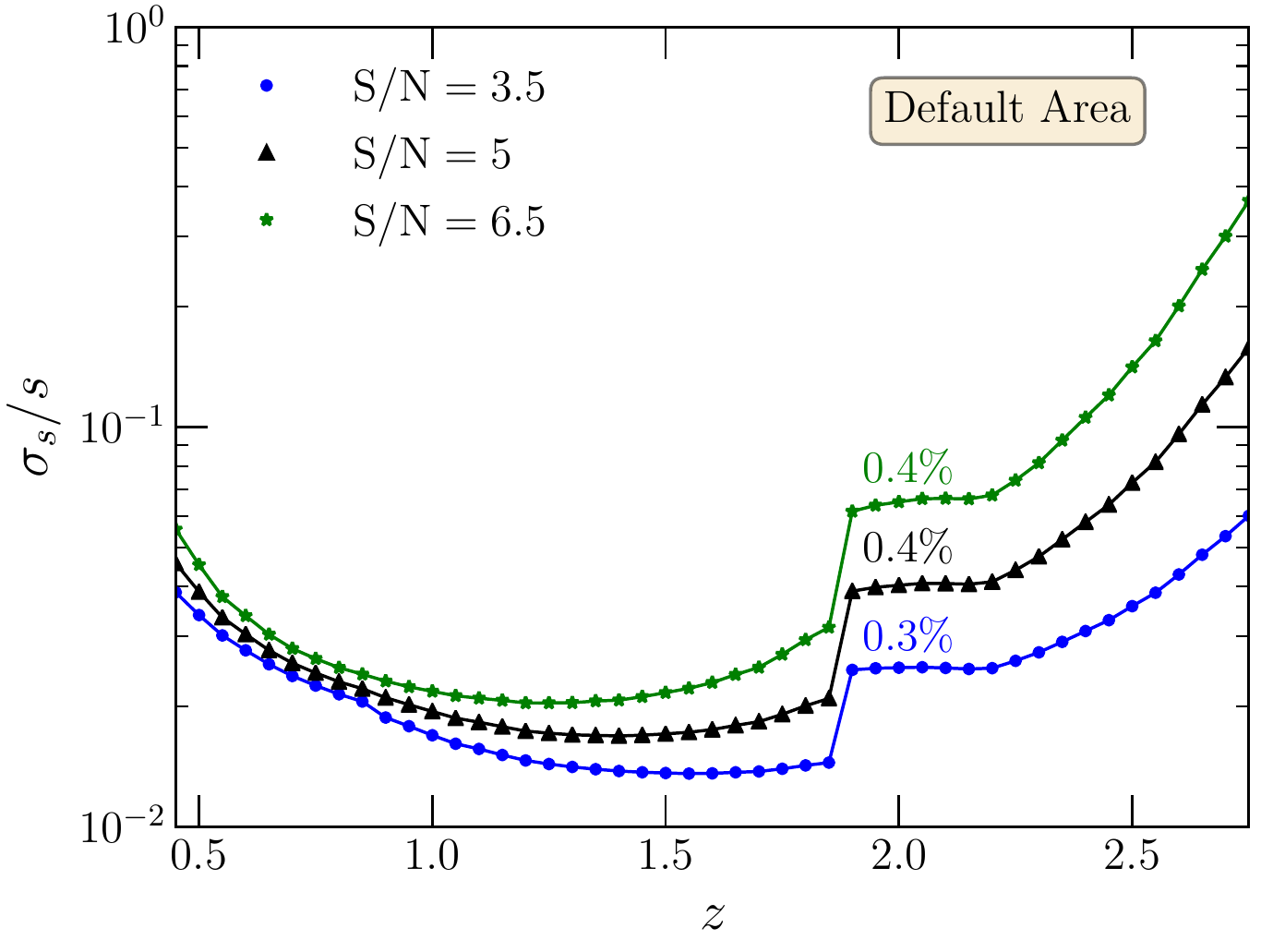}
\includegraphics[width=.45\textwidth]{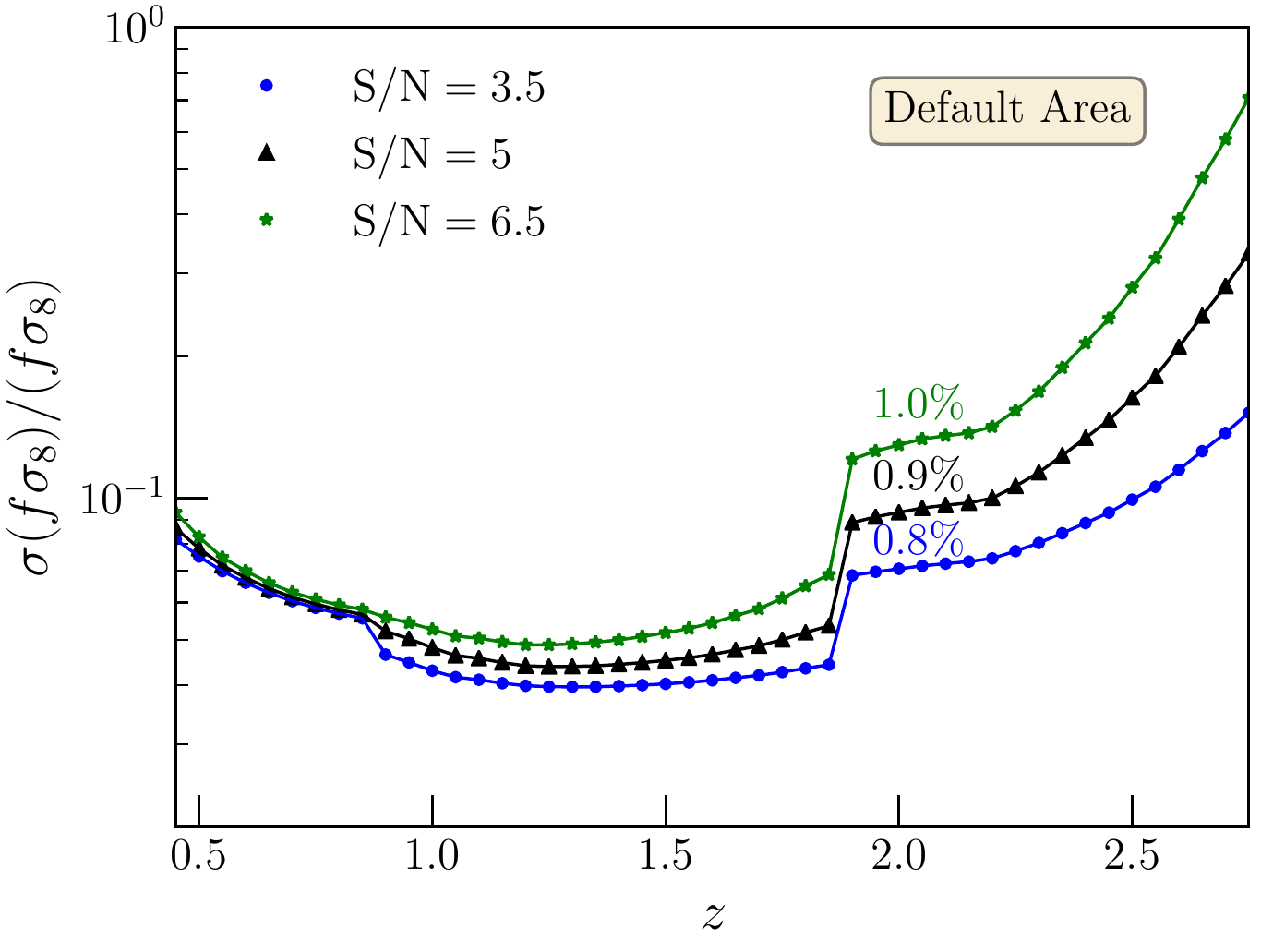}
\includegraphics[width=8.0cm]{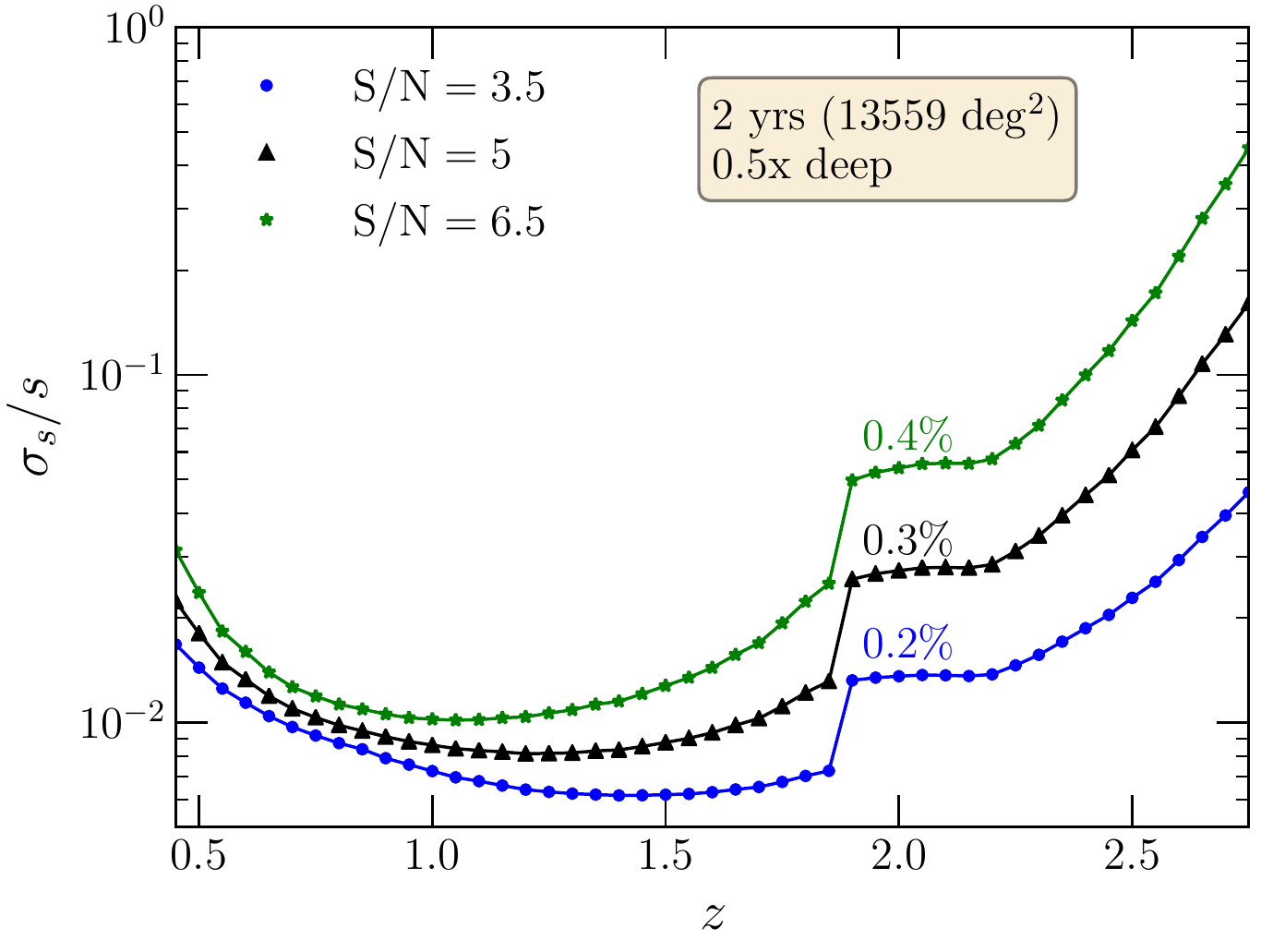}
\includegraphics[width=8.0cm]{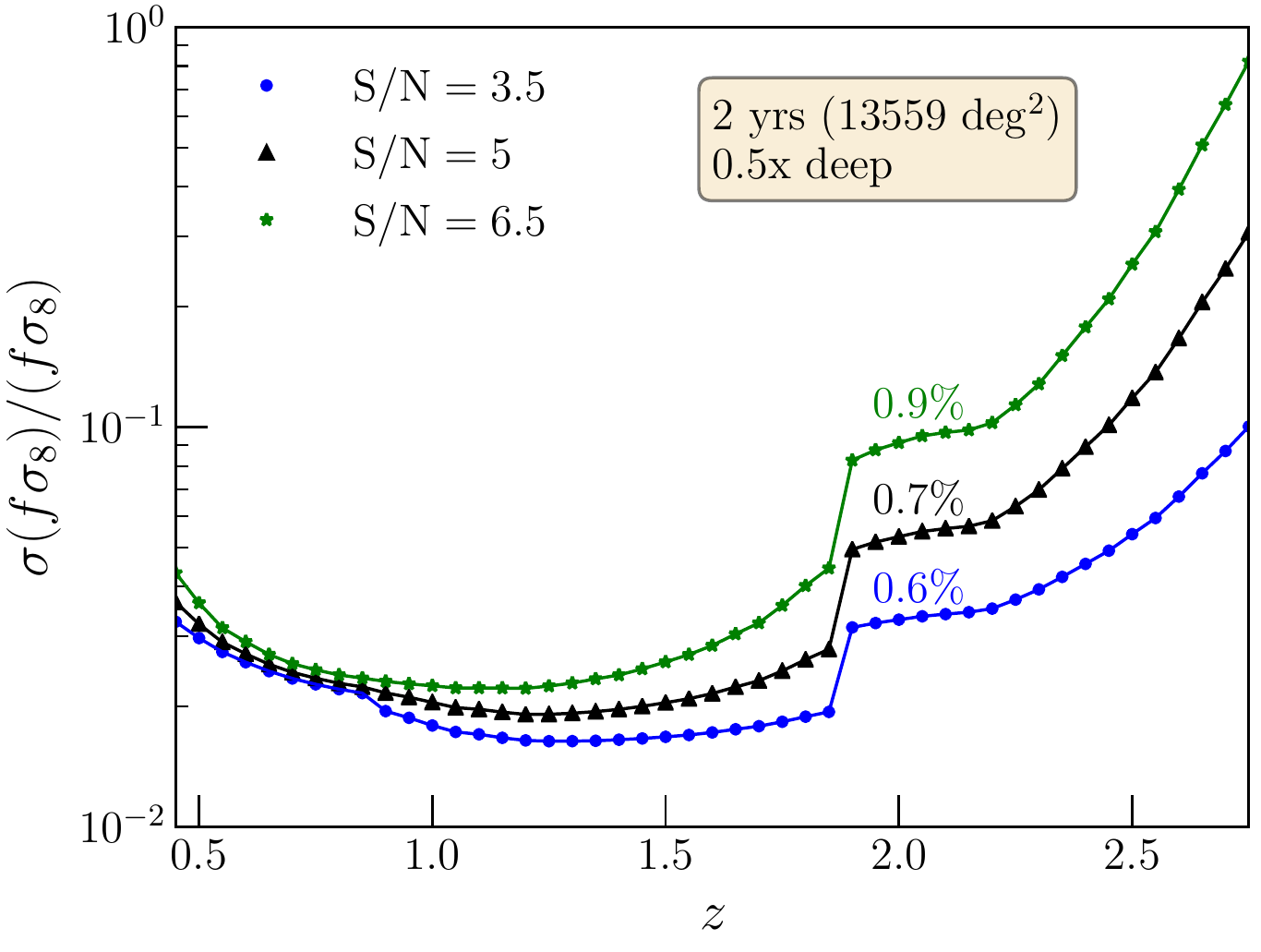}
\caption{For all rows in this plot we show the fractional error on the BAO scale (\textit{left}) and the error on the RSD parameter combination $f \sigma_8$ (\textit{right}) at a  redshift binwidth of $\textup{d}z = 0.05$. The aggregate fractional error over the entire redshift range is indicated near each curve.
\textit{Upper row} shows the results for a 0.6-year HLSS survey of H$_{\alpha}$+[N$_{\rm II}$] and [O$_{\rm III}$] galaxies; 
varying area and depth for a fixed default $S/N$ cutoff of 5. The default scenario (black) has $A =$ 2000 deg$^2$, the wide scenario (green) has twice the area but half the depth, whereas the deep scenario (blue) is twice deeper but half the area. For both the BAO and RSD probes, a wider but shallower survey improves the constraints for $z\lesssim 2$ whereas a deeper but narrower survey improves at $z\gtrsim 2$. \textit{Middle row} shows results when varying the $S/N$ cutoff (3.5, 5, 6.5) for the default area and depth scenario. A lower S/N cutoff yields better constraints everywhere in $z$, with more improvement at higher $z$. \textit{Bottom row} shows results when covering a larger area of 13559 deg$^2$ corresponding to an extended spectroscopic survey time of 2 years at half the default depth. We vary again the $S/N$ cutoff: 3.5, 5 and 6.5. }
\label{fi:grs}
\end{figure*}

In addition to the MCMC analysis in the previous subsection we explore the science return of the HLSS using a Fisher analysis on constraining the BAO scale $s$ and RSD parameter combination $f \sigma_8$ as a function of redshift.

For this analysis we run the ETC in BAO survey mode, using either galaxies observed in H$\alpha$ and [N$_{\rm II}$] (compilation option {\tt -NII}) or in [O$_{\rm III}$] ({\tt -DOIII\_GAL}) as tracers. For the H$\alpha$ and [N$_{\rm II}$] detections, we use model option 992, an average of the three galaxy luminosity functions given in \citet{phg16}, which were derived specifically for Euclid and WFIRST; in all cases, the [N$_{\rm II}$] luminosity function (used to enhance the signal-to-noise of detected galaxies) is assumed to be $0.37$ times the H$\alpha$ luminosity function. For the [O$_{\rm III}$] detections we use model 1992, an average of three luminosity functions: \citet{msc15} and \citet{cta13}, two different analyses of the WFC3 grism, and \citet{ksm15}, based on ground-based narrow-band surveys. In both the H$\alpha\,$+[\,N$_{\rm II}$] and [O$_{\rm III}$] scenarios, we use an updated galaxy size distribution from a mock catalog based on COSMOS data originally based on \citet{jki09}.

The resulting redshift distributions are shown in Fig.~\ref{fig:dndz}, which are used as input for the trade-off studies of the forecast results in this section. In each panel, we vary the survey depth from 0.5x, 1x to 2x the fiducial depth (shown in thick, normal and thin lines respectively), while the different panels show the distributions obtained with different $S/N$ cutoffs (6.5, 5 and 3.5). As expected, the number densities increase significantly when lower $S/N$ cutoffs are chosen. Note that in this section, we explore the impact of survey depth at fixed observation time, so the area of the survey is scaled proportionally in what follows.

Using each of the above-mentioned distributions, we compute the fractional error $\sigma_{p_i}/p_i = [F^{-1}]_{ii}$ on parameter $p_i$, where the Fisher information matrix for parameters $p_i$ and $p_j$ is given by
\be
F_{ij}= \int_{k_{\rm min}}^{k_{\rm max}}
\frac{\partial\ln P_g(\bfk)}{\partial p_i}
\frac{\partial\ln P_g(\bfk)}{\partial p_j}\,
V_{\rm eff}(\bfk)\, \frac{{\rm d} \bfk^3}{2\, (2\pi)^3},
\label{eq:Fisher_BAO}
\ee
assuming spatially constant galaxy density $n$, we have
\bea
V_{\rm eff}(k,\mu)
%&\equiv& \int d\bfr^3 \left[ \frac{n(\bfr) P_g(k,\mu)}
%{ n(\bfr) P_g(k,\mu)+1} \right]^2\nonumber\\
&=&\left[ \frac{ n P_g(k,\mu)}{n P_g(k,\mu)+1} \right]^2 V_{\rm survey} \,.
\eea
There are two separate Fisher matrices, one for the RSD cosntraints on $f\sigma_8$, and another for the BAO constraints on $s$. For the RSD constraint, we follow \cite{2009JCAP...10..007M} (using only one tracer) and model the observed galaxy power spectrum as in Eq. (\ref{eq:Pg_seo}) but without the distance ratios for changing cosmology as we fix the background cosmology
\bea
\label{eq:Pg_rsd_forecast}
P_{g}(k^{\rm ref}_{\perp},k^{\rm ref}_{\parallel}) &=&
 b^2 \left( 1+\beta\, \mu^2 \right)^2 \\
&\times& \left[ \frac{G(z)}{G(z=0)}\right]^2 \, P_m(k, z=0)
e^{-k^2\mu^2\sigma_{r,z}^2} + P_{\rm shot}\,.
\nonumber
\eea
and we marginalize over $\sigma_{r,z} = \sigma_{r,v} (1+z)/H(z)$. We adopt the fiducial value of $\sigma_{r,v} = 0.001$ which is dominated by the observational redshift uncertainty of the grism. Furthermore, for the RSD forecast, we assume perfect reconstruction with $k_* = \infty$.

For the BAO constraints, we calculate errors for the Hubble parameter $H$ and the angular diameter distance $D$ and report their best constrained combination $s$. Again we use Eq. (\ref{eq:Fisher_BAO}) but this time, modeling the galaxy power spectrum as defined in Eq.~\ref{eq:Pg_rsd_forecast}
%Eqs. (\ref{eq:Pg_seo}) and (\ref{eq:dwtransfer}), 
with the following differences: First, the fractional reconstruction capability $p_{\rm NL}$ is set by how well the displacements can be determined given the level of shot noise in the data in linear theory. Second, the nonlinear damping of the BAO wiggles is modelled with a slightly different $k_*^{-1} = 8.0\, \mathrm{Mpc}/h \times (\sigma_8/0.8)\, p_{\rm NL}$. 
%14.9 Mpc * \sigma_8(z) = 8.0 \mathrm{Mpc}/h \times (\sigma_8(z)/0.8)
Finally $\sigma_{r,z}$ is not marginalized for the BAO forecast but is fixed at the same fiducial value mentioned above.

For both BAO and RSD forecasts, we use the inverse galaxy number density for the galaxy shot noise, and the same linear bias model as in~\citet{DESI16} for emission line galaxies (ELG) as is appropriate for the WFIRST GRS: $b_\mr{ELG}(z)D(z) = 0.84$, where $D(z)$ is the growth factor normalized at $z =0$.

% The Fisher matrices are computed at a fixed flat cosmology consistent with Planck 2015 best-fit baseline model 2.6 \cite{aaa16}.
% with matter density $\Omega_m h^2=0.1427$, baryon density $\Omega_b h^2 = 0.02225$, Hubble constant $h_0 = 0.6727$, primordial curvature power spectral amplitude $A_s = 2.207 \times 10^9$, and spectral tilt $n_s = 0.9645$. 

%where $H(z)=\dot{a}/{a}$ (with $a$ denoting the %cosmic scale factor) is
%the Hubble parameter, and $D_A(z)=r(z)/(1+z)$ is the angular diameter distance at $z$,
%with the comoving distance $r(z)$ given by
%\beq
%\label{eq:r(z)}
% r(z)=c\, |\Omega_k|^{-1/2} {\rm sinn}\left[|\Omega_k|^{1/2}\, 
%\int_0^z\frac{dz'}{H(z')}\right],
%\eeq
%where ${\rm sinn}(x)=\sin(x)$, $x$, $\sinh(x)$ for 
%$\Omega_k<0$, $\Omega_k=0$, and $\Omega_k>0$ respectively.
%In addition to the geometrical distortion, this model includes the linear galaxy bias and redshift-space distortion (RSD), nonlinear smearing of the BAO feature, halo shot noise, small-scale peculiar velocities, and redshift errors.

The Fisher matrices are computed at a fixed flat cosmology consistent with Planck 2015 best-fit (baseline model 2.6) \cite{aaa16} and we separately evaluate fractional errors on parameters for the H$_{\alpha}$+ [N$_{\rm II}$] and [O$_{\rm III}$] samples before inverse-variance combining them. In Fig.~\ref{fi:grs} we show the combined fractional error on the BAO scale $s$ (left) and RSD parameter $f\sigma_8$ (right). Note that the H$ _{\alpha}$ is the dominant sample up to $z \approx 1.9$, beyond which the [O$_{\rm III}$] sample becomes the only available sample.

We consider different survey strategies varying depth (top row of Fig.~\ref{fi:grs}) and $S/N$ (middle row) starting from a pilot survey with default area $A = 2000$ deg$^2$ and $S/N$ cutoff 5. We fix the total HLSS observation time to 0.6 years in all cases. In the top panels, we show results for a deeper (twice deeper, half the area) and a wider survey (twice the area, half the depth) compared to the pilot survey. For both $s$ and $f\sigma_8$, the wide survey would improve the low-$z$ constraints, whereas the deep survey is more powerful at higher $z$, as expected. 

Since the aggregate constraint (shown in text beside each curve) is dominated by better errors at low-$z$, the wide survey would improve on the total constraint on parameters compared to the deep survey (e.g. 0.3\% vs 0.4\% for $s$ and 0.7\% vs 1.1\% for $f\sigma_8$). On the other hand, if dark energy behaviour at higher $z$ becomes an important science case, the deep survey improves constraining power by almost a factor of $2-3$ over the wide option. 

In the middle row of Fig.~\ref{fi:grs}, we also show the impact of different $S/N$ cutoffs for galaxy detections at fixed area and depth. We compare our default case of $S/N=5$ with a conservative $S/N=6.5$, and a more optimistic $S/N$ cutoff of 3.5. As expected, a lower $S/N$ cutoff yields better constraints everywhere in $z$, with more improvement at higher $z$ as fainter and distant galaxies are more affected by the cut. There is factor of 2 improvement at high $z$ between the curves at $S/N = 6.5$ and 5. The same is true for 5 and 3.5, we however note that $S/N = 3.5$ is not likely going to be a realistic value for reliable detections.

We perform a similar analysis but for an extended HLSS survey that lasts 2 years instead of 0.6 years and at only half the depth of the pilot survey, which allows us to survey 13,559 deg$^2$ (see bottom row of Fig. \ref{fi:grs}). We show results for 3 different $S/N$ cut and again find unsurprisingly that a $S/N$ cut of 3.5 improves constraining power substantially compared to the more realistic $S/N=5$ and the conservative $S/N=6.5$ cuts.

%%%%%%%%%%%%%%%%%%%%%%%%%%%%%%%%%%%%%%%%%%%%%%%%%%
%%%%%%%%%%%%%%%%%%%%%%%%%%%%%%%%%%%%%%%%%%%%%%%%%%
\section{Conclusions}
\label{sec:conc}
%%%%%%%%%%%%%%%%%%%%%%%%%%%%%%%%%%%%%%%%%%%%%%%%%%
%%%%%%%%%%%%%%%%%%%%%%%%%%%%%%%%%%%%%%%%%%%%%%%%%%

WFIRST's wide-field instrument will join the concert of cosmological endeavors after DESI, LSST, SPHEREx, and Euclid have already made initial measurements. These measurements will inform the design of an optimal WFIRST survey, which can be finalized shortly before launch. The unique versatility of WFIRST's wide-field instrument, ranging from multi-band imaging to high-resolution slitless spectroscopy, in combination with the fact that WFIRST carries enough propellant for at least 10 years of observations with no active cryogens, make it an ideal observatory to flexibly target the most interesting science aspects after its launch in the mid 2020s.

In this paper we study the WFIRST reference survey's science return on dark energy, structure growth, and modified gravity accounting for a variety of observational systematics. We present results for the joint analysis of weak lensing, galaxy clustering (photometric), galaxy cluster number counts, BAO and RSD features in the spectroscopic clustering power spectrum, and combine this with SNIa information from WFIRST \citep[as detailed in][]{hdf18}. We outline strategies for optimizing WFIRST's science return and to identify and retire risks from systematic effects early. 

For each cosmological probe examined in this paper we identify important areas of future research to further increase the level of realism of our WFIRST simulations, to improve the parameterization of systematics, or to shrink the prior range on existing parameterizations. For example, we postpone modeling and mitigation of baryons \citep[e.g.,][]{dsb11,ekd15,crd18,hem19,cmj19} or intrinsic galaxy alignment \citep[e.g.,][]{his04, mhi06, job10, keb16, vcs19, bmt19, sbt19} for lensing based measurements to future studies; a decision that is in part driven by the fact that these uncertainties have different levels of modeling maturity for the different probes considered in this paper. We explore corresponding uncertainties in a companion paper \cite{esk20}, which focusses on 3x2 (weak lensing and photometric galaxy clustering) synergies of WFIRST and LSST.

We impose conservative scale cuts on photometric clustering information due to uncertainties in modeling galaxy bias. Improved galaxy bias modeling for the spectroscopic and photometric galaxy clustering to include small scale information \citep[see e.g.,][]{isz19,sww20,wws20} should become another important area for WFIRST optimization. \cite{kre17} have explored a Halo Occupation Density model to access small scale information in a similarly high-dimensional parameter space (but simulating an LSST 3x2 analysis), and found that tapping into corresponding information is worth the increased modeling complexity.  

Our modeling of the cluster mass observable relation is based on \cite{mon19} but extended to account for possible redshift dependence in the scatter of the mass-richness relation. This again is a conservative choice and tightening priors on the existing parameterization or improving the parameterization itself can significantly change the constraining power from galaxy clusters. Precise modeling of cluster cosmology is an active research field \citep[e.g. see][]{crs19,DESclusters20} and studying multi-wavelength strategies including external data sets will be important.

We quantify all statements in this paper using the well-known FoM metric, however we note that the FoM metric reduces a complex answer to a one-dimensional statement. This compression of information is not lossless, for example the FoM depends on analysis choices: scales considered and excluded in the analysis, redshift distribution binning choices, cosmological parameters and priors, systematics parameterization and priors, which covariance and cross-correlations to include, and how to model the covariance in general, which external data sets to include, are all choices by the analyst. Multiple options are justifiable and for some the impact on the FoM can be significant. 

While the decision on the optimal WFIRST survey strategy can be made shortly before launch, it is critical to develop realistic survey simulation capabilities now in order to characterize the trade space of statistical power and systematic dangers accurately. Some of these systematics will have subdominant uncertainties, which means they can be corrected and need no further parameterization in a likelihood analysis. This type of systematics will hardly change the error bars presented in this paper, it will only move the best-fit value in a likelihood analysis based on data. 

It is important to note that complexity of modeling and covariance code such as the one used in this paper will become a challenge for the community. Increased complexity in a prediction and later in an analysis framework does not automatically increase the precision but it certainly increases the potential for errors. Increased model complexity for systematics must to be rigorously justified by residual uncertainties that are non-negligible, given the constraining power of the survey. This requires a demonstration of the impact of the systematic effect in the presence of a realistic systematics budget overall; it is not sufficient to demonstrate the impact of the systematic as a standalone effect on cosmological parameters. 

This work contributes to developing such a framework for WFIRST, but several extensions are forthcoming in future work. More realistic systematics models, best informed by actual observations and realistic synergy studies across the whole spectrum of multi-messenger astronomy, which includes optical NIR imaging and spectroscopy but also Cosmic Microwave Background, gravitational waves, and radio observations, should be considered to design a survey that fully utilize WFIRST's potential.

\section*{Acknowledgments}
\textcopyright 2020. All rights reserved. This work is supported by NASA ROSES ATP 16-ATP16-0084 and NASA 15-WFIRST15-0008 grants. The Flatiron Institute is supported by the Simons Foundation. Simulations in this paper use High Performance Computing (HPC) resources supported by the University of Arizona TRIF, UITS, and RDI and maintained by the UA Research Technologies department. Part of the research described in this paper was carried out at the Jet Propulsion Laboratory, California Institute of Technology, under a contract with the National Aeronautics and Space Administration. HM has been supported by Grant-in-Aid for Scientific Research from the JSPS Promotion of Science (No.~18H04350, No.~18K13561, and No.~19H05100) and World Premier International Research Center Initiative (WPI), MEXT, Japan.

\label{lastpage}
\bibliographystyle{mnras}
\bibliography{references}

\begin{thebibliography}{}
\makeatletter
\relax
\def\mn@urlcharsother{\let\do\@makeother \do\$\do\&\do\#\do\^\do\_\do\%\do\~}
\def\mn@doi{\begingroup\mn@urlcharsother \@ifnextchar [ {\mn@doi@}
  {\mn@doi@[]}}
\def\mn@doi@[#1]#2{\def\@tempa{#1}\ifx\@tempa\@empty \href
  {http://dx.doi.org/#2} {doi:#2}\else \href {http://dx.doi.org/#2} {#1}\fi
  \endgroup}
\def\mn@eprint#1#2{\mn@eprint@#1:#2::\@nil}
\def\mn@eprint@arXiv#1{\href {http://arxiv.org/abs/#1} {{\tt arXiv:#1}}}
\def\mn@eprint@dblp#1{\href {http://dblp.uni-trier.de/rec/bibtex/#1.xml}
  {dblp:#1}}
\def\mn@eprint@#1:#2:#3:#4\@nil{\def\@tempa {#1}\def\@tempb {#2}\def\@tempc
  {#3}\ifx \@tempc \@empty \let \@tempc \@tempb \let \@tempb \@tempa \fi \ifx
  \@tempb \@empty \def\@tempb {arXiv}\fi \@ifundefined
  {mn@eprint@\@tempb}{\@tempb:\@tempc}{\expandafter \expandafter \csname
  mn@eprint@\@tempb\endcsname \expandafter{\@tempc}}}

\bibitem[\protect\citeauthoryear{{Abazajian} et~al.,}{{Abazajian}
  et~al.}{2016}]{CMBS4}
{Abazajian} K.~N.,  et~al., 2016, arXiv e-prints, \href
  {https://ui.adsabs.harvard.edu/abs/2016arXiv161002743A} {p. arXiv:1610.02743}

\bibitem[\protect\citeauthoryear{Ade et~al.,}{Ade et~al.}{2016}]{aaa16}
Ade P. A.~R.,  et~al., 2016, \mn@doi [\aap] {10.1051/0004-6361/201525830}, 594,
  A13

\bibitem[\protect\citeauthoryear{{Ade} et~al.,}{{Ade} et~al.}{2019}]{SO19}
{Ade} P.,  et~al., 2019, \mn@doi [\jcap] {10.1088/1475-7516/2019/02/056}, \href
  {https://ui.adsabs.harvard.edu/abs/2019JCAP...02..056A} {2019, 056}

\bibitem[\protect\citeauthoryear{{Angulo}, {Baugh}, {Frenk}  \&
  {Lacey}}{{Angulo} et~al.}{2008}]{abf08}
{Angulo} R.~E.,  {Baugh} C.~M.,  {Frenk} C.~S.,   {Lacey} C.~G.,  2008, \mn@doi
  [\mnras] {10.1111/j.1365-2966.2007.12587.x}, \href
  {https://ui.adsabs.harvard.edu/abs/2008MNRAS.383..755A} {383, 755}

\bibitem[\protect\citeauthoryear{{Bernstein} \& {Jarvis}}{{Bernstein} \&
  {Jarvis}}{2002}]{bej02}
{Bernstein} G.~M.,  {Jarvis} M.,  2002, \mn@doi [\aj] {10.1086/338085}, \href
  {http://adsabs.harvard.edu/abs/2002AJ....123..583B} {123, 583}

\bibitem[\protect\citeauthoryear{{Bhattacharya}, {Habib}, {Heitmann}  \&
  {Vikhlinin}}{{Bhattacharya} et~al.}{2013}]{bhh11}
{Bhattacharya} S.,  {Habib} S.,  {Heitmann} K.,   {Vikhlinin} A.,  2013,
  \mn@doi [\apj] {10.1088/0004-637X/766/1/32}, \href
  {http://adsabs.harvard.edu/abs/2013ApJ...766...32B} {766, 32}

\bibitem[\protect\citeauthoryear{{Blazek}, {Vlah}  \& {Seljak}}{{Blazek}
  et~al.}{2015}]{bvs15}
{Blazek} J.,  {Vlah} Z.,   {Seljak} U.,  2015, \mn@doi [\jcap]
  {10.1088/1475-7516/2015/08/015}, \href
  {http://adsabs.harvard.edu/abs/2015JCAP...08..015B} {8, 015}

\bibitem[\protect\citeauthoryear{{Blazek}, {MacCrann}, {Troxel}  \&
  {Fang}}{{Blazek} et~al.}{2019}]{bmt19}
{Blazek} J.~A.,  {MacCrann} N.,  {Troxel} M.~A.,   {Fang} X.,  2019, \mn@doi
  [\prd] {10.1103/PhysRevD.100.103506}, \href
  {https://ui.adsabs.harvard.edu/abs/2019PhRvD.100j3506B} {100, 103506}

\bibitem[\protect\citeauthoryear{{Chisari} et~al.,}{{Chisari}
  et~al.}{2015}]{ccl15}
{Chisari} N.,  et~al., 2015, \mn@doi [\mnras] {10.1093/mnras/stv2154}, \href
  {https://ui.adsabs.harvard.edu/abs/2015MNRAS.454.2736C} {454, 2736}

\bibitem[\protect\citeauthoryear{{Chisari} et~al.,}{{Chisari}
  et~al.}{2018}]{crd18}
{Chisari} N.~E.,  et~al., 2018, \mn@doi [\mnras] {10.1093/mnras/sty2093}, \href
  {https://ui.adsabs.harvard.edu/abs/2018MNRAS.480.3962C} {480, 3962}

\bibitem[\protect\citeauthoryear{{Chisari} et~al.,}{{Chisari}
  et~al.}{2019}]{cmj19}
{Chisari} N.~E.,  et~al., 2019, \mn@doi [The Open Journal of Astrophysics]
  {10.21105/astro.1905.06082}, \href
  {https://ui.adsabs.harvard.edu/abs/2019OJAp....2E...4C} {2, 4}

\bibitem[\protect\citeauthoryear{{Colbert} et~al.,}{{Colbert}
  et~al.}{2013}]{cta13}
{Colbert} J.~W.,  et~al., 2013, \mn@doi [\apj] {10.1088/0004-637X/779/1/34},
  \href {https://ui.adsabs.harvard.edu/abs/2013ApJ...779...34C} {779, 34}

\bibitem[\protect\citeauthoryear{{Costanzi} et~al.,}{{Costanzi}
  et~al.}{2019}]{crs19}
{Costanzi} M.,  et~al., 2019, \mn@doi [\mnras] {10.1093/mnras/stz1949}, \href
  {https://ui.adsabs.harvard.edu/abs/2019MNRAS.488.4779C} {488, 4779}

\bibitem[\protect\citeauthoryear{{DES Collaboration} et~al.,}{{DES
  Collaboration} et~al.}{2020}]{DESclusters20}
{DES Collaboration} et~al., 2020, arXiv e-prints, \href
  {https://ui.adsabs.harvard.edu/abs/2020arXiv200211124D} {p. arXiv:2002.11124}

\bibitem[\protect\citeauthoryear{{DESI Collaboration} et~al.,}{{DESI
  Collaboration} et~al.}{2016}]{DESI16}
{DESI Collaboration} et~al., 2016, arXiv e-prints, \href
  {https://ui.adsabs.harvard.edu/abs/2016arXiv161100036D} {p. arXiv:1611.00036}

\bibitem[\protect\citeauthoryear{{Dodelson} \& {Schneider}}{{Dodelson} \&
  {Schneider}}{2013}]{dos13}
{Dodelson} S.,  {Schneider} M.~D.,  2013, \mn@doi [\prd]
  {10.1103/PhysRevD.88.063537}, \href
  {http://adsabs.harvard.edu/abs/2013PhRvD..88f3537D} {88, 063537}

\bibitem[\protect\citeauthoryear{{Dor{\'e}} et~al.,}{{Dor{\'e}}
  et~al.}{2014}]{dba14}
{Dor{\'e}} O.,  et~al., 2014, arXiv e-prints, \href
  {https://ui.adsabs.harvard.edu/abs/2014arXiv1412.4872D} {p. arXiv:1412.4872}

\bibitem[\protect\citeauthoryear{{Eifler}, {Krause}, {Dodelson}, {Zentner},
  {Hearin}  \& {Gnedin}}{{Eifler} et~al.}{2015}]{ekd15}
{Eifler} T.,  {Krause} E.,  {Dodelson} S.,  {Zentner} A.~R.,  {Hearin} A.~P.,
  {Gnedin} N.~Y.,  2015, \mn@doi [\mnras] {10.1093/mnras/stv2000}, \href
  {https://ui.adsabs.harvard.edu/abs/2015MNRAS.454.2451E} {454, 2451}

\bibitem[\protect\citeauthoryear{{Eifler}, {Simet}, {Krause}, {Hirata}  \&
  {Huang}}{{Eifler} et~al.}{2020}]{esk20}
{Eifler} T.~F.,  {Simet} M.,  {Krause} E.,  {Hirata} C.,   {Huang} H.,  2020,
  preprint

\bibitem[\protect\citeauthoryear{{Einstein}}{{Einstein}}{1917}]{ein17}
{Einstein} A.,  1917, Sitzungsberichte der K{\"o}niglich Preu{\ss}ischen
  Akademie der Wissenschaften (Berlin), \href
  {https://ui.adsabs.harvard.edu/abs/1917SPAW.......142E} {pp 142--152}

\bibitem[\protect\citeauthoryear{{Eisenstein}, {Seo}  \& {White}}{{Eisenstein}
  et~al.}{2007}]{esw07}
{Eisenstein} D.~J.,  {Seo} H.-J.,   {White} M.,  2007, \mn@doi [\apj]
  {10.1086/518755}, \href
  {https://ui.adsabs.harvard.edu/abs/2007ApJ...664..660E} {664, 660}

\bibitem[\protect\citeauthoryear{{Fang}, {Krause}, {Eifler}  \&
  {MacCrann}}{{Fang} et~al.}{2019}]{2019arXiv191111947F}
{Fang} X.,  {Krause} E.,  {Eifler} T.,   {MacCrann} N.,  2019, arXiv e-prints,
  \href {https://ui.adsabs.harvard.edu/abs/2019arXiv191111947F} {p.
  arXiv:1911.11947}

\bibitem[\protect\citeauthoryear{{Foreman-Mackey}, {Hogg}, {Lang}  \&
  {Goodman}}{{Foreman-Mackey} et~al.}{2013}]{fhg13}
{Foreman-Mackey} D.,  {Hogg} D.~W.,  {Lang} D.,   {Goodman} J.,  2013, \mn@doi
  [\pasp] {10.1086/670067}, \href
  {http://adsabs.harvard.edu/abs/2013PASP..125..306F} {125, 306}

\bibitem[\protect\citeauthoryear{Goodman \& Weare}{Goodman \&
  Weare}{2010}]{gow10}
Goodman J.,  Weare J.,  2010, \mn@doi [Communications in Applied Mathematics
  and Computational Science] {10.2140/camcos.2010.5.65}, 5, 65

\bibitem[\protect\citeauthoryear{{Grogin} et~al.,}{{Grogin}
  et~al.}{2011}]{Candels}
{Grogin} N.~A.,  et~al., 2011, \mn@doi [\apjs] {10.1088/0067-0049/197/2/35},
  \href {https://ui.adsabs.harvard.edu/\#abs/2011ApJS..197...35G} {197, 35}

\bibitem[\protect\citeauthoryear{{Hemmati} et~al.,}{{Hemmati}
  et~al.}{2019}]{hcm19}
{Hemmati} S.,  et~al., 2019, \mn@doi [\apj] {10.3847/1538-4357/ab1be5}, \href
  {https://ui.adsabs.harvard.edu/abs/2019ApJ...877..117H} {877, 117}

\bibitem[\protect\citeauthoryear{{Hirata} \& {Seljak}}{{Hirata} \&
  {Seljak}}{2004}]{his04}
{Hirata} C.~M.,  {Seljak} U.,  2004, \mn@doi [\prd]
  {10.1103/PhysRevD.70.063526}, \href
  {http://adsabs.harvard.edu/abs/2004PhRvD..70f3526H} {70, 063526}

\bibitem[\protect\citeauthoryear{{Hirata}, {Gehrels}, {Kneib}, {Kruk},
  {Rhodes}, {Wang}  \& {Zoubian}}{{Hirata} et~al.}{2012}]{hgk12}
{Hirata} C.~M.,  {Gehrels} N.,  {Kneib} J.-P.,  {Kruk} J.,  {Rhodes} J.,
  {Wang} Y.,   {Zoubian} J.,  2012, arXiv:1204.5151, \href
  {https://ui.adsabs.harvard.edu/\#abs/2012arXiv1204.5151H} {}

\bibitem[\protect\citeauthoryear{{Hounsell} et~al.,}{{Hounsell}
  et~al.}{2018}]{hdf18}
{Hounsell} R.,  et~al., 2018, \mn@doi [\apj] {10.3847/1538-4357/aac08b}, \href
  {https://ui.adsabs.harvard.edu/abs/2018ApJ...867...23H} {867, 23}

\bibitem[\protect\citeauthoryear{Huang, Eifler, Mandelbaum  \& Dodelson}{Huang
  et~al.}{2019}]{hem19}
Huang H.-J.,  Eifler T.,  Mandelbaum R.,   Dodelson S.,  2019, \mn@doi [\mnras]
  {10.1093/mnras/stz1714}, 488, 1652

\bibitem[\protect\citeauthoryear{{Ivanov}, {Simonovi{\'c}}  \&
  {Zaldarriaga}}{{Ivanov} et~al.}{2019}]{isz19}
{Ivanov} M.~M.,  {Simonovi{\'c}} M.,   {Zaldarriaga} M.,  2019, arXiv e-prints,
  \href {https://ui.adsabs.harvard.edu/abs/2019arXiv190905277I} {p.
  arXiv:1909.05277}

\bibitem[\protect\citeauthoryear{{Ivezi{\'c}} et~al.,}{{Ivezi{\'c}}
  et~al.}{2019}]{LSST19}
{Ivezi{\'c}} {\v{Z}}.,  et~al., 2019, \mn@doi [\apj]
  {10.3847/1538-4357/ab042c}, \href
  {https://ui.adsabs.harvard.edu/abs/2019ApJ...873..111I} {873, 111}

\bibitem[\protect\citeauthoryear{{Joachimi} \& {Bridle}}{{Joachimi} \&
  {Bridle}}{2010}]{job10}
{Joachimi} B.,  {Bridle} S.~L.,  2010, \mn@doi [\aap]
  {10.1051/0004-6361/200913657}, \href
  {http://adsabs.harvard.edu/abs/2010A%26A...523A...1J} {523, A1}

\bibitem[\protect\citeauthoryear{{Jouvel} et~al.,}{{Jouvel}
  et~al.}{2009}]{jki09}
{Jouvel} S.,  et~al., 2009, \mn@doi [\aap] {10.1051/0004-6361/200911798}, \href
  {https://ui.adsabs.harvard.edu/abs/2009A&A...504..359J} {504, 359}

\bibitem[\protect\citeauthoryear{{Khostovan}, {Sobral}, {Mobasher}, {Best},
  {Smail}, {Stott}, {Hemmati}  \& {Nayyeri}}{{Khostovan} et~al.}{2015}]{ksm15}
{Khostovan} A.~A.,  {Sobral} D.,  {Mobasher} B.,  {Best} P.~N.,  {Smail} I.,
  {Stott} J.~P.,  {Hemmati} S.,   {Nayyeri} H.,  2015, \mn@doi [\mnras]
  {10.1093/mnras/stv1474}, \href
  {https://ui.adsabs.harvard.edu/abs/2015MNRAS.452.3948K} {452, 3948}

\bibitem[\protect\citeauthoryear{{Koekemoer} et~al.,}{{Koekemoer}
  et~al.}{2011}]{candels2}
{Koekemoer} A.~M.,  et~al., 2011, \mn@doi [\apjs] {10.1088/0067-0049/197/2/36},
  \href {https://ui.adsabs.harvard.edu/abs/2011ApJS..197...36K} {197, 36}

\bibitem[\protect\citeauthoryear{{Krause} \& {Eifler}}{{Krause} \&
  {Eifler}}{2017}]{kre17}
{Krause} E.,  {Eifler} T.,  2017, \mn@doi [\mnras] {10.1093/mnras/stx1261},
  \href {http://adsabs.harvard.edu/abs/2017MNRAS.470.2100K} {470, 2100}

\bibitem[\protect\citeauthoryear{{Krause}, {Eifler}  \& {Blazek}}{{Krause}
  et~al.}{2016}]{keb16}
{Krause} E.,  {Eifler} T.,   {Blazek} J.,  2016, \mn@doi [\mnras]
  {10.1093/mnras/stv2615}, \href
  {http://adsabs.harvard.edu/abs/2016MNRAS.456..207K} {456, 207}

\bibitem[\protect\citeauthoryear{Laureijs et~al.,}{Laureijs
  et~al.}{2011}]{laa11}
Laureijs R.,  et~al., 2011, Euclid Definition Study Report (\mn@eprint {arXiv}
  {1110.3193})

\bibitem[\protect\citeauthoryear{{Mandelbaum}, {Hirata}, {Ishak}, {Seljak}  \&
  {Brinkmann}}{{Mandelbaum} et~al.}{2006}]{mhi06}
{Mandelbaum} R.,  {Hirata} C.~M.,  {Ishak} M.,  {Seljak} U.,   {Brinkmann} J.,
  2006, \mn@doi [\mnras] {10.1111/j.1365-2966.2005.09946.x}, \href
  {http://adsabs.harvard.edu/abs/2006MNRAS.367..611M} {367, 611}

\bibitem[\protect\citeauthoryear{{McDonald} \& {Seljak}}{{McDonald} \&
  {Seljak}}{2009}]{2009JCAP...10..007M}
{McDonald} P.,  {Seljak} U.,  2009, \mn@doi [\jcap]
  {10.1088/1475-7516/2009/10/007}, \href
  {https://ui.adsabs.harvard.edu/abs/2009JCAP...10..007M} {2009, 007}

\bibitem[\protect\citeauthoryear{{Mehta} et~al.,}{{Mehta} et~al.}{2015}]{msc15}
{Mehta} V.,  et~al., 2015, \mn@doi [\apj] {10.1088/0004-637X/811/2/141}, \href
  {https://ui.adsabs.harvard.edu/abs/2015ApJ...811..141M} {811, 141}

\bibitem[\protect\citeauthoryear{{Murata} et~al.,}{{Murata}
  et~al.}{2019}]{mon19}
{Murata} R.,  et~al., 2019, \mn@doi [\pasj] {10.1093/pasj/psz092}, \href
  {https://ui.adsabs.harvard.edu/abs/2019PASJ...71..107M} {71, 107}

\bibitem[\protect\citeauthoryear{{Navarro}, {Frenk}  \& {White}}{{Navarro}
  et~al.}{1997}]{nfw97}
{Navarro} J.~F.,  {Frenk} C.~S.,   {White} S.~D.~M.,  1997, \mn@doi [\apj]
  {10.1086/304888}, \href {http://adsabs.harvard.edu/abs/1997ApJ...490..493N}
  {490, 493}

\bibitem[\protect\citeauthoryear{{Padmanabhan}, {Xu}, {Eisenstein}, {Scalzo},
  {Cuesta}, {Mehta}  \& {Kazin}}{{Padmanabhan} et~al.}{2012}]{pxe12}
{Padmanabhan} N.,  {Xu} X.,  {Eisenstein} D.~J.,  {Scalzo} R.,  {Cuesta} A.~J.,
   {Mehta} K.~T.,   {Kazin} E.,  2012, \mn@doi [\mnras]
  {10.1111/j.1365-2966.2012.21888.x}, \href
  {https://ui.adsabs.harvard.edu/abs/2012MNRAS.427.2132P} {427, 2132}

\bibitem[\protect\citeauthoryear{{Perlmutter} et~al.,}{{Perlmutter}
  et~al.}{1999}]{pag99}
{Perlmutter} S.,  et~al., 1999, \mn@doi [\apj] {10.1086/307221}, \href
  {https://ui.adsabs.harvard.edu/abs/1999ApJ...517..565P} {517, 565}

\bibitem[\protect\citeauthoryear{{Planck Collaboration} et~al.,}{{Planck
  Collaboration} et~al.}{2018}]{Planckcosmo18}
{Planck Collaboration} et~al., 2018, arXiv e-prints, \href
  {https://ui.adsabs.harvard.edu/abs/2018arXiv180706209P} {p. arXiv:1807.06209}

\bibitem[\protect\citeauthoryear{{Pozzetti} et~al.,}{{Pozzetti}
  et~al.}{2016}]{phg16}
{Pozzetti} L.,  et~al., 2016, \mn@doi [\aap] {10.1051/0004-6361/201527081},
  \href {https://ui.adsabs.harvard.edu/abs/2016A&A...590A...3P} {590, A3}

\bibitem[\protect\citeauthoryear{{Riess} et~al.,}{{Riess} et~al.}{1998}]{rfc98}
{Riess} A.~G.,  et~al., 1998, \mn@doi [\aj] {10.1086/300499}, \href
  {https://ui.adsabs.harvard.edu/abs/1998AJ....116.1009R} {116, 1009}

\bibitem[\protect\citeauthoryear{{Salcedo}, {Wibking}, {Weinberg}, {Wu},
  {Ferrer}, {Eisenstein}  \& {Pinto}}{{Salcedo} et~al.}{2020}]{sww20}
{Salcedo} A.~N.,  {Wibking} B.~D.,  {Weinberg} D.~H.,  {Wu} H.-Y.,  {Ferrer}
  D.,  {Eisenstein} D.,   {Pinto} P.,  2020, \mn@doi [\mnras]
  {10.1093/mnras/stz2963}, \href
  {https://ui.adsabs.harvard.edu/abs/2020MNRAS.491.3061S} {491, 3061}

\bibitem[\protect\citeauthoryear{{Samuroff} et~al.,}{{Samuroff}
  et~al.}{2019}]{sbt19}
{Samuroff} S.,  et~al., 2019, \mn@doi [\mnras] {10.1093/mnras/stz2197}, \href
  {https://ui.adsabs.harvard.edu/abs/2019MNRAS.489.5453S} {489, 5453}

\bibitem[\protect\citeauthoryear{{Schaan}, {Krause}, {Eifler}, {Dor{\'e}},
  {Miyatake}, {Rhodes}  \& {Spergel}}{{Schaan} et~al.}{2017}]{ske17}
{Schaan} E.,  {Krause} E.,  {Eifler} T.,  {Dor{\'e}} O.,  {Miyatake} H.,
  {Rhodes} J.,   {Spergel} D.~N.,  2017, \mn@doi [\prd]
  {10.1103/PhysRevD.95.123512}, \href
  {https://ui.adsabs.harvard.edu/abs/2017PhRvD..95l3512S} {95, 123512}

\bibitem[\protect\citeauthoryear{{Seljak}}{{Seljak}}{2000}]{sel00}
{Seljak} U.,  2000, \mn@doi [\mnras] {10.1046/j.1365-8711.2000.03715.x}, \href
  {http://adsabs.harvard.edu/abs/2000MNRAS.318..203S} {318, 203}

\bibitem[\protect\citeauthoryear{{Semboloni}, {Hoekstra}, {Schaye}, {van
  Daalen}  \& {McCarthy}}{{Semboloni} et~al.}{2011}]{shs11}
{Semboloni} E.,  {Hoekstra} H.,  {Schaye} J.,  {van Daalen} M.~P.,   {McCarthy}
  I.~G.,  2011, \mn@doi [\mnras] {10.1111/j.1365-2966.2011.19385.x}, \href
  {http://adsabs.harvard.edu/abs/2011MNRAS.417.2020S} {417, 2020}

\bibitem[\protect\citeauthoryear{{Seo} \& {Eisenstein}}{{Seo} \&
  {Eisenstein}}{2003}]{see03}
{Seo} H.-J.,  {Eisenstein} D.~J.,  2003, \mn@doi [\apj] {10.1086/379122}, \href
  {https://ui.adsabs.harvard.edu/abs/2003ApJ...598..720S} {598, 720}

\bibitem[\protect\citeauthoryear{{Seo} \& {Eisenstein}}{{Seo} \&
  {Eisenstein}}{2007}]{see07}
{Seo} H.-J.,  {Eisenstein} D.~J.,  2007, \mn@doi [\apj] {10.1086/519549}, \href
  {https://ui.adsabs.harvard.edu/abs/2007ApJ...665...14S} {665, 14}

\bibitem[\protect\citeauthoryear{{Simpson} et~al.,}{{Simpson}
  et~al.}{2013}]{shp13}
{Simpson} F.,  et~al., 2013, \mn@doi [\mnras] {10.1093/mnras/sts493}, \href
  {https://ui.adsabs.harvard.edu/abs/2013MNRAS.429.2249S} {429, 2249}

\bibitem[\protect\citeauthoryear{{Singh}, {Mandelbaum}  \& {More}}{{Singh}
  et~al.}{2015}]{smm14}
{Singh} S.,  {Mandelbaum} R.,   {More} S.,  2015, \mn@doi [\mnras]
  {10.1093/mnras/stv778}, \href
  {http://adsabs.harvard.edu/abs/2015MNRAS.450.2195S} {450, 2195}

\bibitem[\protect\citeauthoryear{{Spergel} et~al.,}{{Spergel}
  et~al.}{2015}]{sgb15}
{Spergel} D.,  et~al., 2015, arXiv e-prints, \href
  {https://ui.adsabs.harvard.edu/abs/2015arXiv150303757S} {p. arXiv:1503.03757}

\bibitem[\protect\citeauthoryear{{Takada} et~al.,}{{Takada}
  et~al.}{2014}]{tec14}
{Takada} M.,  et~al., 2014, \mn@doi [\pasj] {10.1093/pasj/pst019}, \href
  {https://ui.adsabs.harvard.edu/abs/2014PASJ...66R...1T} {66, R1}

\bibitem[\protect\citeauthoryear{{Takahashi}, {Sato}, {Nishimichi}, {Taruya}
  \& {Oguri}}{{Takahashi} et~al.}{2012}]{tsn12}
{Takahashi} R.,  {Sato} M.,  {Nishimichi} T.,  {Taruya} A.,   {Oguri} M.,
  2012, \mn@doi [\apj] {10.1088/0004-637X/761/2/152}, \href
  {http://adsabs.harvard.edu/abs/2012ApJ...761..152T} {761, 152}

\bibitem[\protect\citeauthoryear{{Taylor}, {Joachimi}  \& {Kitching}}{{Taylor}
  et~al.}{2013}]{tjk13}
{Taylor} A.,  {Joachimi} B.,   {Kitching} T.,  2013, \mn@doi [\mnras]
  {10.1093/mnras/stt270}, \href
  {http://adsabs.harvard.edu/abs/2013MNRAS.432.1928T} {432, 1928}

\bibitem[\protect\citeauthoryear{{Tenneti}, {Singh}, {Mandelbaum}, {Matteo},
  {Feng}  \& {Khandai}}{{Tenneti} et~al.}{2015}]{tsm15}
{Tenneti} A.,  {Singh} S.,  {Mandelbaum} R.,  {Matteo} T.~D.,  {Feng} Y.,
  {Khandai} N.,  2015, \mn@doi [\mnras] {10.1093/mnras/stv272}, \href
  {http://adsabs.harvard.edu/abs/2015MNRAS.448.3522T} {448, 3522}

\bibitem[\protect\citeauthoryear{{Tinker}, {Robertson}, {Kravtsov}, {Klypin},
  {Warren}, {Yepes}  \& {Gottl{\"o}ber}}{{Tinker} et~al.}{2010}]{trk10}
{Tinker} J.~L.,  {Robertson} B.~E.,  {Kravtsov} A.~V.,  {Klypin} A.,  {Warren}
  M.~S.,  {Yepes} G.,   {Gottl{\"o}ber} S.,  2010, \mn@doi [\apj]
  {10.1088/0004-637X/724/2/878}, \href
  {http://adsabs.harvard.edu/abs/2010ApJ...724..878T} {724, 878}

\bibitem[\protect\citeauthoryear{{Troxel} \& {Ishak}}{{Troxel} \&
  {Ishak}}{2014}]{tri14}
{Troxel} M.~A.,  {Ishak} M.,  2014, preprint, \href
  {http://adsabs.harvard.edu/abs/2014arXiv1407.6990T} {} (\mn@eprint {arXiv}
  {1407.6990})

\bibitem[\protect\citeauthoryear{{Vlah}, {Chisari}  \& {Schmidt}}{{Vlah}
  et~al.}{2019}]{vcs19}
{Vlah} Z.,  {Chisari} N.~E.,   {Schmidt} F.,  2019, arXiv e-prints, \href
  {https://ui.adsabs.harvard.edu/abs/2019arXiv191008085V} {p. arXiv:1910.08085}

\bibitem[\protect\citeauthoryear{{Wang}, {Chuang}  \& {Hirata}}{{Wang}
  et~al.}{2013}]{wch13}
{Wang} Y.,  {Chuang} C.-H.,   {Hirata} C.~M.,  2013, \mn@doi [\mnras]
  {10.1093/mnras/stt068}, \href
  {https://ui.adsabs.harvard.edu/abs/2013MNRAS.430.2446W} {430, 2446}

\bibitem[\protect\citeauthoryear{{Wibking}, {Weinberg}, {Salcedo}, {Wu},
  {Singh}, {Rodr{\'\i}guez-Torres}, {Garrison}  \& {Eisenstein}}{{Wibking}
  et~al.}{2020}]{wws20}
{Wibking} B.~D.,  {Weinberg} D.~H.,  {Salcedo} A.~N.,  {Wu} H.-Y.,  {Singh} S.,
   {Rodr{\'\i}guez-Torres} S.,  {Garrison} L.~H.,   {Eisenstein} D.~J.,  2020,
  \mn@doi [\mnras] {10.1093/mnras/stz3423}, \href
  {https://ui.adsabs.harvard.edu/abs/2020MNRAS.492.2872W} {492, 2872}

\bibitem[\protect\citeauthoryear{{Zentner}, {Semboloni}, {Dodelson}, {Eifler},
  {Krause}  \& {Hearin}}{{Zentner} et~al.}{2013}]{zsd13}
{Zentner} A.~R.,  {Semboloni} E.,  {Dodelson} S.,  {Eifler} T.,  {Krause} E.,
  {Hearin} A.~P.,  2013, \mn@doi [\prd] {10.1103/PhysRevD.87.043509}, \href
  {http://adsabs.harvard.edu/abs/2013PhRvD..87d3509Z} {87, 043509}

\bibitem[\protect\citeauthoryear{{de Jong}}{{de Jong}}{2019}]{4MOST19}
{de Jong} R.~S.,  2019, \mn@doi [Nature Astronomy] {10.1038/s41550-019-0808-x},
  \href {https://ui.adsabs.harvard.edu/abs/2019NatAs...3..574D} {3, 574}

\bibitem[\protect\citeauthoryear{{van Daalen}, {Schaye}, {Booth}  \& {Dalla
  Vecchia}}{{van Daalen} et~al.}{2011}]{dsb11}
{van Daalen} M.~P.,  {Schaye} J.,  {Booth} C.~M.,   {Dalla Vecchia} C.,  2011,
  \mn@doi [\mnras] {10.1111/j.1365-2966.2011.18981.x}, \href
  {http://adsabs.harvard.edu/abs/2011MNRAS.415.3649V} {415, 3649}

\makeatother
\end{thebibliography}

\end{document}